\documentclass[12pt]{article}
\usepackage{epsfig,amssymb,amsmath,psfrag}

\catcode`\@=11
\newcommand{\insertfig}[2]{\mbox{\epsfxsize=#1cm \epsfbox{#2.eps}}}
\font\cmss=cmss12 
\def\1{\hbox{{1}\kern-.25em\hbox{l}}}
\def\bfZ{\relax{\hbox{\cmss Z\kern-.4em Z}}}
\def \be  {\begin{equation}}
\def \ee  {\end{equation}}
\def \ba  {\begin{eqnarray}}
\def \ea  {\end{eqnarray}}
\def \baa {\begin{eqnarray*}}
\def \eaa {\end{eqnarray*}}
\def \bb  {\begin {thebibliography} }
\def \eb  {\end{thebibliography}}
\def \lab #1 {\label{#1}}

\newcommand\re[1]{(\ref{#1})}

\def \qqqquad {\qquad\qquad}
\def \matrix #1 {\left(\begin{array}{cc} #1 \end{array}\right)}
\def \S {\mathop{\rm\bf S}}

\def \tr {\mathop{\rm tr}\nolimits}
\def \Im {\mathop{\rm Im}\nolimits}

\def \e  {\mathop{\rm e}\nolimits}

\newcommand \vev [1] {\langle{#1}\rangle}

\newcommand \ket [1] {|{#1}\rangle}

\newcommand{\as}{\ifmmode\alpha_{\rm s}\else{$\alpha_{\rm s}$}\fi}
\newcommand{\asbar}{\ifmmode\bar{\alpha}_{\rm s}\else{$\bar{\alpha}_{\rm s}$}\fi}

\newcommand{\bit}[1]{\mbox{\boldmath$#1$}}
\newcommand{\ft}[2]{{\textstyle\frac{#1}{#2}}}

\font\cmss=cmss12 
\def\inbar{\,\vrule height1.5ex width.4pt depth0pt}
\def\IC{\relax\hbox{$\inbar\kern-.3em{\rm C}$}}
\def\IZ{\relax{\hbox{\cmss Z\kern-.4em Z}}}
\def\IR{{\hbox{{\rm I}\kern-.2em\hbox{\rm R}}}}

\def\IP{{\hbox{{\rm I}\kern-.2em\hbox{\rm P}}}}
\def\II{\hbox{{1}\kern-.25em\hbox{l}}}

\def\numberbysection{\@addtoreset{equation}{section}
                     \def\theequation{\thesection.\arabic{equation}}}
\numberbysection

\newbox\lett\newdimen\lheight\newdimen\lwidth
\def\ontop#1#2{\setbox\lett=\hbox{#2}\lheight\ht\lett
\multiply\lheight by 12 \divide\lheight by 10\relax%
\lwidth\wd\lett \multiply\lwidth by 8 \divide\lwidth by 10\relax #2\kern-\lwidth%
\raise\lheight\hbox{{$\scriptstyle #1$}}\kern.1ex}

\def\Xint#1{\mathchoice
   {\XXint\displaystyle\textstyle{#1}}%
   {\XXint\textstyle\scriptstyle{#1}}%
   {\XXint\scriptstyle\scriptscriptstyle{#1}}%
   {\XXint\scriptscriptstyle\scriptscriptstyle{#1}}%
   \!\int}
\def\XXint#1#2#3{{\setbox0=\hbox{$#1{#2#3}{\int}$}
     \vcenter{\hbox{$#2#3$}}\kern-.5\wd0}}

\def\dashint{\Xint-}

\psfrag{dots}[cc][cc]{$...$}
\psfrag{Z1}[cc][cc]{$\scriptstyle Z_1$}
\psfrag{Z2}[cc][cc]{$\scriptstyle Z_2$}
\psfrag{Z3}[cc][cc]{$ $}
\psfrag{ZN-1}[cc][cc]{$ $}
\psfrag{ZN}[cc][cc]{$\scriptstyle Z_L$}

\begin{document}

\begin{titlepage}
\begin{flushright}
\begin{tabular}{l}
LPT--Orsay--05--65
\end{tabular}
\end{flushright}

\vskip2cm

\centerline{\large \bf Integrability of two-loop dilatation operator in gauge theories}

\vspace{1cm}

\centerline{\sc A.V. Belitsky$^a$, G.P. Korchemsky$^b$, D. M\"uller$^a$}

\vspace{10mm}

\centerline{\it $^a$Department of Physics and Astronomy, Arizona State University}
\centerline{\it Tempe, AZ 85287-1504, USA}

\vspace{3mm}

\centerline{\it $^b$Laboratoire de Physique Th\'eorique\footnote{Unit\'e
                    Mixte de Recherche du CNRS (UMR 8627).},
                    Universit\'e de Paris XI}
\centerline{\it 91405 Orsay C\'edex, France}

\def\thefootnote{\fnsymbol{footnote}}%
\vspace{1cm}

\centerline{\bf Abstract}

\vspace{5mm}

We study the two-loop dilatation operator in the noncompact $SL(2)$ sector of QCD
and supersymmetric Yang-Mills theories with $\mathcal{N}=1,2,4$ supercharges. The
analysis is performed for Wilson operators built from three quark/gaugino fields
of the same helicity belonging to the fundamental/adjoint representation of the
$SU(3)$/$SU(N_c)$ gauge group and involving an arbitrary number of covariant
derivatives projected onto the light-cone. To one-loop order, the dilatation
operator inherits the conformal symmetry of the classical theory and is given in
the multi-color limit by a local Hamiltonian of the Heisenberg magnet with the
spin operators being generators of the collinear subgroup of full
(super)conformal group. Starting from two loops, the dilatation operator depends
on the representation of the gauge group and, in addition, receives corrections
stemming from the violation of the conformal symmetry. We compute its
eigenspectrum and demonstrate that to two-loop order integrability survives the
conformal symmetry breaking in the aforementioned gauge theories, but it is
violated in QCD by the contribution of nonplanar diagrams. In SYM theories with
extended supersymmetry, the $\mathcal{N}-$dependence of the two-loop dilatation
operator can be factorized (modulo an additive normalization constant) into a
multiplicative c-number. This property makes the eigenspectrum of the two-loop
dilatation operator alike in all gauge theories including the maximally
supersymmetric theory. Our analysis suggests that integrability is only tied to
the planar limit and it is sensitive neither to conformal symmetry nor
supersymmetry.

\end{titlepage}

\setcounter{footnote} 0

\thispagestyle{empty}

\newpage

\pagestyle{empty} {\small\tableofcontents}

\newpage

\pagestyle{plain} \setcounter{page} 1

%%%%%%%%%%%%%%%%%%%%%%%%%%%%%%%%%%%%%%%%%%%%%%%%%%%%%%%%%%%%%%%%%%%%%
\section{Introduction}
%%%%%%%%%%%%%%%%%%%%%%%%%%%%%%%%%%%%%%%%%%%%%%%%%%%%%%%%%%%%%%%%%%%%%

There is a growing amount of evidence that four-dimensional gauge theories
possess a hidden symmetry. The latter is not seen at the level of the classical
Lagrangian and manifests itself through remarkable integrability properties of
effective dynamics of the quantum theory. The symmetry has been first discovered
in studies of high-energy asymptotics of scattering amplitudes in multi-color
QCD \cite{Lip94,FadKor94} which led to the identification of the evolution
equation for the partial waves with the Schr\"odinger equation for the celebrated
Heisenberg magnet with spins being generators of the $SL (2, \mathbb{C})$ group.
Later, similar integrable structures have been observed in the renormalization
group (Callan-Symanzik) equations for multi-particle distributions in QCD
\cite{BraDerMan98,Bel99,DerKorMan00}. These distributions are given by
(nonperturbative) matrix elements of composite Wilson operators built from
elementary quark and gluon fields and an arbitrary number of covariant derivatives
in the totally symmetric representations of the conformal group $SO(4,2)$. The scale
dependence of the distribution amplitudes is driven by the anomalous dimensions of
these Wilson operators which, in their turn, are defined as eigenvalues of the
dilatation operator in QCD. It was found that the one-loop dilatation operator
in QCD in the sector of the so-called maximal-helicity operators can be mapped
in the multi-color limit into a Hamiltonian of the Heisenberg magnet with the
spin operators in the chain sites defined by the generators of the `collinear'
$SL(2 ,\mathbb{R})$ subgroup of the full conformal group
\cite{BraDerMan98,Bel99,DerKorMan00}. The phenomenon of integrability is not
merely of academic interest as it has a number of important phenomenological
applications in QCD (for a review see \cite{BelBraGorKor04}).

The above mentioned integrable structures are not specific to QCD. They are
also present in generic gauge theories including supersymmetric Yang-Mills
models with $\mathcal{N}=1,2,4$ supercharges. Supersymmetry enhances the
phenomenon by extending integrability to a larger class of observables.
In this context, the maximally supersymmetric $\mathcal{N}=4$ Yang-Mills
theory is of a special interest with regards to the AdS/CFT correspondence
\cite{Mal97}. The gauge/string duality establishes the correspondence between
this superconformal Yang-Mills theory and the type IIB string theory on the
AdS$_5 \times$S$^5$ background \cite{Mal97}. In particular, the anomalous
dimensions of composite Wilson operators in $\mathcal{N}=4$ gauge theory are
mapped into the energies of string excitations in the dual string picture
\cite{Pol01,BerMalNas02,GubKlePol03,FroTse03}. The string sigma models on
the anti-de Sitter space are known to possess an infinite set of conserved
currents and thus turn out to be completely integrable. This was shown
classically to be the case for the critical AdS$_5 \times$S$^5$ string action
\cite{ManSurWad02,BeRoiPol03} as well as for noncritical sigma models with
the anti-de Sitter as a factor of the target space \cite{Pol04}, which are
dual to particular (non-)supersymmetric Yang-Mills theories. The gauge/string
duality hints that these structures should manifest themselves through hidden
symmetries of the spectra of anomalous dimensions of Wilson operators in dual
gauge theories to all orders of perturbation theory.

At present, all studies firmly point out that the one-loop dilatation operator
in certain sectors of QCD and supersymmetric Yang-Mills (SYM) theories with
$\mathcal{N} < 4$ supercharges possesses integrability properties. In maximally
supersymmetric theory the phenomenon gets extended to all Wilson operators
\cite{MinZar02}. A natural question arises whether integrability persists to
higher loops as suggested by the gauge/string duality. Recently a great deal
of activity has been devoted to the test of higher loop integrability in the
$\mathcal{N} = 4$ theory and significant evidence has been gathered supporting
its presence in various closed `compact' subsectors specific to this particular
theory (see Ref.\ \cite{Bei04} for a review and Ref.\ \cite{Ple05} for recent
developments). These sectors embrace composite operators with a nontrivial
$R-$charge for which the corresponding mixing matrices have a finite size
determined by the number of constituent fields which build up the former.
They do not comprise however Wilson operators involving an arbitrary number
of covariant derivatives.%
\footnote{The exception is the study in Ref.\ \cite{Ede04} of spin-three scalar
operators with three covariant derivatives.} For such operators, the mixing
matrices could have an arbitrary large size and they define the `noncompact'
sector. In distinction with the compact sectors, which are intrinsic to models
with a nontrivial $R-$symmetry group, the noncompact ones are present in all
gauge theories including nonsupersymmetric ones. This offers one an opportunity
to compare the properties of the dilatation operators in various gauge theories
in the same noncompact sector and identify the origin of the symmetry responsible
for the existence of integrability.

In this paper we will study the dilatation operator in the noncompact $SL(2)$
sector for the aligned-helicity fermionic operators in QCD and supersymmetric
Yang-Mills theories with $\mathcal{N}=1,2,4$ supercharges.%
\footnote{A preliminary account of this work was published in a letter
\cite{BelKorMul04}.} These operators are constructed from quark/gaugino fields of
helicity $+\frac{1}{2}$ belonging to the fundamental/adjoint representation of
the gauge group and an arbitrary number of covariant derivatives projected onto
the light-cone. To one-loop order, both in QCD and all SYM models in the
multi-color limit, the mixing matrix for such operators can be mapped into a
Hamiltonian of the Heisenberg spin chain. The length of the spin chain equals the
number of fermions which form a given Wilson operator and the spin operators are
determined by the generators of the collinear $SL(2;\mathbb{R})$ subgroup of full
(super)conformal group.

It is well-known that the conformal symmetry is broken in QCD and SYM theories
with $\mathcal{N} < 4$ supercharges while in the maximally supersymmetric
$\mathcal{N}=4$ model it survives on the quantum level. However the conformal
anomaly modifies anomalous dimensions starting from two loops only and,
therefore, the one-loop dilatation operator inherits the conformal symmetry of
the classical theory. Starting from two-loop order, the dilatation operator in
the $SL(2)$ sector acquires several new features. First, it receives conformal
symmetry breaking corrections arising both due to a nonzero beta-function and
a subtle symmetry-violating effect induced by the regularization procedure.%
\footnote{It worth mentioning that in distinction with the compact sectors, the
dilatation operator of gauge theories in the $SL(2)$ sector defines one of the
component of the spin operator on the spin chain side.} Second, the form of the
dilatation operator starts to depend on the representation of the fermion fields,
i.e., fundamental $SU(3)$ in QCD and adjoint $SU(N_c)$ in SYM theories. The
difference between the two is that it is only in the latter case that one can
select planar diagrams by going over to the multi-color limit, while in the
former case the large-$N_c$ counting is inapplicable and the two-loop dilatation
operator receives equally important contributions from both planar and nonplanar
Feynman graphs. Thus, by studying the two-loop dilatation operator in the $SL(2)$
sector we are able to identify what intrinsic properties of gauge theories are
responsible for the existence of the integrability phenomenon per se:
\begin{itemize}
\item Conformal symmetry

\item Supersymmetry

\item Planar limit
\end{itemize}
We are able to give definite answers to all of the raised questions by performing
the two-loop calculation of the anomalous dimensions of the aforementioned
aligned-helicity fermionic operators in QCD and all SYM theories. The computation
is performed within two different renormalization schemes based on dimensional
regularization and dimensional reduction. The latter is required for preserving
supersymmetry in SYM theories. Actually, we find that the anomalous dimensions
in the two schemes only differ by an additive constant indicating that the
integrability phenomenon (if present) is scheme independent.

Our finding can be summarized as follows. According to Liouville, the dynamical
model is integrable provided it acquires enough integrals of motion to match the
number of degrees of freedom. For an all-loop dilatation operator $\mathbb{H}
(\lambda)$, depending on 't Hooft coupling constant $\lambda$ and acting on a
Wilson operator built from $L$ constituent fields and an arbitrary number of
covariant derivatives, integrability would require, in general, the existence of
$L$ conserved charges. Two of the charges---the light-cone component of the
total momentum of $L$ fields and the scaling dimension of the operator---follow
immediately from Lorentz covariance of the gauge theory.%
\footnote{The two charges commute for the operators with zero total light-cone
momentum only, the so-called forward limit.}
However, the identification of the remaining charges $q_k(\lambda)$ with
$k=3,\ldots,L$ is an extremely nontrivial task. The eigenvalues of the charges
$q_k$ define the complete set of quantum numbers parameterizing the eigenspectrum
of the dilatation operator. Integrability imposes a nontrivial analytical
structure of anomalous dimensions of Wilson operators and implies the double
degeneracy of eigenvalues with the opposite parity
\cite{FadKor94,GraMat95,BraDerMan98}. To the lowest order of perturbation theory
this property has been verified in Refs.~\cite{BraDerMan98,Bel99} with the help
of the Quantum Inverse Scattering Method \cite{TakFad79}. In particular, it was
found there that breaking of integrability leads to lifting of the degeneracy
in the eigenspectrum of the one-loop dilatation operator.%
\footnote{One should mention however that, in general, integrability does not
necessary imply degeneracy in the eigenspectrum. The simplest example is provided
by integrable open spin chains which determine the one-loop dilatation operator
in multi-color QCD in the sector of mixed antiquark-gluon-quark operators
\cite{Bel99,DerKorMan00}.} We argue that the same relation between integrability
and degeneracy of the eigenstates holds true to two loops and allows one to
identify high loop corrections to the conserved charges.

Going over through diagonalization of the two-loop dilatation operator, we find
that the desired pairing of eigenvalues occurs for three-gaugino operators in SYM
theories with $\mathcal{N}=1,2,4$ supercharges and the $SU(N_c)$ gauge group. In
that case, nonplanar diagrams provide vanishing contribution and, as a
consequence, the exact anomalous dimensions coincide with those in the
multi-color limit. The three-quark (baryonic) operators in QCD are well-defined
for the $SU(3)$ gauge group only. Due to different color representation, the
two-loop dilatation operator in QCD receives corrections from certain nonplanar
diagrams which yield breaking of pairing of eigenvalues. To trace the origin of
breaking of integrability we split the expression for the two-loop anomalous
dimensions into the sum of two gauge invariant terms with one of them depending
on the beta-function. The latter term provides conformal symmetry breaking
contribution to the dilatation operator both in QCD and SYM theories with
$\mathcal{N}=1,2$. As follows from our analysis, the pairing of the eigenstates
persists in QCD for terms depending on the beta function, but does not hold for
the remaining ``conformal'' piece.

Starting from two loops, the dilatation operator in SYM theories depends on the
number of supercharges $\mathcal{N}$. This dependence comes about through the
contribution of $2(\mathcal{N}-1)$ real scalars and $\mathcal{N}$ gaugino fields
propagating inside loops. We find that both contributions are irrelevant for
enhancement or insuring of integrability since the $\mathcal{N}-$dependence of
the two-loop dilatation operator can be factored out (modulo an additive
normalization factor) into a multiplicative c-number. This property makes the
eigenspectrum of the two-loop dilatation operator alike in all gauge theories
including the $\mathcal{N}=4$ SYM. Thus the main conclusion of our analysis is
that the phenomenon of integrability is only tied to the planar limit and is
sensitive neither to conformal symmetry nor supersymmetry.

In our subsequent consideration we thoroughly elaborate on this subject. The
presentation is organized as follows. In the following section we describe gauge
theories under study---QCD and a generic Yang-Mills theory with fermions and
scalars belonging to the adjoint representation of the gauge group,---and specify
three-particle operators whose renormalization is scrutinized here. Then in
Sect.~\ref{RepsDilOpSection} we work out two equivalent representations of the
dilatation operator---as a mixing matrix on the space of local Wilson operators
and as an integral operator in the representation of momentum fractions. The
former representation is convenient for discussing symmetry properties whereas
the latter one is the most appropriate for higher-loop calculations. To start
with, we describe in Sect.~\ref{DilOp1LoopSection} the main features of the
approach on a simpler example of the one-loop dilatation operator. Then, we
move on to the two-loop case and compute the irreducible components of the two-loop
three-particle kernel term by term in Sect.~\ref{DilOpTwoLoop}. The reader,
who is not interested in intermediate details but only wants to see the final
result, is directed straightforwardly to Sect.~\ref{TwoLoopResultTotal}. Having
computed the two-loop dilatation operator, we turn to the study of its
eigenspectrum in Sect.~\ref{SectionEigenspectrum}. At first, we address in
Sect.~\ref{EigenspectrumTw2Section} the sector of two-particle (twist-two)
operators and diagonalize the dilatation operator at one loop making use of the
conformal invariance. Then, we describe the constraints on the form of the
dilatation operator imposed by conformal symmetry breaking in gauge theories. In
Sect.~\ref{TwistTwoTwoLoopADs} we calculate the two-loop anomalous dimensions of
twist-two operators in SYM theories with an arbitrary number of supersymmetries
$\mathcal{N}$ and demonstrate that all of them are expressed in terms of the
beta-function of the underlying gauge theory and the so-called ``universal''
anomalous dimension. Finally, we turn in Sect.~\ref{3Pspectrum} to the
three-particle dilatation operator. We review the relation between integrability
of the one-loop dilatation operator and pairing of its eigenstates and demonstrate
that the double degeneracy observed in QCD and all SYM theories to one loop implies
the existence of a conserved charge. We extend this consideration to two loops in
Sect.~\ref{CriteriumSection} and explicitly construct the set of operators
commuting with the dilatation operator and among themselves. We present the
two-loop spectra of anomalous dimensions in Sect.~\ref{NumericsSection} for QCD
and all SYM theories and draw conclusions about two-loop integrability. We close
this section with presenting two limiting cases in which we are able to
reconstruct the all-loop dilatation operator and identify the corresponding
integrable structures. Section \ref{DiscussionConcl} contains concluding remarks.
Two appendices provide technical details of the calculational procedures used in
our analysis.

%%%%%%%%%%%%%%%%%%%%%%%%%%%%%%%%%%%%%%%%%%%%%%%%%%%%%%%%%%%%%%%%%%%%%
\section{Case of study}
\label{Lagrangians}
%%%%%%%%%%%%%%%%%%%%%%%%%%%%%%%%%%%%%%%%%%%%%%%%%%%%%%%%%%%%%%%%%%%%%

The goal of the present study is to reveal hidden integrability in
four-dimensional Yang-Mills theory describing the coupling of gauge fields
to fermions/scalars. Depending on the representation in which the latter
fields are defined one can distinguish two different types of the gauge
theories: QCD and supersymmetric extensions of Yang-Mills theory. In the
first case, the gauge fields are coupled to quarks in the fundamental
representation of the $SU(N_c)$ gauge group and the QCD Lagrangian is
given by a well-known expression
\be
{\cal L}_{\scriptscriptstyle\rm QCD}
=
- \ft12 {\rm tr} \, F_{\mu\nu} F^{\mu\nu} + i \bar\psi \gamma_\mu D^\mu \psi \, .
\label{L-QCD}
\ee
Here quarks are described by four-component Dirac fermions $\psi=
(\lambda_\alpha ,\bar\chi^{\dot\alpha})$ and the gauge field strength
$F_{\mu\nu}=\frac{i}{g}[D_\mu,D_\nu]$ is determined in terms of the covariant
derivatives $D_\mu = \partial_\mu - i g A_\mu^a t^a$ with generators
$t^a$ in the fundamental representation of the $SU(N_c)$ normalized
conventionally as ${\rm tr} \, (t^a t^b) = \ft12 \delta^{ab}$. In the second
case, the gauge fields are coupled to fermions and scalars belonging to the
adjoint representation of the $SU(N_c)$ group. The Lagrangian of the
corresponding Yang-Mills theory takes the form
\ba
{\cal L}_{\rm adj} = {\rm tr} \, \Big\{ \!\!\!\!&-&\!\!\! \ft12 F_{\mu\nu}
F^{\mu\nu} + 2 i \bar\lambda_{\dot\alpha A} \sigma^{\dot\alpha \beta}_\mu D^\mu
\lambda^A_\beta + \ft12 \left( D_\mu \phi^{AB} \right) \left( D^\mu
\bar\phi_{AB} \right)
\nonumber\\
&-&\!\!\! h_{\rm Y} \, \lambda^{\alpha A} [\bar\phi_{AB}, \lambda_\alpha^B] +
h_{\rm Y} \,
\bar\lambda_{\dot\alpha A}
[\phi^{AB}, \bar\lambda^{\dot\alpha}_B] - h_{4} \, [\phi^{CD},
\bar\phi_{AB}] [\phi^{AB} , \bar\phi_{CD}] \Big\} \, ,
\label{L-SYM}
\ea
with the gauge fields, gauginos and scalars, $X = \{ A_\mu , \lambda^A,
\phi^{AB} \}$, all in the adjoint representation of the gauge group, $X =
X^at^a$ and ${\scriptstyle A,B}=1,\ldots,\mathcal{N}$. The covariant derivative
in the adjoint representation is given by $D_\mu=\partial_\mu - i g [A_\mu,\ ]$.
The gauginos are described by the Weyl fermion $\lambda^A$ which belongs to
the fundamental representation of an internal $SU(\mathcal{N})$ symmetry group
with its complex conjugate $\bar \lambda_A=(\lambda^A)^*$. The scalars are
assembled into the antisymmetric tensor $\phi^{AB} = - \phi^{BA}$, with its
complex conjugate $( \phi^{AB} )^\ast = \bar\phi_{AB}$. The supersymmetric
Yang-Mills theories with $\mathcal{N}=1,2$ and $4$ supercharges are obtained
from the Lagrangian \re{L-SYM} by adjusting the values of the coupling
constants and the number of gaugino and scalar species, $n_f$ and $n_s$,
respectively:
\begin{itemize}
\item $\mathcal{N} = 1$ SYM: there is a single specie of the gaugino, $n_f=1$,
and no scalars, $n_s=0$, so that
\be
h_{\rm Y} = 0 \, , \qquad h_{4} = 0 \, .
\ee

\item $\mathcal{N} = 2$ SYM: there are two species of gaugino, $n_f=2$, and two
real scalars, $n_s=2$,
\be
h_{\rm Y} = g \, , \qquad h_{4} = \ft{1}{16} g^2 \, , \qquad \phi^{AB} =
\sqrt{2} \varepsilon^{AB} \phi \, .
\ee

\item $\mathcal{N} = 4$ SYM: there are four species of gaugino $n_f=4$, and six
real scalars, $n_s=6$,
\be
h_{\rm Y} = \sqrt{2} g \, , \qquad h_{4} = \ft{1}{8} g^2 \, , \qquad \phi^{AB}
= \ft12 \varepsilon^{ABCD} \bar\phi_{CD} \, .
\ee
\end{itemize}
The above relations between the coupling constants are imposed by supersymmetry.
It also relates to each other the number of gauginos and scalars
\be
n_f = 1 + \frac{n_s}2 = \mathcal{N}\,.
\label{n_s}
\ee
One of our findings is that, as we will show in Sect.~\ref{Scalars}, integrability
is not tied to supersymmetry. In other words, the phenomenon persists in the generic
Yang-Mills theory \re{L-SYM} for arbitrary  $\mathcal{N}$, to two loop order at least.

The classical Lagrangians of the Yang-Mills theories \re{L-QCD} and \re{L-SYM}
enjoy the conformal symmetry which will play an important r\^ole in our
analysis. On the quantum level this symmetry is anomalous (except of
$\mathcal{N}=4$ SYM) due to running of the gauge coupling constant,
\be
\label{DefBetaExpansion}
\frac{d\ln g^2}{d\ln\mu}
=
\beta (g^2) = - \beta_0 \frac{g^2}{8 \pi^2} -
\beta_1
\left(\frac{g^2}{8 \pi^2}\right)^2
+ \mathcal{O}(g^6)
\, .
\ee
The beta-function in QCD is given to two loop order by the well-known expression
\be
\beta_{\scriptscriptstyle\rm QCD} (g^2)
=
- \frac{g^2}{8 \pi^2}
\left(\frac{11}3 N_c -\frac23 n_f\right)
-
\left(\frac{g^2}{8 \pi^2} \right)^2
\left( \frac{17}{3} N_c^2 - C_F n_f-\frac{5}3 N_c n_f \right)
+
\mathcal{O}(g^6)
\, ,
\label{beta-QCD}
\ee
where $C_F=(N_c^2-1)/(2N_c)$ is the Casimir operator in the fundamental
representation of the $SU(N_c)$ group and $n_f$ the number of the quark
flavors. In supersymmetric Yang-Mills theory with $\mathcal{N}=1,2,4$
supercharges the two-loop beta-function is given by
\be
\beta_{\scriptscriptstyle\rm SYM} (g^2)
=
-
\frac{g^2 N_c}{8 \pi^2}  (4 - \mathcal{N})
-
\left(\frac{g^2 N_c }{8 \pi^2}\right)^2 (2 - \mathcal{N})(4 - \mathcal{N})
+
\mathcal{O}(g^6)
\, .
\label{beta-SYM}
\ee
For $\mathcal{N}=4$ the beta-function vanishes to all orders of perturbation
theory and the gauge theory remains conformal on the quantum level. For
$\mathcal{N}<4$ the beta-function is nonvanishing and, as a consequence, the
conformal symmetry is broken. For $\mathcal{N}=2$ the beta-function receives
contribution to one-loop only while for $\mathcal{N}=1$ the beta-function
coincides with the QCD expression \re{beta-QCD} upon substitution $n_f \to N_c$
and $C_F \to N_c$.

The central object of our study is the scale dependence of the gauge invariant
nonlocal operators given by the product of elementary fields $X (z n_\mu)$
``living'' on the light-ray defined by a light-like vector $n_\mu$ (so that
$n_\mu^2=0$). The explicit form of these operators depends on the representation
of the $SU(N_c)$ gauge group to which these fields belong. For elementary fields
$X=X^a t^a$ in the adjoint representation one can construct single-trace operators
of the length $L$
\be
\mathbb{O}_{\rm adj} (z_1, z_2, \dots , z_L) = {\rm tr} \left\{ X (z_1 n) X
(z_2 n) \cdots X (z_L n) \right\} \, ,
\label{O-adj}
\ee
where the gauge invariance is restored by inserting the Wilson lines in between
each pair of fields,
\be
[z_j, z_{j+1}] = P \exp\left( i g \int_{z_{j+1}}^{z_j} d z \, n^\mu  A_\mu (z n)
\right) \, .
\label{gauge link}
\ee
Later on we will adopt the light-light gauge $(n\cdot A)\equiv A_+ = 0$, so that
the gauge links all shrink to the unit matrix. In what follows we do not display
them for brevity. For fields $X^j$ (quarks) in the fundamental representation, we
can introduce a baryon operator
\be
\mathbb{O}_{\rm fun} (z_1, z_2, \dots , z_{N_c})
=
\varepsilon_{j_1 j_2 \dots j_{N_c}} \,
X^{j_1} (z_1 n) X^{j_2}(z_2 n) \cdots X^{j_{N_c}} (z_{N_c} n) \, .
\label{O-fun}
\ee
It is built from exactly $N_c$ fields and is completely antisymmetric with
respect to permutation of any pair of them.

The nonlocal light-cone operators \re{O-adj} and \re{O-fun} have the meaning of
generating functions of local gauge invariant Wilson operators of the maximal
Lorentz spin. The latter operators can be obtained from Taylor expansion of
$\mathbb{O}_{\rm adj} (z_1, z_2, \dots , z_L)$ and $\mathbb{O}_{\rm fun} (z_1,
z_2, \dots , z_{N_c})$ in powers of $z_i$. Each field entering \re{O-adj} and
\re{O-fun} can be decomposed into different helicity components and, as a
consequence, the nonlocal light-cone operators can be classified according to
their helicity content. Among them one can distinguish light-cone operators with
all helicities aligned. In what follows we shall refer to them as maximal
helicity operators.

It is well-known that local Wilson operators mix under renormalization and
satisfy the evolution (Callan-Symanzik) equations. The same is true for the
light-cone operators \re{O-adj} and \re{O-fun} although the explicit form of the
evolution kernels driving the scale dependence of these operators is different
due to their nonlocal form. In particular, for nonlocal light-cone operators
\re{O-adj} built from the so-called ``good'' components of fundamental fields the
evolution equation takes the following form%
\footnote{This equation has been written under tacit assumption that the light-cone
operators \re{O-adj} evolve autonomously in the multicolor limit. In general,
$\mathbb{H}_L$ has a matrix form as the light-cone operators with different
partonic content could mix with each other.} \cite{BukFroKurLip85,BalBra88}
\be
\left(
\mu \frac{\partial}{\partial \mu} + \beta(g^2) \frac{\partial}{\partial g^2}
\right) \mathbb{O}_{\rm adj} (z_1, z_2,\ldots, z_L)
=
- [ \mathbb{H}_L \cdot \mathbb{O}_{\rm adj}] (z_1, z_2,\ldots, z_L)
+
\mathcal{O}(1/N_c^2)\, ,
\label{RG}
\ee
where $\mathcal{O}(1/N_c^2)$ denotes terms corresponding to transitions involving
splitting of the single trace into multi trace operators, i.e., $ \mathbb{O}_{\rm adj}
(z_1, z_2, \dots , z_L) \to \mathbb{O}_{\rm adj} (z_1, z_2) \mathbb{O}_{\rm adj}
(z_3, \dots , z_L)$. The evolution equation \re{RG} follows from the Ward identity
under dilatation transformations in the underlying gauge theory.

The integral operator $\mathbb{H}_L$ entering \re{RG} defines a representation of
the dilatation operator $\mathbb{D}$ on the space spanned by nonlocal light-cone
operators \re{O-adj}. To the lowest order of perturbation theory, the explicit form
of $\mathbb{H}_L$ in Yang-Mills theories with an arbitrary number of supercharges
has been found in Ref.~\cite{BelDerKorMan04}. It turns out that to one-loop order,
the evolution kernel has a number of remarkable properties in the multi-color limit:
\begin{itemize}
\item The length $L$ of the single trace operator is preserved and, as a
consequence, $\mathbb{H}_L$ can be realized as a quantum mechanical Hamiltonian
for a system of $L$ particles with nearest-neighbor interaction,
$\mathbb{H}_L=H_{12}+H_{23} + \ldots + H_{L1}$;
\item  $\mathbb{H}_L$ reveals a hidden integrability symmetry in the sector
of maximal helicity operators both in QCD and in supersymmetric Yang-Mills
theories: it possesses exactly $L$ conserved charges and, therefore, is
completely integrable;
\item $\mathbb{H}_L$ can be mapped into a Hamiltonian of the Heisenberg spin chain of
length $L$ and spin operators being generators of the $SL(2|\mathcal{N})$ group
with $\mathcal{N}$ counting the number of supercharges.
\end{itemize}
The main question that we would like to address in the present paper is whether
the integrability phenomenon just described is preserved to two-loop order and
beyond it.

To this end, we shall evaluate two-loop corrections to the evolution kernel
$\mathbb{H}_L$ and study its properties. Due to different partonic content in
Yang-Mills theories with different number of supercharges, the number of maximal
helicity light-cone operators varies with $\mathcal{N}$. Calculation of the
two-loop evolution kernel for any of these operators is an extremely tedious
task. Supersymmetry implies that the evolution kernels corresponding to different
maximal helicity operators of the same length $L$ are related to each other.
Making use of this property, we shall restrict our analysis to maximal helicity
operators built from the helicity $+\frac{1}{2}$ component of the gaugino field
$\lambda_\alpha^A$ (with $\alpha=1,2$)
\be
\lambda_{+ \alpha}^A \equiv \ft12
\bar\sigma^-{}_{\alpha\dot\beta} \, \sigma^{+ \; \dot\beta\gamma}
\lambda_\gamma^A =  \left(\lambda^A \atop 0 \right)\,.
\label{1/2-SYM}
\ee
It actually possesses a single nonvanishing component $\lambda^A=\lambda^{aA}
\,t^a$ (with $a=1,\ldots,N_c^2-1$ and ${\scriptstyle A}=1,\ldots,\mathcal{N}$)
which is charged with respect to internal $SU(\mathcal{N})$ isotopic group. Then,
one decomposes the product of $L$ fields over irreducible $SU(\mathcal{N})$
components and selects the one with the maximal $R$-charge
\be
\mathbb{O}(z_1, z_2,\ldots, z_L) = \S_{A_1A_2\ldots A_L} {\rm tr} \left\{
\lambda^{A_1} (z_1 n) \lambda^{A_2} (z_2 n)\ldots \lambda^{A_L}(z_L n) \right\}
\, ,
\label{O-1/2}
\ee
where the operation {\bf S} stands for symmetrization over the $SU(\mathcal{N})$
indices. The counter-part of the operators \re{O-1/2} in QCD is the baryon
operator \re{O-fun} built from helicity $\pm \frac{1}{2}$ ``good'' component of
the quark fields
\be
\psi_{+ \uparrow}
= \ft14 ( 1 - \gamma_5 ) \gamma_- \gamma_+ \psi=\left(
\begin{array}{c}
q \\
0 \\
0 \\
0 \\
\end{array}\right)
\, , \qquad \psi_{+ \downarrow} = \ft14 ( 1 + \gamma_5 ) \gamma_- \gamma_+
\psi=\left(
\begin{array}{c}
0 \\
0 \\
0 \\
\bar{q} \\
\end{array}\right)
\label{1/2-QCD}
\ee
which have a single nonvanishing component transforming in the fundamental
representation of $SU(N_c)$. Later in the paper, we will encounter the operators
of length $L=2$ and $L=3$:
\begin{itemize}
\item For $L=2$, the generating functions of maximal-helicity operators in SYM and
QCD read, respectively,
\ba
\mathbb{O}_{\rm adj} (z_1, z_2)
\!\!\!&=&\!\!\!
\tr [ \lambda^{\{A} (z_1 n) \lambda^{B\}} (z_2 n)]
\, ,
\label{O2-SYM}
\\
\mathbb{O}_{\rm fun} (z_1, z_2)
\!\!\!&=&\!\!\!
\ft12 \bar\psi_{+ \downarrow} (z_1 n) \gamma_+ \bar\gamma_\perp \psi_{+ \uparrow} (z_2 n)
=
\bar q^\dagger_j (z_1 n) q^j (z_2 n) \, ,
\label{O2-QCD}
\ea
where $q^j$ and $\bar q^j$ are independent helicity $\pm \frac{1}{2}$ components
of the quark Dirac field $\psi_+$ and $\bar\gamma_\perp = (\gamma_1 - i \gamma_2)/
\sqrt{2}$ is the antiholomorphic Dirac matrix.

\item For $L=3$, analogous generating functions look like
\ba
\mathbb{O}_{\rm adj} (z_1, z_2, z_3)
\!\!\!&=&\!\!\! \S_{ABC} {\rm tr} \left\{ \lambda^A (z_1
n) \lambda^B (z_2 n) \lambda^C (z_3 n) \right\}
\, , \label{O3-SYM}
\\
\mathbb{O}_{\rm fun} (z_1, z_2, z_3)
\!\!\!&=&\!\!\!
\varepsilon_{jkl} \, q^j (z_1 n) q^k (z_2 n) q^l (z_3 n)
\, . \label{O3-QCD}
\ea
\end{itemize}
We remind that for fermions in the fundamental representation of the $SU(N_c)$
group, the length of the operator \re{O-fun} ought to be $N_c$. Therefore, the
operator $\mathbb{O}_{\rm fun} (z_1, z_2, z_3)$ is well defined only for $N_c=3$.
The operators $\mathbb{O}_{\rm fun} (z_1, z_2)$ and $\mathbb{O}_{\rm fun} (z_1,
z_2, z_3)$ have a direct phenomenological significance: their matrix elements
determine the transversity distributions in the nucleon~\cite{RalSop79} and the
distribution amplitude of the delta-isobar~\cite{BroLep79}, respectively.

We would like to stress that the choice of the maximal helicity operator is
ambiguous. The reason why we prefer to work with the operators built from
helicity $+\frac{1}{2}$ fermions (gaugino, quarks) is that the number and complexity
of contributing Feynman diagrams is gradually reduced and more importantly we
can make use of available two-loop QCD calculations. In addition, as we will
show in Sect.~\ref{DilOpTwoLoop}, the two-loop dilatation operator acts
elastically on the maximal helicity operators built from the fermion fields
\re{1/2-SYM} and \re{1/2-QCD}. In other words, the operators \re{O-1/2} evolve
autonomously and the corresponding evolution kernel $\mathbb{H}_L$ can be
realized as a quantum mechanical Hamiltonian. To two-loop order, $\mathbb{H}_L$
is given by the sum over two- and three-particle kernels, $H_{k,k+1}$ and
$H_{k,k+1,k+2}$, respectively. To identify the explicit form of these kernels
it suffices to consider maximal helicity operators of length $L = 3$,
Eqs.~\re{O3-SYM} and \re{O3-QCD}. Our strategy will be to evaluate the two-loop
evolution kernels for the operators \re{O3-SYM} and \re{O3-QCD} and, then,
draw conclusions about their symmetry properties.

%%%%%%%%%%%%%%%%%%%%%%%%%%%%%%%%%%%%%%%%%%%%%%%%%%%%%%%%%%%%%%%%%%%%%
\section{Representations of the dilatation operator}
\label{RepsDilOpSection}
%%%%%%%%%%%%%%%%%%%%%%%%%%%%%%%%%%%%%%%%%%%%%%%%%%%%%%%%%%%%%%%%%%%%%

Before we proceed with actual calculations we would like to summarize general
properties of the evolution kernels entering the renormalization group equation
for the light-cone operators \re{O3-SYM} and \re{O3-QCD}
\be
\left( \mu \frac{\partial}{\partial \mu} + \beta (g^2) \frac{\partial}{\partial
g^2} \right) \mathbb{O} (z_1, z_2, z_3) = - [ \mathbb{H} \cdot \mathbb{O}] (z_1,
z_2, z_3)
\, ,
\label{RG-3}
\ee
where $\mathbb{H} = \mathbb{H}_{L = 3}$. The evolution kernel $\mathbb{H}$ admits
the perturbative expansion
\be
\mathbb{H} = \lambda\, \mathbb{H}^{(0)} + \lambda^2 \mathbb{H}^{(1)} +
\mathcal{O} (\lambda^3) \, , \label{H-dec}
\ee
with $\lambda = C g^2/8 \pi^2$ and the color factor $C$ depending on the representation
of the fermion fields. Its value will be specified in Sect.~\ref{OneLoopDilOper}.

The central object of our study is the spectral problem for the dilatation
operator
\be
\mathbb{H} \,\Psi_{\bit{\scriptstyle q}}(z_i) = \gamma_{\bit{\scriptstyle
q}}(\lambda)\, \Psi_{\bit{\scriptstyle q}}(z_i) \, .\label{Sch-eq}
\ee
Here the eigenfunctions $\Psi_{\bit{\scriptstyle q}}(z_i)$ depend on the
light-cone coordinates, $\gamma_{\bit{\scriptstyle q}}(\lambda)$ define the
eigenvalues of the dilatation operator and $\bit{q}$ denotes the complete set of
quantum numbers to be specified later on.  As we will show in
Sect.~\ref{3Pspectrum}, hidden integrability of the dilatation operator leads to
a peculiar regularity in the spectrum of $\gamma_{\bit{\scriptstyle
q}}(\lambda)$. In gauge theories with unbroken conformal symmetry,
$\beta(g^2)=0$, the solution to the evolution equation \re{RG-3} looks like
\be
\mathbb{O} (z_1, z_2, z_3) = \sum_{\bit{\scriptstyle q}} \Psi_{\bit{\scriptstyle
q}} (z_1,z_2,z_3)\, \mathcal{O}_{\bit{\scriptstyle q}} (0) \, ,
\label{B-expansion}
\ee
with $\mathcal{O}_{\bit{\scriptstyle q}} (0)$ being {\sl local} Wilson operators.
Substituting this relation into \re{RG-3} one finds that the operators
$\mathcal{O}_{\bit{\scriptstyle q}}(0)$ have an autonomous scale dependence and
their anomalous dimensions are given by $\gamma_{\bit{\scriptstyle q}}
(\lambda)$. In gauge theories with $\beta(g^2)\neq 0$, the conformal anomaly
induces the mixing between the operators $\mathcal{O}_{\bit{\scriptstyle q}}(0)$
starting from two loops. It is driven by the additional term in the evolution
equation $\beta (g^2){\partial}\Psi_{\bit{\scriptstyle q}}(z_i)/{\partial g^2}$
proportional to the beta-function. It is straightforward to write down a solution
to the evolution equation in that case but its explicit form is more complicated
as compared to the one with $\beta(g^2)=0$.

%%%%%%%%%%%%%%%%%%%%%%%%%%%%%%%%%%%%%%%%%%%%%%%%%%%%%%%%%%%%%%%%%%%%%
\subsection{Mixing matrix}
\label{MixingMatrix}
%%%%%%%%%%%%%%%%%%%%%%%%%%%%%%%%%%%%%%%%%%%%%%%%%%%%%%%%%%%%%%%%%%%%%

We remind that the nonlocal light-cone operators \re{O3-SYM} and \re{O3-QCD}
generate infinite towers of local Wilson operators. In particular, for the
operator \re{O3-SYM} one finds
\be
\mathbb{O} (z_1, z_2, z_3) = \sum_{k_1, k_2, k_3 \geq 0}^\infty \frac{(- i
z_1)^{k_1}}{k_1 !} \frac{(- i z_2)^{k_2}}{k_2 !} \frac{(- i z_3)^{k_3}}{k_3 !}
\mathcal{O}_{k_1 k_2 k_3} (0) \, , \label{B-dec}
\ee
with the Wilson operators
\be
\label{Local3ParticleOper}
\mathcal{O}_{k_1 k_2 k_3} (0) =\S_{ABC} {\rm tr} \left\{ (i D_+)^{k_1} \lambda^A
(0) (i D_+)^{k_2} \lambda^B (0) (i D_+)^{k_3} \lambda^C (0) \right\}
\ee\\[0.2mm]
and $D_+ = (n \cdot D) = \partial_+$ in the light-like axial gauge $(n\cdot
A)=0$. Substituting \re{B-dec} into the evolution equation \re{RG-3} one recovers
a conventional renormalization group equation for three-particle Wilson operators
\be
\left(
\mu \frac{\partial}{\partial \mu} + \beta(g^2) \frac{\partial}{\partial g^2}
\right) \mathcal{O}_{k_1 k_2 k_3}(0)
=
- \sum_{l_1, l_2, l_3 \ge 0}
\mathbb{V}_{k_1 k_2 k_3}^{l_1 l_2 l_3} \, \mathcal{O}_{l_1 l_2 l_3}(0) \, .
\label{RG-local}
\ee
Here the mixing matrix defines a representation of the dilatation operator
\re{H-dec} in the basis of polynomials $\ket{k_1,k_2,k_3} =  {(-i z_1)^{k_1}}
{(-i z_2)^{k_2}} {(-i z_3)^{k_3}}/({k_1!}{k_2!}{k_3!})$
\be
\mathbb{H}\, \ket{l_1,l_2,l_3} = \sum_{k_1,k_2,k_3\ge 0}\mathbb{V}_{k_1 k_2
k_3}^{l_1 l_2 l_3} \,\ket{k_1,k_2,k_3}\,.
\label{H=V}
\ee
To find the spectrum of the dilatation operator \re{Sch-eq} one has to solve the
spectral problem for the evolution kernel or equivalently diagonalize the mixing
matrix
\be
\sum_{l_1,l_2,l_3\ge 0}\mathbb{V}_{k_1 k_2 k_3}^{l_1 l_2 l_3}\,c_{l_1 l_2 l_3}
(\bit{q})
=\gamma_{\bit{\scriptstyle q}}(\lambda)\,c_{k_1 k_2 k_3}(\bit{q})\,.
\label{V-left}
\ee
The coefficients $c_{k_1 k_2 k_3}{(\bit{q})}$ define the expansion of the
eigenstates $\Psi_{\bit{\scriptstyle q}}(z_i)$ over the basis \re{H=V}
\be
\Psi_{\bit{\scriptstyle q}}(z_1,z_2,z_3) = \sum_{k_1 k_2 k_3}   {c_{k_1 k_2
k_3} (\bit{q})} \, %\ket{k_1,k_2,k_3}
\frac{(-i z_1)^{k_1}}{k_1!} \frac{(-i z_2)^{k_2}}{k_2!} \frac{(-i
z_3)^{k_3}}{k_3!} \,.
\label{Psi-polynomial}
\ee
Let us take into account that the mixing can occur between the Wilson operators
$\mathcal{O}_{k_1 k_2 k_3} (0)$ with the same canonical dimension. This implies
that the mixing matrix $\mathbb{V}_{k_1 k_2 k_3}^{l_1 l_2 l_3}$ has nonvanishing
entries only for $k_1+k_2+k_3=l_1+l_2+l_3$. In terms of the evolution kernel,
Eq.~\re{H=V}, this property states that the evolution kernel preserves the overall
power of the polynomials $\ket{k_1,k_2,k_3}$, or equivalently
\be
[ \mathbb{H} , z_1 \partial_{z_1} + z_2 \partial_{z_2} + z_3 \partial_{z_3} ]
= 0
\, . \label{dilat}
\ee
In addition, due to Poincar\'e covariance, the anomalous dimension of the
operators $\mathcal{O}_{k_1 k_2 k_3} (x)$ does not depend on the coordinate
$x$. In terms of the evolution kernel, this implies that $\mathbb{H}$ has
to be translation invariant
\be
[ \mathbb{H} ,  \partial_{z_1} +  \partial_{z_2} +   \partial_{z_3} ]
= 0
\, . \label{trans}
\ee
Since the dilatations and translations do not commute with each other, the
eigenstates can not diagonalize the two charges simultaneously unless the total
light-cone momentum equals zero, $(\partial_{z_1} +  \partial_{z_2} +
\partial_{z_3})\Psi_{\bit{\scriptstyle q}}(z_i)=0$.

Putting $z_1=z_2=z_3=0$ in \re{B-dec} one obtains a local operator with no
derivatives
\be
\mathbb{O} (0, 0, 0) = \mathcal{O}_{0 0 0} (0) = \frac{i}4f^{abc} \S_{ABC}
\lambda^{aA}(0)\lambda^{bB}(0)\lambda^{cC}(0)\,. \label{B0}
\ee
This operator evolves autonomously and according to \re{RG-local} and \re{H=V}
its anomalous dimension defines the eigenvalue of the dilatation operator
\be
\mathbb{H} \ket{0,0,0} = 3 \Gamma(\lambda) \ket{0,0,0}\,, \label{Gamma}
\ee
where the additional factor $3$ was inserted for the later convenience. As we
will show in Sect.~\ref{NumericsSection}, the anomalous dimension
\be
\label{GammaPertExpansion} \Gamma(\lambda) = \lambda \Gamma^{(0)} +
\lambda^2  \Gamma^{(1)} + \mathcal{O}(\lambda^3)
\ee
sets up the lower/upper bound in the spectrum of three-particle dilatation
operators \re{H-dec}.

%%%%%%%%%%%%%%%%%%%%%%%%%%%%%%%%%%%%%%%%%%%%%%%%%%%%%%%%%%%%%%%%%%%%%
\subsection{Momentum representation}
%%%%%%%%%%%%%%%%%%%%%%%%%%%%%%%%%%%%%%%%%%%%%%%%%%%%%%%%%%%%%%%%%%%%%

The evolution operator $\mathbb{H}$ acts along the light-cone direction $n_\mu$
in Minkowski space-time. Yet another representation of $\mathbb{H}$ can be
obtained by going over from the configuration to the reciprocal, momentum space.
To this end, one performs Fourier transformation of the light-cone operators
\be
\mathbb{O} (z_1, z_2, z_3) = \int_{-\infty}^\infty  {d u_1} \int_{-\infty}^\infty
{d u_2} \int_{-\infty}^\infty  {d u_3} \, {\rm e}^{i u_1 z_1 + i u_2 z_2 + i u_3 z_3}
\widetilde{\mathbb{O}} (u_1, u_2, u_3)
\label{B-Fourier}
\ee
with $u_i$ having the meaning of light-cone momenta of particles. Substituting
\re{B-dec} into this relation one finds that the local Wilson operators are given
by the moments of $\widetilde{\mathbb{O}} (u_1, u_2, u_3)$
\be
\mathcal{O}_{k_1k_2k_3}(0) = \int d u_1 du_2 du_3 \, u_1^{k_1} u_2^{k_2}
u_3^{k_3}\, \widetilde{\mathbb{O}} (u_1, u_2, u_3)\,.
\ee
In the momentum representation, the Callan-Symanzik equation is known in QCD
as the Brod\-sky-Lepage equation \cite{BroLep79}
\be
\left( \mu \frac{\partial}{\partial \mu} + \beta (g^2) \frac{\partial}{\partial
g^2} \right) \widetilde{\mathbb{O}} (u_1, u_2, u_3) = \int [dv]_3 \mathbb{V}
(u_1, u_2, u_3| v_1, v_2, v_3) \widetilde{\mathbb{O}} (v_1, v_2, v_3) \, ,
\label{BL}
\ee
where the integration measure is
\be
[dv]_3 \equiv dv_1 dv_2 dv_3 \delta \left( \sum_{j = 1}^3 v_j - \sum_{j = 1}^3
u_j \right) \, .
\label{3-measure}
\ee
The evolution kernel in the momentum representation, $\mathbb{V}$, is obtained
from $\mathbb{H}$ through the Fourier transform
\be
\mathbb{H} \cdot \e^{-i\sum_{k} v_k z_k} = - \int [du]_3 \mathbb{V}
(\mbox{\boldmath $u$}| \mbox{\boldmath $v$} )\e^{-i\sum_{k} u_k z_k}
\, ,
\label{H-plane waves}
\ee
with $[d u]_3$ determined by Eq.\ \re{3-measure} with $u_i$ and $v_i$ interchanged.
According to \re{trans}, the evolution kernel preserves the total momentum
$v_1+v_2+v_3=u_1+u_2+u_3$. This property is automatically preserved in \re{BL}.
In addition, it follows from \re{dilat} that the evolution kernel is a homogeneous
function of the momentum fractions
\be
\mathbb{V}
(\lambda u_1,\lambda u_2,\lambda u_3|\lambda v_1,\lambda v_2,\lambda v_3)
=
{\lambda}^{-2}\, \mathbb{V} (u_1, u_2, u_3| v_1, v_2, v_3)\,.
\label{rescale}
\ee
Obviously, the perturbative structure of the evolution kernel $\mathbb{V}$ is
identical to the one introduced for $\mathbb{H}$, Eq.~\re{H-dec}.

Combining together \re{H-plane waves} and \re{H=V} one obtains the following
representation for the mixing matrix
\be
\int [du]_3 \mathbb{V} (\mbox{\boldmath $u$}| \mbox{\boldmath $v$} ) u_1^{k_1}
u_2^{k_2}u_3^{k_3} = - \sum_{l_1,l_2,l_3\ge 0\atop \sum k_i=\sum l_j}
\mathbb{V}_{k_1k_2k_3}^{l_1l_2l_3}\,  v_1^{l_1} v_2^{l_2} v_3^{l_3}
\label{V-moments}
\ee
that is, positive integer moments of the evolution kernel with respect to
$u-$variables are polynomials in $v-$variables of the same degree $k_1+k_2+k_3
=l_1+l_2+l_3$. Below this property will play an important role in our analysis
and we shall refer to it as the {\sl polynomiality} condition.

By construction, the eigenvalues of the evolution kernel $\mathbb{V} (\bit{u}
| \bit{v})$ defined in \re{H-plane waves} coincide with $\gamma_{\bit{\scriptstyle q}}
(\lambda)$, Eq.~\re{Sch-eq}. To see this, one considers the spectral problem for
the mixing matrix
\be
\sum_{k_1,k_2,k_3\ge 0}w^{k_1 k_2 k_3} (\bit{q})
\,\mathbb{V}_{k_1 k_2 k_3}^{l_1 l_2 l_3}
=
\gamma_{{\bit{\scriptstyle q}}} (\lambda) \,w^{l_1 l_2 l_3}(\bit{q})\,.
\label{V-right}
\ee
Comparing this relation with \re{V-left} one observes that the coefficients
$w^{l_1 l_2 l_3}(\bit{q})$ and $c_{k_1 k_2 k_3}(\bit{q})$ are, respectively,
the left and right eigenstates of the same mixing matrix corresponding to the
same eigenvalue $\gamma_{\bit{\scriptstyle q}} (\lambda)$. Introducing the
polynomial
\be
P_{{\bit{\scriptstyle q}}} (u_i)
=
\sum_{k_1,k_2,k_3\ge 0} w^{k_1 k_2 k_3}(\bit{q}) u_1^{k_1} u_2^{k_2}u_3^{k_3}
\label{P-polynomial}
\ee
one finds from \re{V-moments} that it diagonalizes the evolution kernel
\be
\int [du]_3 \mathbb{V} ( \bit{u}| \bit{v}) \, P_{\bit{\scriptstyle q}}(u_i)
=
- \gamma_{\bit{\scriptstyle q}} (\lambda)\,P_{\bit{\scriptstyle q}} (v_i)
\, . \label{spectral}
\ee
The left and right eigenstates of the mixing matrix are orthogonal to each other.
Being written in terms of the polynomials \re{Psi-polynomial} and
\re{P-polynomial} this condition reads
\be\label{ortho}
P_{\bit{\scriptstyle q}} (\partial_{z_1},\partial_{z_2},\partial_{z_3})
\Psi_{{\bit{\scriptstyle q}}'}(z_1,z_2,z_3)\big|_{z_i=0}
\sim
\delta_{{\bit{\scriptstyle q}}{\bit{\scriptstyle q}}'}\,.
\ee
Combining this relation together with \re{B-expansion} one concludes that the
$P-$polynomials determine the explicit form of local Wilson operators entering
\re{B-expansion}
\be\label{O_q}
\mathcal{O}_{\bit{\scriptstyle q}}(0) = P_{\bit{\scriptstyle q}}
(\partial_{z_1},\partial_{z_2},\partial_{z_3})
\mathbb{O}(z_1,z_2,z_3)\big|_{z_i=0} \,.
\ee
Moreover, it follows from \re{rescale} that the eigenstates of the evolution
kernel $\mathbb{V} ( \bit{u}| \bit{v})$ have to be homogeneous polynomials
\be
P_{\bit{\scriptstyle q}} (\lambda u_1,\lambda u_2,\lambda u_3)
=
\lambda^N P_{\bit{\scriptstyle q}} ( u_1,u_2, u_3)
\,.
\label{homo}
\ee
Here, integer $N\ge 0$ counts the total number of light-cone derivatives in the
expression for the Wilson operator \re{O_q}. It also defines the Lorentz spin of
$\mathcal{O}_{\bit{\scriptstyle q}} (0)$. The mixing between the operators with
different $N$ is protected by Lorentz symmetry and, therefore, $N$ can be
identified as one of the quantum numbers $\bit{q}$.

In this section we described the dilatation operator in two different
representations---as a mixing matrix for local Wilson operators and
as an integral operator in the representation of momentum fractions.
Obviously, the spectrum of the dilatation operator does not depend on the
particular representation and its choice is a question of convenience. As we will
demonstrate in Sect.~\ref{DilOpTwoLoop}, the momentum representation has a number
of advantages as far as the two-loop calculation of the evolution kernels is
concerned. Firstly, by virtue of Lorentz covariance, the eigenvalues of the
evolution kernel $\mathbb{V} (\bit{u}|\bit{v} )$ do not depend on the total
light-cone momenta $\sum_j u_j=\sum_j v_j$. This allows one to put its value
equal to zero. From the point of view of nonlocal operators, this corresponds
to assuming translation invariance of $\mathbb{O}(z_1,z_2,\ldots,z_L)$ along
the light-cone, or, equivalently, to considering forward matrix elements of the
nonlocal light-cone operators in \re{B-Fourier}. In this way, one automatically
eliminates the contribution of local operators with total derivatives in
\re{B-dec}, which in turn simplifies significantly the form of the mixing matrix
\re{RG-local}. Secondly, there exist well developed techniques for calculating
the renormalization group kernels in the momentum representation. We shall
profit from higher-order results available in the literature.

%%%%%%%%%%%%%%%%%%%%%%%%%%%%%%%%%%%%%%%%%%%%%%%%%%%%%%%%%%%%%%%%%%%%%
\section{Dilatation operator: 1-loop}
\label{DilOp1LoopSection}
%%%%%%%%%%%%%%%%%%%%%%%%%%%%%%%%%%%%%%%%%%%%%%%%%%%%%%%%%%%%%%%%%%%%%

Let us now turn to the computation of the evolution kernel $\mathbb{H}$ for
the maximal helicity operators \re{O3-SYM} and \re{O3-QCD} both in QCD and
supersymmetric Yang-Mills theory. To start with, we review in this section
the one-loop calculation of $\mathbb{H}$. It will serve as an illustration
of the general formalism which will then be used for more involved two-loop
analysis.

%%%%%%%%%%%%%%%%%%%%%%%%%%%%%%%%%%%%%%%%%%%%%%%%%%%%%%%%%%%%%%%%%%%%%
\subsection{Regularization scheme}
\label{RegularizationSection}
%%%%%%%%%%%%%%%%%%%%%%%%%%%%%%%%%%%%%%%%%%%%%%%%%%%%%%%%%%%%%%%%%%%%%

Performing the calculation we shall adopt the light-like axial gauge $(n\cdot
A(x)) = A_+(x) =0$. Since the evolution operator $\mathbb{H}$ is a gauge-invariant
quantity, any choice of gauge fixing is possible. In covariant gauges, the
nonlocal light-cone operators \re{O-adj} involve additional gauge links
\re{gauge link} between the elementary constituent fields. This leads to
proliferation of the number of Feynman diagrams contributing to $\mathbb{H}$.
The light-like axial gauge $A_+(x) = 0$ is advantageous as the number of relevant
graphs is much smaller, especially at higher loops. Yet another advantage of this
choice is that we can employ certain results for one- and two-loop evolution
kernels in QCD available in the literature.

The Yang-Mills theory in the light-like axial gauge has a number of subtleties
\cite{Bas91}. The gauge field propagator $\vev{0|T A_\mu^a(x) A_\nu^b(y)|0}=
(- i)\delta^{ab} D_{\mu\nu}(x-y)$ is given in the momentum representation by
\be
D_{\mu\nu}(k) = \frac{d_{\mu\nu} (k)}{k^2 + i0}
\, , \qquad
d_{\mu\nu} (k)
=
g_{\mu\nu} - \frac{k_\mu n_\nu + k_\nu n_\mu}{k_+}
\, ,
\label{axial}
\ee
with $k_+ = (k \cdot n)$. It has a spurious singularity at $k_+=0$ which indicates
that the gauge ambiguity is not fixed completely for the choice $A_+(x)=0$. To give
the meaning to the gluon propagator \re{axial}, one has to specify the prescription
for integrating around the pole $k_+=0$ inside momentum Feynman integrals. The
only available residual gauge fixing which is consistent with causality properties
of Feynman integrals is the one due to Mandelstam and Leibbrandt \cite{Man83,Lei83}
\be
{\frac1{[k_+]}}_{\scriptscriptstyle\rm ML} = \frac{k_-}{k_+ k_- + i0}\,.
\label{ML}
\ee
It has been checked for various gauge invariant quantities that to two-loop
accuracy this prescription leads to the same results as the Feynman gauge.
However, the calculations within the Mandelstam-Leibbrandt prescription are
rather involved and there exist another popular choice, the so-called principal
value prescription, defined as \cite{CurFurPet80}
\be
\label{PrincipalValuePrescription}
{\frac1{[k_+]}}_{\scriptscriptstyle\rm PV}
=
\frac12 \left(\frac1{k_+ + i\delta} + \frac1{k_+-i\delta} \right)
\, ,
\ee
with $\delta\to 0$. This prescription has proved to be viable and robust in
higher order QCD calculations of the evolution kernels and we will accept it
for two-loop calculations in our subsequent analysis.

To isolate divergences of Feynman integrals we will employ in parallel two
different regularization schemes---dimensional reduction (DRED) and dimensional
regularization (DREG). In both schemes one regularizes Feynman integrals by
setting the space-time dimension to $D = 4 - 2 \varepsilon$ and taking the
limit $\varepsilon\to 0$ afterwards. In the DRED scheme one keeps the number
of all tensor fields components (gauge bosons and fermions) to be the same
as in four dimensions \cite{Sie79} while in the DREG scheme this number is
$\varepsilon-$dependent. The two schemes are equally suitable for two-loop
calculations but the usage of the DRED is mandatory as long as one wants to
preserve the supersymmetry of Yang-Mills theory.

To determine the evolution kernel $\mathbb{H}$ we shall examine perturbative
corrections to the matrix elements of nonlocal light-cone operators \re{O3-SYM}
and \re{O3-QCD}. For $D \neq 4$, the ultraviolet divergences manifest themselves
as poles in $\varepsilon$ and we shall renormalize ``bare'' nonlocal light-cone
operators using the modified minimal subtraction procedure. Depending on the
choice of the DREG or DRED regularization schemes, it will be denoted as
$\overline{\rm MS}$ and $\overline{\rm DR}$, respectively. The renormalized
operator is defined as
\begin{equation}
\mathbb{O}^{\scriptscriptstyle\rm R}(z_1,z_2,z_3) = \mathbb{Z} \cdot
\mathbb{O}^{\rm (bare)}(z_1,z_2,z_3) \, , \qquad \mathbb{Z}(1/\varepsilon,g^2) =
1 + \sum_{n = 1}^\infty \frac{\mathbb{Z}^{[n]}(g^2)}{\varepsilon^n} \, ,
\label{O-renorm}
\end{equation}
with the subtraction ``constants'' $\mathbb{Z}^{[n]}(g^2)$ chosen in such a
manner that the Green functions $\vev{0|\mathbb{O}^{\scriptscriptstyle\rm R}
(z_1, z_2, z_3) \ldots|0}$ remain finite for $\varepsilon\to 0$.%
\footnote{It worth mentioning that the renormalization constants entering
\re{O-renorm} are certain integral operators acting on the light-cone coordinates
of $\mathbb{O}^{\rm (bare)}(z_1,z_2,z_3)$.} Substituting \re{O-renorm} into the
evolution equation \re{RG} one finds that the evolution operator is related to
the residue of the simple $1/\varepsilon$ pole
\begin{equation}
\mathbb{H} = - \frac{d\ln \mathbb{Z}(1/\varepsilon,g^2)}{d \ln \mu}
=
\frac{d}{d \ln g} \mathbb{Z}^{[1]}(g^2)
\, . \label{H from Z}
\end{equation}
Here in the second relation we used the fact that the gauge coupling constant
acquires a nontrivial dimension for $\varepsilon\neq 0$ and, as a consequence, the
beta-function in the $D$-dimensional Yang-Mills theory possesses an additional
$\varepsilon-$dependent term
\begin{equation}\label{beta-epsilon}
\frac{d\ln g^2}{d\ln\mu}
=
\beta_\varepsilon (g^2)
=
- 2 \varepsilon + \beta (g^2)
\, .
\end{equation}
As we will explain in detail in section \ref{ConfSymmBreakSection},
%\ref{ConformalConstraints},
the presence
of this term has far reaching consequences for the properties of the dilatation
operator. In particular, it implies that in Yang-Mills theory with $\beta(g^2)=0$
the conformal symmetry is broken for $\varepsilon\neq 0$.

We would like to stress that the renormalization constants in the  $\overline{\rm
MS}-$  and $\overline{\rm DR}-$renorma\-li\-za\-tion schemes are different but
they are related to each other though a finite renormalization. As a consequence,
the evolution operator depends on the renormalization scheme and one has to
distinguish the operators $\mathbb{H}_{\scriptscriptstyle\overline{\rm MS}}$ and
$\mathbb{H}_{\scriptscriptstyle\overline{\rm DR}}$. As we will show in
Sect.~\ref{SchemeDependence}, these operators differ starting from two-loop order
only and the difference amounts to a c-number. As a result, if the dilatation
operator possesses a hidden symmetry it will hold independently of the chosen
renormalization procedure.

%%%%%%%%%%%%%%%%%%%%%%%%%%%%%%%%%%%%%%%%%%%%%%%%%%%%%%%%%%%%%%%%%%%%%
%            Figure
%%%%%%%%%%%%%%%%%%%%%%%%%%%%%%%%%%%%%%%%%%%%%%%%%%%%%%%%%%%%%%%%%%%%%
\begin{figure*}[t]
\begin{center}
\mbox{
\begin{picture}(0,72)(200,0)
\psfrag{V}[cc][cc]{$\mathbb{V}$}
\put(0,-8){\insertfig{13}{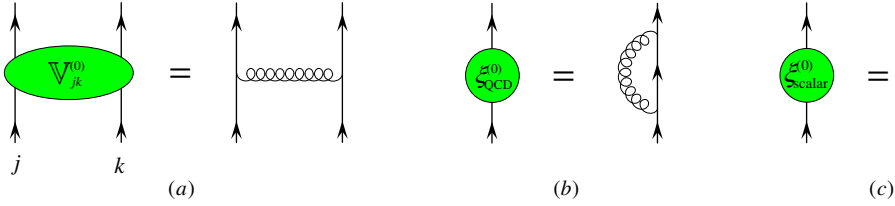}}
\end{picture}
}
\end{center}
\caption{\label{oneloopkernel} One-loop diagrams in the light-cone gauge
contributing to the pair-wise kernel ($a$) and the self-energy due to gluons
($b$) and scalars ($c$).}
\end{figure*}
%%%%%%%%%%%%%%%%%%%%%%%%%%%%%%%%%%%%%%%%%%%%%%%%%%%%%%%%%%%%%%%%%%%%%

%%%%%%%%%%%%%%%%%%%%%%%%%%%%%%%%%%%%%%%%%%%%%%%%%%%%%%%%%%%%%%%%%%%%%
\subsection{One-loop}
\label{OneLoopDilOper}
%%%%%%%%%%%%%%%%%%%%%%%%%%%%%%%%%%%%%%%%%%%%%%%%%%%%%%%%%%%%%%%%%%%%%

Examining Feynman diagrams contributing to $\mathbb{O} (z_1, z_2, z_3)$ it is
easy to see that to the $k^{\rm th}$ order in $\lambda$ the interaction could
occur between $(k+1)$ particles at most. As a consequence, to one-loop order
$\mathbb{H}^{(0)}$ receives contribution from pair-wise interactions as well as
their self-energies. To determine the one-loop evolution kernel, one has to
calculate one-loop corrections to the nonlocal light-cone operator $\mathbb{O}
(z_1,z_2,z_3)$, renormalize it according to \re{O-renorm} and, finally, evaluate
$\mathbb{H}^{\scriptscriptstyle (0)}$ using \re{H from Z}. We sketch below the
intermediate steps while the details of the calculation can be found in
Refs.~\cite{BelDerKorMan04,BelRad05}.

Since to one-loop order the interaction can only occur between two fermions,
the dilatation operator has a pair-wise structure
\be
\label{LOkernel}
\mathbb{H}^{\scriptscriptstyle (0)}_{\phantom{i}}
=
\mathbb{H}_{12}^{\scriptscriptstyle (0)}
+
\mathbb{H}_{23}^{\scriptscriptstyle (0)}
+
\mathbb{H}_{31}^{\scriptscriptstyle (0)} + 3\Gamma^{(0)} \, .
\ee
Here the two-particle kernel $\mathbb{H}_{jk}^{\scriptscriptstyle (0)}$ acts
locally on the $j^{\rm th}$ and $k^{\rm th}$ fields while $\Gamma^{(0)}$
receives contribution from the self-energy corrections to the fermion. They are
determined by the Feynman diagrams shown in Fig.\ \ref{oneloopkernel}. Notice
that in the SYM theory, the self-energy diagrams involve both gauge field and
scalar exchanges while the diagram for $\mathbb{H}_{jk}^{\scriptscriptstyle (0)}$
contains solely the gluon exchange. The reason for this is the following. One
can straightforwardly deduce from the Lagrangian \re{L-SYM} that the scalar
exchange corresponds to the transition $\lambda^{\{A} \lambda^{B\}} \to
\bar\lambda_C \bar\lambda_D$ (symmetrized over $\scriptstyle A$ and $\scriptstyle B$)
involving the scalar propagator $\vev{\phi^{AC} \phi^{BD}}$. As follows from
\re{L-SYM}, this propagator vanishes for $\mathcal{N}=1$ and $\mathcal{N}=2$,
while for $\mathcal{N}=4$ it is proportional to $\sim\varepsilon^{ABCD}$ and
vanishes upon symmetrization over the isotopic indices $\scriptstyle A$ and
$\scriptstyle B$.

The contribution of the diagram in Fig.\ \ref{oneloopkernel}(a) to the one-loop
dilatation operator in QCD and SYM only differs by an overall color factor. For
gauginos in the adjoint representation, the color factor equals $T_j^a T_k^a$
with $(T_j)^a_{bc} = i f^{abc}$ being the $SU(N_c)$ generators in the adjoint
representation acting on the color indices of the $j^{\rm th}$ particle. Since
the three-fermion operators have zero total color charge, one has $\sum_{j=1,2,3}
T^a_j=0$ and, therefore,
\be
\label{LOColorStructureAdj}
T_j^a T_k^a
=
\frac12 (T_l^a)^2 -\frac12(T_j^a )^2 -\frac12( T_k^a)^2 = -\frac 12 N_c\,,
\ee
with $T_l^a=-(T_j + T_k)^a$ and $(T_j^a )^2=N_c$ being the Casimir operator in
the adjoint representation of the $SU(N_c)$. For quarks in the fundamental
representation of the $SU(N_c)$, the color factor equals \
\be
\label{LOColorStructureFun}
\varepsilon_{i_1\ldots i'_j\ldots i'_k \ldots i_{N_c}} (t^a)_{i'_ji_j}
(t^a)_{i'_ki_k} =\varepsilon_{i_1\ldots i_j\ldots i_k \ldots
i_{N_c}}\left(-\frac{N_c+1}{2N_c}\right)\,,
\ee
where one takes into account the identity $(t^a)_{jk} (t^a)_{nm} = \delta_{jm}
\delta_{kn}/2- \delta_{jk}\delta_{nm}/(2N_c)$. We remind that the baryon operator
\re{O3-QCD} is only well-defined for $N_c=3$ in which case
$-(N_c+1)/(2N_c)=-2/3$. As a result, the contribution of the diagram in Fig.\
\ref{oneloopkernel}(a) to the one-loop evolution kernel for the light-cone
operators \re{O3-SYM} and \re{O3-QCD} is accompanied by the prefactors
\be
\label{CouplingConstant}
\lambda_{\rm fun} = \frac{g^2}{8 \pi^2} \frac{4}{3} \, , \qqqquad \lambda_{\rm adj}
= \frac{g^2}{8 \pi^2} N_c \, ,
\ee
correspondingly. In what follows we shall use $\lambda_{\rm fun}$ and $\lambda_{\rm
adj}$ as parameters of the perturbative expansion of the dilatation operator
\re{H-dec}. In this case, the two-particle kernel $\mathbb{H}_{jk}^{(0)}$ takes
the same form for the light-cone operators in the adjoint, Eq.~\re{O3-SYM}, and
fundamental, Eq.~\re{O3-QCD}, representations
\be
\label{LightConeKernel} [ \mathbb{H}_{12}^{\scriptscriptstyle (0)} \cdot
\mathbb{O} ] (z_1, z_2, z_3) = \int_0^1 \frac{d \alpha}{\alpha} \bar\alpha^{2 j -
1} {}\Big[ 2 \mathbb{O} (z_1, z_2, z_3) - \mathbb{O} (\bar\alpha z_1 + \alpha
z_2, z_2, z_3) - \mathbb{O} (z_1, \alpha z_1 + \bar\alpha z_2, z_3) {}\Big],
\ee
where $\bar\alpha \equiv 1 - \alpha$ and $j = 1$ is the conformal spin of the
quark/gaugino. For the later use, we shall display the dependence of the
evolution kernels on the conformal spin $j$. According to
Eq.~\re{LightConeKernel}, the operator $\mathbb{H}_{nk}^{(0)}$ has a transparent
physical interpretation: it displaces $n^{\rm th}$ and $k^{\rm th}$ particles
along the light-cone in the direction of each other.

The gauge field contribution to the self-energy shown in Fig.\
\ref{oneloopkernel}(b) is proportional to the quadratic Casimir operator:
$C_A=N_c$ for the gaugino in the adjoint representation of the $SU(N_c)$ and
$C_F=(N_c^2-1)/(2N_c)=4/3$ for the quark in the fundamental representation of the
$SU(N_c=3)$ group. In addition, in the SYM theory one has to include the
contribution of scalars to gaugino self-energy (see in Fig.\
\ref{oneloopkernel}(c)) which is proportional to $\sim n_s =2(\mathcal{N}-1)$.
Combining together the two contributions one obtains the following expressions
for the normalization constant $\Gamma^{(0)}$ in \re{LOkernel}
\be
\Gamma^{(0)}_{\scriptscriptstyle\rm QCD} = \frac{1}{2} \,,\qqqquad
\Gamma^{(0)}_{\scriptscriptstyle\rm SYM}= \frac{1}{2}  \mathcal{N} \,.
\label{Gamma0}
\ee
Notice that $\Gamma^{(0)}_{\scriptscriptstyle\rm SYM}$ coincides with the QCD
expression for $\mathcal{N}=1$. Combining together \re{LOkernel},
\re{LightConeKernel} and \re{Gamma0}, one obtains the one-loop dilatation
operator for the three-particle quark and gaugino operators in QCD and SYM
theories, respectively, in the coordinate representation. As follows from
\re{LightConeKernel}, the two-particle kernel annihilates the local operator
$\mathbb{O}(0,0,0)$ and, therefore, $3\Gamma^{(0)}_{\scriptscriptstyle\rm SYM}$
determines the anomalous dimension of the local gaugino operator \re{B0}, in
agreement with \re{Gamma}.

Let us transform the obtained expression for $\mathbb{H}^{\scriptscriptstyle
(0)}_{\phantom{i}}$ into the momentum representation. Applying \re{H-plane waves}
and performing the Fourier transformation one finds that $\mathbb{V}
(\mbox{\boldmath $u$}| \mbox{\boldmath $v$} )$ has the same structure as
\re{LOkernel}
\ba
\label{PairWiseLOkernel}
\mathbb{V}^{(0)} (\mbox{\boldmath $u$}| \mbox{\boldmath $v$} )
\!\!\!&=&\!\!\!
\left[ \mathbb{V}^{(0)} ( u_1, u_2 | v_1, v_2 ) \right]_+
\\
&+&\!\!\! \left[ \mathbb{V}^{(0)} ( u_2, u_3 | v_2, v_3 ) \right]_+ + \left[
\mathbb{V}^{(0)} ( u_3, u_1 | v_3, v_1 ) \right]_+ - 3\Gamma^{(0)}
\delta(u_1-v_1)\delta(u_2-v_2) \, , \nonumber
\ea
with the two-particle kernel given by a well-known expression
\cite{BukFroKurLip85,BelMul97}
\be
\label{OneLoopMomDO}
\mathbb{V}^{(0)} ( u_1, u_2 | v_1, v_2 )
=
\bigg[
\left( \frac{u_1}{v_1} \right)^{2 j - 1} \frac{\Theta (u_1, v_1)}{v_1 - u_1}
+
\left( \frac{u_2}{v_2} \right)^{2 j - 1} \frac{\Theta (u_2, v_2)}{v_2 - u_2}
\bigg]
\delta (u_1 + u_2 - v_1 - v_2)
\, . \nonumber
\ee
Here the notation was introduced for a generalized ``step'' function
\be
\label{step-function} \Theta (u_n, v_n) = \theta(u_n) \theta (v_n - u_n) -
\theta(- u_n) \theta (u_n - v_n) \, .
\ee
The symbol $[\ldots]_+$ stands for the plus-distribution
\be
\label{SinglePlusDistr}
\left[ \varphi(u,v) \right]_+ \equiv \varphi(u,v) - \delta (u - v) \int d u'
\varphi(u',v) \, ,
\ee
with $\varphi(u,v)$ being a test function. This distribution regularizes the
end-point singularity of \re{OneLoopMomDO} for $v_i \to u_i$ in such a way that
the evolution kernel \re{PairWiseLOkernel} has finite moments \re{V-moments}.
Notice that due to the total momentum conservation, $\sum u_n = \sum v_n$,
Eq.~\re{3-measure}, the two-particle evolution kernel \re{OneLoopMomDO} preserves
the momentum of the third particle, $u_3=v_3$. By the same token, the constant
term in the evolution kernel \re{PairWiseLOkernel} preserves the momenta of all
particles, $u_n=v_n$ with $n=1,2,3$.

The variables $v_i$ and $u_i$ have the meaning of the light-cone momenta of
particles. Calculating the moments of the evolution kernel \re{V-moments}, one
can assign to $v_i$ arbitrary real values. For given $v_i$, the integration
region over the $u-$variables is not arbitrary and it is determined by the step
functions \re{step-function}. It is instructive to consider a special case, the
so-called Brodsky-Lepage (BL) limit \cite{BroLep79},
\be
0 \le v_1,\ v_2,\ v_3 \le 1\,,\qquad v_1+v_2+v_3=1\,.
\label{simplex}
\ee
It is easy to see that the step function reduces in this limit to $\Theta (u_i,
v_i) \stackrel{\rm BL}{=} \theta(u_i) \theta(v_i - u_i)$ and, as a consequence,
\ba
\label{OneLoopMomDOBL}
\left[ \mathbb{V}^{(0)} ( u_1, u_2 | v_1, v_2 ) \right]_+
\!\!\!&\stackrel{\rm BL}{=}&\!\!\!
\bigg[
\left( \frac{u_1}{v_1} \right)^{2j - 1}
\frac{\theta(u_1)\theta(v_1 - u_1)}{v_1 - u_1}
\\
&&\!\!\!\!+
\left( \frac{u_2}{v_2} \right)^{2j - 1}
\frac{\theta(u_2)\theta(v_2 - u_2)}{v_2 - u_2}
\bigg]_+ \delta (u_1 + u_2 - v_1 - v_2)
\, . \nonumber
\ea
Substituting this expression into \re{PairWiseLOkernel}, one verifies that the
possible values of the $u-$variables are restricted to the same simplex as in
\re{simplex}
\be
0 \le u_1,\ u_2,\ u_3 \le 1\,,\qquad u_1+u_2+u_3=1\,.
\label{u-simplex}
\ee
This property has a simple physical interpretation. It implies that for all three
particles moving forward in time along the light-cone direction $n_\mu$, the
interaction described by the evolution kernel \re{PairWiseLOkernel} can turn none
of the particles to propagate backward in time. Remarkably enough, this property
is general enough to hold to all order of perturbation theory~\cite{Rad84,Jaf83}.
Below in Sect.~\ref{DilOpTwoLoop} we will demonstrate its validity to two loops.
In the following exposition we will stick to the BL-representation since, on the
one hand, it simplifies the step-function structure of the momentum space dilatation
operator and, on the other hand, suffices for evaluation of anomalous dimension
mixing matrix. A simple procedure \cite{MulDitRobGeyHor98} for reconstruction of
the complete evolution operator without the restriction to the simplex
\re{simplex} is given in Appendix~\ref{ForwardLimitAppendix}.

%%%%%%%%%%%%%%%%%%%%%%%%%%%%%%%%%%%%%%%%%%%%%%%%%%%%%%%%%%%%%%%%%%%%%
\section{Dilatation operator: 2-loops}
\label{DilOpTwoLoop}
%%%%%%%%%%%%%%%%%%%%%%%%%%%%%%%%%%%%%%%%%%%%%%%%%%%%%%%%%%%%%%%%%%%%%

In the previous section we have calculated the one-loop evolution kernel for the
three-particle gaugino/quark operators \re{O3-SYM} and \re{O3-QCD}. We
demonstrated that it is given by the same, universal expression
\re{LightConeKernel} both in QCD and in SYM theories with arbitrary number of
supercharges $\mathcal{N}$. The fact that the fermions are defined in different
representations of the gauge group manifests itself in the different color
factors accompanying the coupling constant, Eq.~\re{CouplingConstant}. Let us now
extend our analysis to two loops. As we will see, the above mentioned
universality gets lost beyond leading order and one has to consider separately
the gaugino and quark operators. In the former case, we will start with the
$\mathcal{N}=1$ SYM theory and will gradually increase the diversity of the
particle's content by going over to $\mathcal{N}=2$ and $\mathcal{N}=4$ theories.
In this way, we will be able to clearly identify whether emerging integrability
phenomena are sensitive to the representation in which the fermions fields are
defined as well as to the presence of scalars and, on top of it, supersymmetry.

Our analysis will be entirely formulated in the momentum space, i.e., we will
calculate the two-loop kernel $\mathbb{V}$ of the evolution equation (\ref{BL}).
It is straightforward to translate the obtained expressions from the momentum
to the configuration space by applying \re{H-plane waves}. As we will argue,
to two-loop order, the three-fermion operators \re{O3-SYM} and \re{O3-QCD}
evolve autonomously and the perturbative expansion of the evolution kernel
reads in the momentum representation
\be
\mathbb{V} (\bit{u} | \bit{v})
=
\lambda \, \mathbb{V}^{(0)} (\bit{u} | \bit{v})
+
\lambda^2 \, \mathbb{V}^{(1)} (\bit{u} | \bit{v})
+
\mathcal{O} (\lambda^3)
\, ,
\ee
where depending on the representation of the fermion fields, the coupling
constant $\lambda$ is given by the corresponding expression in
(\ref{CouplingConstant}). The one-loop result was given earlier in Eq.\
(\ref{PairWiseLOkernel}). In addition, to simplify the calculation we shall apply
the Brodsky-Lepage limit and assume that the $v-$ and $u-$variables take
nonnegative values and belong to the simplices \re{simplex} and \re{u-simplex},
respectively. This does not lead to a loss of generality since the resulting
expression for $\mathbb{V} (\bit{u} | \bit{v})$ can be uniquely continued to
arbitrary values of the scaling variables. As before, the calculation will be
performed in the light-like axial gauge $A_+(x)=0$ within the $\overline{{\rm
DR}}-$ and $\overline{{\rm MS}}-$renormalization schemes.

The two-loop Feynman diagrams contributing to $\mathbb{V}^{(1)}$ in the
light-cone gauge can be separated into three different sets sorted according
to the number of fermions involved in the interaction, i.e., one-, two- and
three-particle irreducible contributions. Their contribution to the evolution
kernel can be written as
\be
\label{TwoLoopThreeParticle}
\mathbb{V}^{(1)}(\bit{u} | \bit{v}) = \mathbb{V}^{(1)}_{123} +
\mathbb{V}^{(1)}_{231} + \mathbb{V}^{(1)}_{312} + \mathbb{V}^{(1)}_{12} +
\mathbb{V}^{(1)}_{23} + \mathbb{V}^{(1)}_{31}
+ \mathbb{V}^{(1)}_{1} + \mathbb{V}^{(1)}_{2} + \mathbb{V}^{(1)}_{3}
\, ,
\ee
where the additional terms have been added to ensure the invariance of the
evolution kernel under cyclic permutations of quarks. Denoting the kernel
of the operators $\mathbb{V}^{(1)}_{123}$ and $\mathbb{V}^{(1)}_{12}$
as $\mathbb{V}^{(1)}(u_1,u_2,u_3|v_1,v_2,v_3)$ and $\mathbb{V}^{(1)}
(u_1,u_2|v_1,v_2)$, respectively, one has
\be
\mathbb{V}^{(1)}_{ijk} =\mathbb{V}^{(1)}(u_i,u_j,u_k|v_i,v_j,v_k)
\,,\qquad
\mathbb{V}^{(1)}_{jk} =\mathbb{V}^{(1)}(u_j,u_k|v_j,v_k)
\,.
\label{V-symmetry}
\ee
The three-particle irreducible kernel $\mathbb{V}^{(1)}_{123}$ receives
contributions from the diagrams shown in Fig.~\ref{gluonthreeparticle}. The set
of two-loop diagrams contributing to the two-particle kernel
$\mathbb{V}^{(1)}_{12}$ is displayed in Figs.~\ref{twoloopkernel} and
\ref{scalarvertex}. Among these diagrams one can distinguish those without
scalars (see Fig.\ \ref{twoloopkernel}). They are exactly the same as involved in
renormalization of twist-two maximal-helicity operators in
QCD~\cite{Vog97,BelMulFre00}. However, as we will see momentarily, the important
difference with the twist-two operators is that for three-fermion operators
\re{O3-SYM} and \re{O3-QCD} each pair of particles has a nonvanishing color
charge and this affects the color factors accompanying individual diagrams.
Finally, the self-energy corrections to gaugino/quark fields induce c-number
corrections $\mathbb{V}^{(1)}_i$, see Fig.\ \ref{GauginoSEgluons2loop}, which
have the structure
\be
\label{V-one-to-xi}
\mathbb{V}^{(1)}_i = \xi^{(1)}_i \delta (u_1 - v_1) \delta (u_2 - v_2) \, ,
\ee
with $\xi^{(1)}_i$ being the residue of the renormalization constant of $i^{\rm th}$
fermion line to two loop order.

Let us consider three different contributions to $\mathbb{V}^{(1)}(\bit{u} | \bit{v})$
one after another.

%%%%%%%%%%%%%%%%%%%%%%%%%%%%%%%%%%%%%%%%%%%%%%%%%%%%%%%%%%%%%%%%%%%%%
\subsection{Three-particle contributions}
\label{ThreeParticleIrrContribution}
%%%%%%%%%%%%%%%%%%%%%%%%%%%%%%%%%%%%%%%%%%%%%%%%%%%%%%%%%%%%%%%%%%%%%

%%%%%%%%%%%%%%%%%%%%%%%%%%%%%%%%%%%%%%%%%%%%%%%%%%%%%%%%%%%%%%%%%%%%%
%            Figure
%%%%%%%%%%%%%%%%%%%%%%%%%%%%%%%%%%%%%%%%%%%%%%%%%%%%%%%%%%%%%%%%%%%%%
\begin{figure*}[t]
\begin{center}
\mbox{
\begin{picture}(0,65)(160,0)
\psfrag{V}[cc][cc]{$\mathbb{V}$}
\put(0,-8){\insertfig{11}{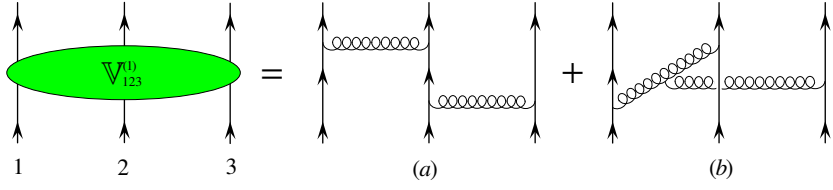}}
\end{picture}
}
\end{center}
\caption{\label{gluonthreeparticle} The three-particle contributions (with mirror
symmetrical graphs tacitly implied) to the three-particle dilatation operator.}
\end{figure*}
%%%%%%%%%%%%%%%%%%%%%%%%%%%%%%%%%%%%%%%%%%%%%%%%%%%%%%%%%%%%%%%%%%%%%

The three-particle irreducible kernel $\mathbb{V}^{(1)}_{123}$ receives
contribution from Feynman diagrams which either represent an iteration
of a single-particle exchange, or contain a triple gluon vertex (see
Figs.\ \ref{gluonthreeparticle} (a) and (b), respectively). In QCD and
$\mathcal{N}=1$ SYM theory, such kind of diagrams can only involve gluon
exchanges while in the $\mathcal{N}=2$ and $\mathcal{N}=4$ theories
scalars may, in principle, propagate instead of gluons in internal lines.
As it was explained in Sect.~\ref{OneLoopDilOper}, by virtue of the maximal
$R-$charge of the three-particle operator \re{O3-SYM}, the scalar fields do not
contribute to the two-particle kernel $\mathbb{H}_{12}^{\scriptscriptstyle (0)}$
to one loop order. The very same argument applies to the three-particle
irreducible kernel $\mathbb{V}^{(1)}_{123}$. In other words, the diagrams
with scalar exchanges vanish identically and the evolution kernel
$\mathbb{V}^{(1)}_{123}$ takes the same form in SYM theories
with $\mathcal{N}=1,2,4$.

Another interesting feature of the kernel $\mathbb{V}^{(1)}_{123}$ is that the
diagram in Fig.\ \ref{gluonthreeparticle} (b) involving the three-gluon vertex
vanishes identically. This property holds true independently on the
representation in which the fermion fields are defined and, therefore, it is
valid both in QCD and SYM theories. Namely, for the quarks in the fundamental
representation of the $SU(3)$ the color factor $C$ corresponding to the diagram
in Fig.\ \ref{gluonthreeparticle} (b) equals
\be
\varepsilon^{i_1i_2i_3}f^{abc} (t^a)_{i_1j_1} (t^b)_{i_2j_2} (t^c)_{i_3j_3}  =
C\, \varepsilon^{j_1j_2j_3}\,.
\ee
The two sides of this relation have an opposite parity under permutations $j_1
\rightleftarrows j_2$ leading to $C=0$. In the similar manner, for the gaugino in
the adjoint representation of the $SU(N_c)$ the color factor equals
\be
f^{abc} T^a_1 T^b_2 T^c_3  = - f^{abc} T^a_1 T^b_2 (T^c_1 + T^c_2) = 0 \, ,
\label{zero}
\ee
where $(T^a_j)_{bc} = i f^{abc}$ is the color charge of $j^{\rm th}$ particle and
$f^{abc}=-f^{acb}$ are the $SU(N_c)$ structure constants. Here in the first
relation we used the color neutrality of the single-trace three-particle
operator, i.e., $\sum_{j = 1, 2, 3} T^a_j = 0$, and applied the identity $f^{abc}
T^a_j T^b_j = -\frac{i}2 N_c T^c_j$. In arriving at \re{zero}, it was crucial
that the color-singlet operator \re{O3-SYM} is built from $L=3$ particles. For $L
> 3$, the contribution of the diagram in Fig.\ \ref{gluonthreeparticle} (b) to the
multi-particle single-trace operator is different from zero but it is suppressed
in the multi-color limit as $\sim 1/N_c^2$ since the diagram is nonplanar.

We conclude that the three-particle irreducible kernel $\mathbb{V}^{(1)}_{123}$
is determined by the Feynman diagram shown in Fig.~\ref{gluonthreeparticle} (a)
and its mirror image. The two diagrams describe iteration of one-gluon exchange
and the corresponding color factor is merely the square of the one-loop color
factor. For gluinos in the adjoint representation it yields $ T_1^a T_2^a T_2^b
T_3^b = \left( - \frac{N_c}{2} \right)^2 \, , $ where we have used the identity
(\ref{LOColorStructureAdj}) twice. For fundamental quarks, one uses in the same
fashion Eq.\ (\ref{LOColorStructureFun}). In both cases the color factors can be
absorbed into the redefinition of the coupling constants \re{CouplingConstant}
and, as a result, $\mathbb{V}^{(1)}_{123}$ has the same form in QCD and in all
SYM theories.

\begin{figure*}[t]
\begin{center}
\mbox{
\begin{picture}(0,185)(100,0)
\put(0,-3){\insertfig{7}{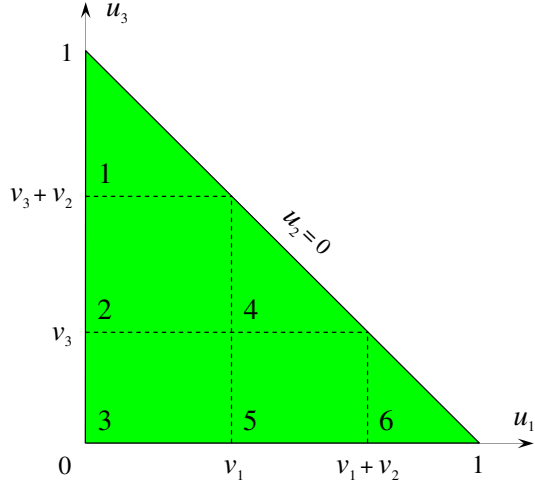}}
\end{picture}
}
\end{center}
\caption{\label{region} Phase-space regions of three-particle irreducible
kernel.}
\end{figure*}

The evolution kernel $\mathbb{V}^{(1)}_{123}(\bit{u} | \bit{v})$ depends on two
sets of momenta. As was already mentioned, we restrict our analysis to the values
of variables belonging to the simplices (\ref{simplex}) and \re{u-simplex}. For
given values of the $v-$variables and $u_1+u_2+u_3=1$, the three-particle
irreducible kernel $\mathbb{V}^{(1)}_{123}(\bit{u} | \bit{v})$ has support inside
the triangle $0\le u_1,u_3\le 1$ and $u_1+u_3\le 1$ shown in Fig.\ \ref{region}.
Three sides of the triangle corresponds to $u_1=0$, $u_2=0$ and $u_3=0$. Due to a
larger variety of possibilities of momentum redistributions between three interacting
fields, $\mathbb{V}^{(1)}_{123}(\bit{u} | \bit{v})$ has a more complex structure
as compared with the two-particle kernel \re{OneLoopMomDOBL}. A simple analysis
demonstrates that there exist six distinct regions in the phase space
displayed in Fig.\ \ref{region}: the diagram in Fig.\ \ref{gluonthreeparticle}
(a) contains two internal gluon lines carrying the light-cone momenta $v_1 - u_1$
and $u_3 - v_3$ and the virtual fermion line with the momentum $v_1 + v_2 - u_1$.
For the mirror symmetric diagram the latter momentum reads $v_3 + v_2 - u_3$. In
distinction with the external $u-$variables, the momenta flowing through internal
lines are not necessarily positive in the Brodsky-Lepage limit. For $v_1 > u_1$ we
can readily identify two regions: region 3 for $v_3 > u_3$ and combined regions 1
and 2 for $v_3 < u_3$. In these kinematical regions the internal fermionic line
does not bring in any new restrictions since the momentum flow is always positive
$v_1 + v_2 - u_1 > 0$. Considering now the situation of the negative momentum flow
in the first gluon line, i.e., $v_1 < u_1$, one immediately finds that the virtual
quark/gluino line can have either positive ($v_1 + v_2 > u_1$) or negative ($v_1
+ v_2 < u_2$) momentum flow, which is laid over the restriction from the second
gluon line $v_3 \gtrless u_3$. From these we find three regions in the phase
space labelled as 4, 5 and 6. Notice that the boundary $v_1 + v_2 = u_1$ is
reflected from the boundary $u_2 = 0$ of the support region into the boundary
$u_3 = v_3$. The consideration of the mirror symmetric diagram to Fig.\
\ref{gluonthreeparticle} (a) allows one to further separate the regions 1 and 2
from the positivity/negativity of the momentum flow of the internal fermion line.
The above analysis covers all regions in the support region of the two-loop
three-particle irreducible kernel.

Summarizing, one decomposes the three-particle irreducible kernel as follows
\begin{eqnarray}
\label{Def-Vcon}
\mathbb{V}_{123}^{(1)} (\bit{u} | \bit{v}) \!\!\!&=&\!\!\! \theta (u_3 -
\bar{v}_1) \mathbb{F}_1 (\bit{u} | \bit{v}) + \theta(\bar{v}_1 - u_3) \theta(v_1
- u_1)  \theta(u_3 - v_3) \mathbb{F}_2 (\bit{u} | \bit{v})
\\
&+&\!\!\!
\theta (v_1 - u_1)\theta (v_3 - u_3) \mathbb{F}_3 (\bit{u} | \bit{v})
+
\theta (u_1 - v_1)\theta (u_3 - v_3) \mathbb{F}_4 (\bit{u} | \bit{v})
\nonumber\\
&+&\!\!\!
\theta(v_3 - \bar{u}_1) \mathbb{F}_5 (\bit{u} | \bit{v})
+
\theta(\bar{u}_1 - v_3) \theta(u_1 - v_1) \theta(v_3 - u_3)
\mathbb{F}_6 (\bit{u} | \bit{v})
\, , \nonumber
\end{eqnarray}
where $\bar u_j=1-u_j$, $\bar v_j=1-v_j$ and the function $\mathbb{F}_i(\bit{u}
| \bit{v})$ denotes the contribution of the $i^{\rm th}$ region on the phase
diagram in Fig.\ \ref{region}. It is tacitly assumed that the $u-$ and
$v-$momenta belong to the simplices \re{u-simplex} and \re{simplex},
respectively. The sum of the diagrams shown in Fig.~\ref{gluonthreeparticle} is
symmetric under the interchange of the $1^{\rm st}$ and $3^{\rm rd}$ lines.
Therefore the kernel \re{Def-Vcon} has to be invariant under the interchange
of the corresponding momenta $\{u_1,u_3 | v_1,v_3\} \leftrightarrow \{u_3,u_1 |
v_3,v_1\}$, which yields the following relations
\begin{eqnarray}
\label{SymmetryRelations}
&&\mathbb{F}_3 (u_1,u_2,u_3 | v_1,v_2,v_3) = \mathbb{F}_3 (u_3,u_2,u_1 | v_3,v_2,v_1)
\, , \\
&&\mathbb{F}_4 (u_1,u_2,u_3 | v_1,v_2,v_3) = \mathbb{F}_4 (u_3,u_2,u_1 | v_3,v_2,v_1)
\, , \nonumber\\
&&\mathbb{F}_5 (u_1,u_2,u_3 | v_1,v_2,v_3) = \mathbb{F}_2 (u_3,u_2,u_1 | v_3,v_2,v_1)
, \\
&&\mathbb{F}_6 (u_1,u_2,u_3 | v_1,v_2,v_3) = \mathbb{F}_1 (u_3,u_2,u_1 |
v_3,v_2,v_1) \, , \nonumber
\end{eqnarray}
so that the three-particle kernel \re{Def-Vcon} only involves four nontrivial
functions. Moreover, as we will see in a moment, one of these functions vanishes.

%%%%%%%%%%%%%%%%%%%%%%%%%%%%%%%%%%%%%%%%%%%%%%%%%%%%%%%%%%%%%%%%%%%%%
%            Figure
%%%%%%%%%%%%%%%%%%%%%%%%%%%%%%%%%%%%%%%%%%%%%%%%%%%%%%%%%%%%%%%%%%%%%
%%%%%%%%%%%%%%%%%%%%%%%%%%%%%%%%%%%%%%%%%%%%%%%%%%%%%%%%%%%%%%%%%%%%%

To determine the functions $\mathbb{F}_i$ one has to evaluate the three-particle
connected diagrams displayed in Fig.\ \ref{gluonthreeparticle} using
particular renormalization scheme to treat ultraviolet divergences and
extract $\mathbb{V}_{123}^{(1)} (\bit{u} | \bit{v})$ from a single pole
$\sim 1/\varepsilon$ contribution using \re{H from Z} . The calculation
can be straightforwardly performed either in the covariant or light-cone
formalism. We found that the contribution of diagrams in Fig.\
\ref{gluonthreeparticle} to $\mathbb{V}_{123}^{(1)} (\bit{u} | \bit{v})$
is the same in the $\rm \overline{DR}-$ and $\rm \overline{MS}-$schemes.
It leads to the following expressions for the $\mathbb{F}-$functions
\begin{eqnarray}
\label{Def-F1}
\mathbb{F}_1 \!\!\!&=&\!\!\!
\ft12 f(u_1, v_1) f(\bar{v}_1 - u_3, v_2)
\ln \frac{(u_1 - v_1)(u_3 - v_3)}{(\bar{v}_1 - u_3)(\bar{u}_1 - v_3)}
\\
&+&\!\!\!
\ft12 f(u_1, v_1) f(u_2, \bar{u}_1 - v_3)
\ln \frac{u_1 (\bar{u}_1 - v_3)^2}{u_2 \bar{v}_3 (u_3 - v_3)}
+
\ft12 f(u_2, \bar{v}_1 - u_3) f(\bar{v}_1 - u_3, v_2)
\ln \frac{v_1 (\bar{u}_1 - v_3)}{\bar{v}_3 (v_1 - u_1)}
\, , \nonumber\\
\label{Def-F2}
\mathbb{F}_2 \!\!\!&=&\!\!\!
\ft12 f(u_1, v_1) f(\bar{v}_1 - u_3, v_2)
\ln \frac{\bar{u}_3 (u_3 - v_3)}{u_1 (\bar{u}_1 - v_3)}
\\
&+&\!\!\!
\ft12 f(u_1, v_1) f(u_2, \bar{u}_1 - v_3)
\ln \frac{u_1 (\bar{u}_1 - v_3)^2}{u_2 \bar{v}_3 (u_3 - v_3)}
+
\ft12 f(u_2, \bar{u}_1 - v_3) f(\bar{u}_1 - v_3, v_2)
\ln \frac{\bar{u}_3 (\bar{u}_1 - v_3)}{u_2 \bar{v}_3}
\, , \nonumber\\
\label{Def-F3}
\mathbb{F}_3 \!\!\!&=&\!\!\! 0
\, , \\
\label{Def-F4}
\mathbb{F}_4 \!\!\!&=&\!\!\!
\ft12 f(u_2, \bar{u}_1 - v_3) f(\bar{u}_1 - v_3, v_2)
\ln\frac{\bar{u}_1 \bar{u}_3 (\bar{v}_1 - u_3)(\bar{u}_1 - v_3)}{u_2^2 \bar{v}_1 \bar{v}_3}
\\
&+&\!\!\!
\ft12 f(u_1, v_1) f(u_2, \bar{u}_1 - v_3)
\ln\frac{u_1 v_2}{\bar{v}_3 (u_1 - v_1)}
+
\ft12 f(u_1, v_1) f(\bar{v}_1 - u_3, v_2)
\ln \frac{\bar{u}_3 (u_1 - v_1)}{u_1 (\bar{v}_1 - u_3)}
\nonumber\\
&+&\!\!\! \ft12 f(u_2, \bar{v}_1 - u_3) f(u_3, v_3) \ln \frac{u_3 v_2}{\bar{v}_1
(u_3 - v_3)} + \ft12 f (\bar{u}_1 - v_3, v_2) f (u_3, v_3) \ln \frac{\bar{u}_1
(u_3 - v_3)}{u_3 (\bar{u}_1 - v_3)} \, , \nonumber
\end{eqnarray}
with $\bar v_i=1-v_i$ and $\bar u_i=1-u_i$. The remaining functions can be
deduced making use of the symmetry relations (\ref{SymmetryRelations}). Here the
notation was introduced for the function
\be
f (u, v) = \frac{u}{v}\frac{1}{v - u} \, ,
\label{f-def}
\ee
which we already encountered at one loop, see Eq.~\re{OneLoopMomDOBL}.

A few comments are in order. According to \re{Def-F3}, the third region in
Fig.~\ref{region} produces a vanishing contribution to the three-particle kernel.
This comes about as a result of cancellation of the contribution of the diagram
shown in Fig.\ \ref{gluonthreeparticle} (a) against the one coming from a mirror
symmetric diagram.

It is known from the QCD calculations that, in general, two-loop evolution
kernels involve special functions (dilogarithms ${\rm Li}_2(x)$). Still, the
two-loop result for $\mathbb{V}_{123}^{(1)} (\bit{u} | \bit{v})$ is represented
in terms of elementary functions only: the $\mathbb{F}-$functions are given by a
product of one-loop elements \re{f-def} dressed by logarithmic dependence on the
light-cone momenta.

The function $f(u,v)$ has a pole at $u=v$. In the expression for the one-loop
evolution kernel, Eq.~\re{OneLoopMomDOBL}, this pole is regularized by the
`+'--prescription. To two loops, the $\mathbb{F}-$functions involve the product
of two $f-$functions and develop poles at $u_1=v_1$ and $u_3=v_3$. It turns out
that these poles cancel in the sum \re{Def-Vcon} in such a way that the evolution
kernel $\mathbb{V}_{123}^{(1)} (\bit{u} | \bit{v})$ only has integrable
singularities at $u_1=v_1$ and $u_3=v_3$. It is easy to see that the two
singularities correspond to the limit when the internal gluons in the diagram
shown in Fig.~\ref{gluonthreeparticle} (a) have vanishing light-cone momenta,
$k_+ = 0$. We remind that, in the light-cone axial gauge $A_+=0$, the gauge field
propagator \re{axial} is not well-defined for $k_+=0$ and has to be supplemented
by a particular prescription, Eqs.~\re{ML} and \re{PrincipalValuePrescription}.
The fact that the two-loop kernel \re{Def-Vcon} is integrable at $u_1=v_1$ and
$u_3=v_3$ implies that $\mathbb{V}_{123}^{(1)} (\bit{u} | \bit{v})$ is not
sensitive to the choice of the axial gauge prescription and, most importantly,
it has finite moments $\int [du]_3 \mathbb{V}_{123}^{(1)} ( \bit{u}| \bit{v} )
u_1^{k_1} u_2^{k_2}u_3^{k_3}$.

We remind that the entire two-loop kernel \re{TwoLoopThreeParticle} has to fulfill
the polynomiality condition, that is, its moments ought to be polynomial in the
$v-$variables, Eq.~\re{V-moments}. It turns out that the three-particle kernel
$\mathbb{V}_{123}^{(1)} (\bit{u} | \bit{v})$ alone does not obey this condition.
Its moments are given by rational functions of $v_{1,2,3}$ decorated by logarithms
and dilogarithms depending on the ratio $v_1/v_3$ and $v_2/v_3$. As we will
demonstrate in Sect.~\ref{2pKernelFull}, the ``unwanted'' terms violating the
polynomiality of $\mathbb{V}_{123}^{(1)} (\bit{u} | \bit{v})$ cancel in the
right-hand side of Eq.~\re{TwoLoopThreeParticle} against similar terms coming
from the moments of two-particle irreducible kernels $\mathbb{V}^{(1)}_{jk}$.

To separate the ``unwanted'' terms it proves convenient to introduce the
double plus--dis\-tri\-bu\-tion. It represents a natural generalization of
\re{SinglePlusDistr} for three-particle irreducible kernels and is defined as
\ba\nonumber
&& \int_0^1 [du]_3 \varphi(u_1,u_3)\left[\mathbb{V}_{123}^{(1)} (\bit{u} |
\bit{v})\right]_{++} \\
&& \qqqquad = \int_0^1 [du]_3\,
\left[\varphi(u_1,u_3)-\varphi(v_1,u_3)-\varphi(u_1,v_3)+\varphi(v_1,v_3)\right]
\mathbb{V}_{123}^{(1)} (\bit{u} | \bit{v})\,,
\ea
where $\varphi(u_1,u_3)$ is a test function. Since the kernel $\mathbb{V}_{123}^{(1)}
(\bit{u} | \bit{v})$ lives on the simplices \re{simplex} and \re{u-simplex}, as
exhibited by the step-function structure, the constraint on the $u-$momentum
fractions arising from the integration measure can be omitted, i.e., $\theta (1 -
u_1 - u_3) \mathbb{V}_{123}^{(1)} (\bit{u} | \bit{v}) = \mathbb{V}_{123}^{(1)}
(\bit{u} | \bit{v})$. One can verify then that the kernel $[\mathbb{V}_{123}^{(1)}
(\bit{u} | \bit{v})]_{++}$ defined in this manner satisfies the polynomiality
condition. By definition,
\begin{eqnarray}
\label{Dec-Vcon}
\mathbb{V}^{(1)}_{123} (\bit{u} | \bit{v})  \!\!\!&=&\!\!\!
\left[\mathbb{V}^{(1)}_{123} (\bit{u} | \bit{v})\right]_{++} + \delta(u_1 - v_1)
\int_0^1 d u_1 \mathbb{V}^{(1)}_{123} (\bit{u} | \bit{v})
\\
&+&\!\!\! \delta(u_3 - v_3)  \int_0^1 d u_3 \mathbb{V}^{(1)}_{123} (\bit{u} | \bit{v})
-
\delta(u_1 - v_1) \delta(u_3 - v_3)  \int_0^1 [du]_3 \,
\mathbb{V}^{(1)}_{123} (\bit{u} | \bit{v}) \, , \nonumber
\end{eqnarray}
where the integrals entering the subtraction terms are well-defined and can be
evaluated with the help of \re{Def-Vcon} (see Eqs.~\re{Def-DelF1}--\re{Def-DelF3}
in Appendix~\ref{DoublePlusAppendix}). We conclude that the polynomiality is
violated by the subtraction terms in the right-hand side of \re{Dec-Vcon}. The
first two subtraction terms preserve the momenta of a single particle and take
the form of two-particle irreducible kernels. The last subtraction term has the
structure of a single-particle contribution with the only difference that it
involves the integral $\int [du]_3 \, \mathbb{V}^{(1)}_{123} (\bit{u} | \bit{v})$
which is a rather complicated function of all three momentum fractions,
$v_{1,2,3}$. Later in this section, we will demonstrate that the subtraction
terms are compensated by similar polynomiality breaking terms coming from
genuine two-particle evolution kernels.

%%%%%%%%%%%%%%%%%%%%%%%%%%%%%%%%%%%%%%%%%%%%%%%%%%%%%%%%%%%%%%%%%%%%%
\subsection{Two-particle contributions}
\label{TwoLoopTwoParticleContrSection}
%%%%%%%%%%%%%%%%%%%%%%%%%%%%%%%%%%%%%%%%%%%%%%%%%%%%%%%%%%%%%%%%%%%%%

%%%%%%%%%%%%%%%%%%%%%%%%%%%%%%%%%%%%%%%%%%%%%%%%%%%%%%%%%%%%%%%%%%%%%
%            Figure
%%%%%%%%%%%%%%%%%%%%%%%%%%%%%%%%%%%%%%%%%%%%%%%%%%%%%%%%%%%%%%%%%%%%%
\begin{figure*}[t]
\begin{center}
\mbox{
\begin{picture}(0,70)(245,0)
\psfrag{V}[cc][cc]{$\mathbb{V}$}
\put(0,-8){\insertfig{17.28}{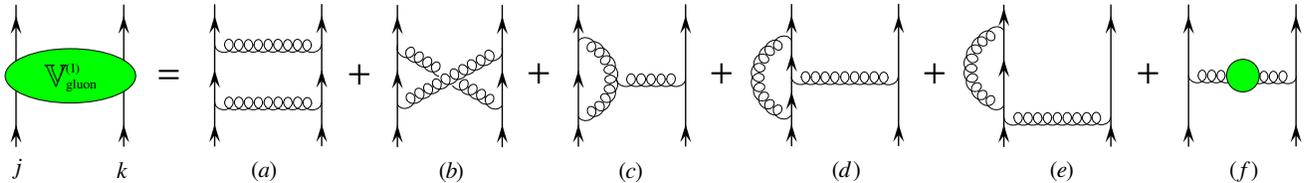}}
\end{picture}
}
\end{center}
\caption{\label{twoloopkernel} Two-loop corrections to the two-particle evolution
kernel involving only gluons in internal lines. The mirror symmetrical diagrams
are not displayed. The gluon polarization insertion is displayed in Fig.\
\ref{GluonPolarizationOneLoop}.}
\end{figure*}
%%%%%%%%%%%%%%%%%%%%%%%%%%%%%%%%%%%%%%%%%%%%%%%%%%%%%%%%%%%%%%%%%%%%%

%%%%%%%%%%%%%%%%%%%%%%%%%%%%%%%%%%%%%%%%%%%%%%%%%%%%%%%%%%%%%%%%%%%%%
%            Figure
%%%%%%%%%%%%%%%%%%%%%%%%%%%%%%%%%%%%%%%%%%%%%%%%%%%%%%%%%%%%%%%%%%%%%
\begin{figure*}[t]
\begin{center}
\mbox{
\begin{picture}(0,37)(170,0)
\put(0,-8){\insertfig{12}{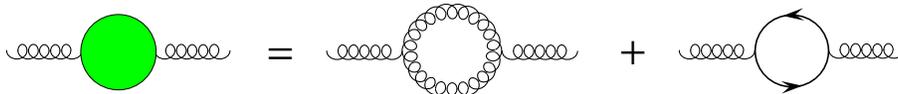}}
\end{picture}
}
\end{center}
\caption{\label{GluonPolarizationOneLoop} One-loop gluon polarization contributing
as an insertion into the one-loop gluon exchange diagram. The gluon and fermion
bubbles in Fig.\ \ref{gluonthreeparticle} (f) are be denoted as (f)$_{\rm gl}$
and (f)$_{\rm fer}$, respectively.}
\end{figure*}
%%%%%%%%%%%%%%%%%%%%%%%%%%%%%%%%%%%%%%%%%%%%%%%%%%%%%%%%%%%%%%%%%%%%%

\begin{figure*}[t]
\begin{center}
\mbox{
\begin{picture}(0,75)(242,0)
\psfrag{V}[cc][cc]{$\mathbb{V}$}
\put(0,-9){\insertfig{17}{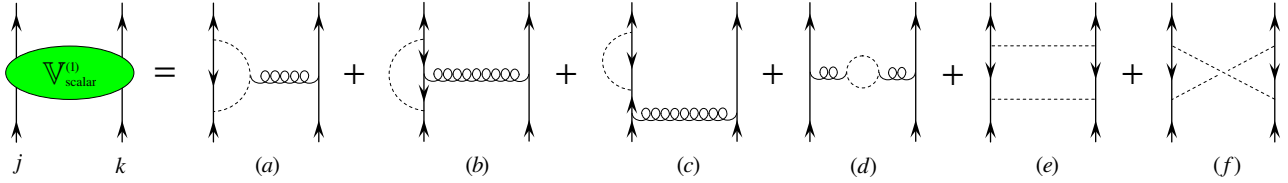}}
\end{picture}
}
\end{center}
\caption{\label{scalarvertex} Two-particle irreducible contribution involving
scalars in internal lines.}
\end{figure*}

Let us now consider the two-particle irreducible contribution to the two-loop
evolution kernel \re{TwoLoopThreeParticle}. It is determined by Feynman diagrams
shown in Figs.~\ref{twoloopkernel} and \ref{scalarvertex}. In distinction with
the three-particle kernel, the two-particle kernel $\mathbb{V}^{(1)}_{12}$ is not
universal and depends on the gauge theory under consideration. For the sake of
simplicity we will distinguish between QCD-like diagrams, that is, those involving
only gluons, and diagrams with scalars. The former are relevant for both QCD and
SYM theories whereas the latter only appear in $\mathcal{N}=2$ and $\mathcal{N}=4$
theories. Let us analyze the two sets of diagrams in turn.

%%%%%%%%%%%%%%%%%%%%%%%%%%%%%%%%%%%%%%%%%%%%%%%%%%%%%%%%%%%%%%%%%%%%%
\subsubsection{Gluon contribution}
\label{TwoLoopGluonContributionSection}
%%%%%%%%%%%%%%%%%%%%%%%%%%%%%%%%%%%%%%%%%%%%%%%%%%%%%%%%%%%%%%%%%%%%%

The diagrams with gluon exchanges are shown in Fig.\ \ref{twoloopkernel}. As was
already mentioned, similar diagrams define the two-loop contribution to the
evolution kernel for the twist-two maximal helicity operators in QCD defined in
\re{O2-QCD}. The important difference between the three-particle operators,
Eqs.~\re{O3-QCD} and \re{O3-SYM}, and the two-particle operators,
Eqs.~\re{O2-QCD} and \re{O2-SYM}, is that in the latter case the pair of fermions
is in the color-singlet state. However, this only affects the color factors
accompanying the individual Feynman diagrams shown in Fig.\ \ref{twoloopkernel},
not their momentum structure. Let us examine these color factors and relate to
each other the two-particle contribution to the dilatation operators for the two-
and three-particle operators.

To start with, we present the result for the two-loop evolution kernel for the
twist-two operators \re{O2-QCD} in QCD. In the momentum representation, it has
the following form for $v_1+v_2=u_1+u_2=1$
\be
\mathbb{V}_{\scriptscriptstyle\rm QCD}(u_1, u_2 | v_1, v_2)
=
\frac{g^2 C_F}{4\pi^2} \mathbb{V}^{(0)}_{\scriptscriptstyle\rm QCD} (u_1, v_1)
+
\left(\frac{g^2 C_F}{4 \pi^2} \right)^2
\mathbb{V}^{(1)}_{\scriptscriptstyle\rm QCD} (u_1, v_1)
\label{V-QCD}
\ee
where $C_F={(N_c^2 - 1)}/{(2 N_c)}$ is the Casimir operator in the fundamental
representation of the $SU(N_c)$. To one-loop order, the evolution kernel is
determined by the universal expression \re{OneLoopMomDOBL}
\be
\label{BLqcd}
\mathbb{V}^{(0)}_{\scriptscriptstyle\rm QCD} (u, v)
=
\frac{u}{v} \frac{\theta (v - u)}{v - u}
+
\left\{ u \to \bar u \atop v\to \bar v \right\}
\, .
\ee
To two-loop order, the kernel $\mathbb{V}^{(1)}_{\scriptscriptstyle\rm QCD} (u,
v)$ has been first calculated in Ref.~\cite{Vog97} in the so-called forward
limit, for $v_1+v_2=u_1 + u_2 = 0$. It was later reconstructed for $u_1 + u_2
\neq 0$, making use of the conformal symmetry constraints, and yielded the
result \cite{BelMulFre00}
\begin{equation}
\label{Def-ERBL-tra}
\mathbb{V}^{(1)}_{\scriptscriptstyle\rm QCD} (u, v)
=
\mathbb{V}^{(1)}_F (u, v)
+
\frac{C_F - \frac12N_c}{C_F} \mathbb{V}^{(1)}_A (u, v)
+
\frac{\beta_{{\scriptscriptstyle\rm QCD},0}}{C_F} \mathbb{V}^{(1)}_\beta (u, v)
+
\left\{ u \to \bar u \atop v\to \bar v \right\}
\, ,
\end{equation}
where $\beta_{{\scriptscriptstyle\rm QCD},0}=11N_c/3-2n_f/3$ is the one-loop
beta-function in QCD, Eq.~\re{beta-QCD}, and
\begin{eqnarray}
\mathbb{V}^{(1)}_F (u, v)
\!\!\!&=&\!\!\!
\frac{1}{2v} \ln \bar{u} \ln u
\\
&+&\!\!\! \theta (v - u)
\left[
f (u, v)
\left(
\frac{2}{3} - \zeta(2) - \frac{3}{4} \ln\frac{u}{v}
-
\frac{1}{2} \ln\frac{u}{v} \ln\frac{v - u}{u}
\right)
+
\frac{1}{2} f (\bar{u}, \bar{v}) \ln\frac{u}{v} \ln\frac{v - u}{v}
\right]
\, , \nonumber\\
\label{V-A}
\mathbb{V}^{(1)}_A (u, v)
\!\!\!&=&\!\!\!
\theta (v - u)
\left[
f (u, v)
\left(
- \frac{2}{3} - {\rm Li}_2 (u) - {\rm Li}_2 (\bar{v}) \right)
+
f (\bar{u}, \bar{v}) \ln\bar{u} \ln v + \frac{u}{2 v} \right]
\\
&-&\!\!\!
\theta (v - \bar{u})
\bigg[
f (u, v)
\left(
{\rm Li}_2 \left( 1 - \frac{u}{v} \right)
+
{\rm Li}_2 (\bar{v})
+
\frac{1}{2} \ln^2 v
-
\ln u \ln v \right)
\nonumber\\
&&\qquad\qquad\qquad\qquad\qquad\qquad\quad\!
+
f (\bar{u}, \bar{v})
\left(
{\rm Li}_2 (\bar{u}) - {\rm Li}_2 \left( 1 - \frac{u}{v} \right)
-
\frac{1}{2} \ln^2 v
\right)
-
\frac{\bar{u}}{2 v} \bigg]
\, , \nonumber\\
\mathbb{V}^{(1)}_\beta (u, v)
\!\!\!&=&\!\!\! \theta (v - u)
\left[ \frac{1}{4} \ln\frac{u}{v} + \frac{5}{12} \right] f (u, v)
\, .
\label{V-beta}
\end{eqnarray}
Here $\zeta(n)$ is the Riemann zeta function, ${\rm Li}_2(x) = - \int_0^x
\frac{dt}{t}\ln(1-t)$ is the Euler dilogarithm and the $f-$function is defined in
\re{f-def}.

Comparing the QCD expression \re{V-QCD} with Eqs.\ (\ref{H-dec}),
(\ref{CouplingConstant}) and (\ref{OneLoopMomDOBL}), we see that to one-loop
order the evolution kernel $\mathbb{V}_{12}$ for three-particle operators can be
deduced from \re{V-QCD} through substitution of the color factors
\be
\label{LOsunstitutionColorCF}
C_F \to \frac{2}{3} \, , \qqqquad C_F \to \frac{N_c}{2} \, ,
\ee
for fermions in the fundamental and adjoint representations, respectively.
Unfortunately, this simple property gets lost starting from two loops. To see
this, one examines the color factors corresponding to the Feynman diagrams
shown in Fig.~\ref{twoloopkernel} in four different cases: twist-two operator
\re{O2-QCD} in QCD, its counter-part \re{O2-SYM} in SYM theory, three-quark
operator \re{O3-QCD} in QCD and three-gaugino \re{O3-SYM} operator in SYM.

\renewcommand{\arraystretch}{1.5}
\begin{center}
\begin{tabular}[pos]{||c|c|c|c|c|c|c|c||}
\hline
\hline
$\mathbb{O}$
& (a)
& (b)
& (c)
& (d)
& (e)
& (f)$_{\rm gl}$
& (f)$_{\rm fer}$
\\
\hline \hline $\bar{q}_i q^i$ & $C_F^2$ & $C_F (C_F-\frac{1}{2} N_c)$ & $\ft12
C_F N_c$ & $C_F (C_F-\frac{1}{2} N_c)$ & $C_F^2$ & $\ft12 C_F N_c$ & $\ft12 C_F
n_f$
\\
\hline $\varepsilon_{ijk} q^i q^j q^k$ & $\frac49$ & $-\frac59$ & $1$ &
$-\frac19$ & $\frac89$ & $1$ & $\frac13 n_f$
\\
\hline \hline ${\rm tr} \{ \lambda \lambda \}$ & $N_c^2$ & $\ft12 N_c^2$ & $\ft12
N_c^2$ & $\ft12 N_c^2$ & $N_c^2$ & $\frac{1}{2} N_c^2$ & $\frac{1}{2} N_c^2
\mathcal{N}$
\\
\hline ${\rm tr} \{ \lambda \lambda \lambda \}$ & $\frac{1}{4} N_c^2$ & 0 &
$\frac{1}{4} N_c^2$ & $\frac{1}{4} N_c^2$ & $\frac{1}{2} N_c^2$ & $\frac{1}{4}
N_c^2$ & $\frac{1}{4} N_c^2\mathcal{N}$
\\ \hline \hline
\end{tabular}
\end{center}

\medskip

\noindent We notice that for two-particle operators, going over from QCD to SYM
amounts to replacing $C_F$ by the Casimir operator in the adjoint representation
$C_F \to N_c$ and substituting the number of quark flavors as $n_f \to N_c
\mathcal{N}$ (see Eq.~\re{n_s}). Let us now turn to three-particle operators and
substitute the color factors in the second row of the table according to
\re{LOsunstitutionColorCF} (we recall that the three-quark operators are
well-defined for $N_c=3$ only). It is easy to see that this leads to the correct
expressions for the color factors in the third and fifth rows except for the
columns (d) and (e). Thus, in order to reconstruct the two-particle contribution
to the evolution kernels for the three-quark and three-gaugino operators from the
QCD expression \re{Def-ERBL-tra} with the help of \re{LOsunstitutionColorCF}, it
suffices to supplement \re{Def-ERBL-tra} with the contribution of two additional
diagrams in QCD shown in Fig.\ \ref{twoloopkernel} (d) and (e). More precisely,
one only needs the abelian part of their contribution, i.e., the one proportional
to $C_F^2$. In what follows, we denote the corresponding kernel as
$\mathbb{V}^{(1)}_{\scriptscriptstyle\rm SV}$. A straightforward calculation
leads to
\begin{eqnarray}
\label{SelEneVer-tra}
\mathbb{V}^{(1)}_{\scriptscriptstyle\rm SV} (u, v)
\!\!\!&=&\!\!\!
\theta (v - u)
\Bigg[
f (u,v)
\Bigg(
-
\frac{3}{4} \ln\frac{u}{v}
-
\frac{1}{4} \ln^2\frac{u}{v}
-
\frac{1}{2} \ln\frac{u}{v} \ln\frac{(v - u) \bar{u}}{v \bar{v}}
\\
&&\qquad\qquad\qquad
+
\frac{1}{2} {\rm Li}_2 \left( \frac{u - v}{\bar{v}} \right)
-
\frac{1}{2} {\rm Li}_2 \left( \frac{v - u}{v} \right)
+
\frac{1}{2} {\rm Li}_2 (u)
-
\frac{1}{2} {\rm Li}_2 (v)
-
\frac{1}{2} \zeta (2)
\Bigg)
\nonumber\\
&&\qquad\quad
+
f (\bar{u}, \bar{v})
\Bigg(
\frac{1}{2} \ln\frac{v-u}{v}\ln\frac{u}{v}
-
\frac{1}{4} \ln^2\frac{\bar{u}}{\bar{v}}
-
\frac{1}{2} \ln\bar{u} \ln v
\nonumber\\
&&\qquad\qquad\qquad
+
\frac{1}{2} {\rm Li}_2 (u)
-
\frac{1}{2} {\rm Li}_2 (v)
+
\frac{1}{2} {\rm Li}_2\left(\frac{v - u}{v}\right)
-
\frac{1}{2} {\rm Li}_2\left(\frac{u - v}{\bar{v}}\right)
-
\frac{1}{2} \zeta(2)
\Bigg)
\Bigg]
\nonumber\\
\!\!\!&+&\!\!\!
\left\{u \to \bar u \atop v\to \bar v\right\}
\, .
\nonumber
\end{eqnarray}
We recall that \re{V-QCD} is defined for $v_1+v_2=u_1+u_2=1$ while for
three-particle operator the two-particle evolution kernel, say
$\mathbb{V}^{(1)}_{12}$, the momentum conservation implies that
$v_1+v_2=u_1+u_2=1-v_3$. To verify this condition, one can rescale the
momentum $u$ and $v$ in \re{V-QCD} making use of \re{rescale}.

In this way, we obtain the following expression for the gauge field contribution
to the two-particle kernel, $\mathbb{V}_{12}^{(1)}$, coming from the diagrams
shown in Fig.~\ref{twoloopkernel}
\be
\label{Def-FurSubCon}
\mathbb{V}_{\rm gluon}^{(1)} (u_1, u_2 | v_1, v_2) = \frac{\delta(u_1 + u_2 - v_1
- v_2)}{u_1 + u_2} \left( \mathbb{V}^{(1)}_{\scriptscriptstyle\rm QCD} +
\mathbb{V}^{(1)}_{\scriptscriptstyle\rm SV} \right) \left( \frac{u_1}{u_1 + u_2},
\frac{v_1}{v_1 + v_2} \right) \, ,
\ee
where $\mathbb{V}^{(1)}_{\scriptscriptstyle\rm SV} $ is defined in
\re{SelEneVer-tra} and $\mathbb{V}^{(1)}_{\scriptscriptstyle\rm QCD}$ is given by
the QCD expression \re{Def-ERBL-tra} with the color factors substituted as
\be
\begin{array}{ll}
\bullet ~~{\rm QCD}: & \qquad C_F \to \frac{2}{3} \, , \qquad N_c \to 3 \\[2mm]
\bullet ~~{\rm SYM}: & \qquad C_F \to {N_c}/{2} \, , \qquad n_f \to N_c \mathcal{N} \\
\end{array}
\label{SYM-sub}
\ee
Notice that the $\mathbb{V}^{(1)}_A-$term, Eqs.~\re{Def-ERBL-tra} and \re{V-A},
only contributes in QCD and vanishes for gaugino in the adjoint representation.
As we will show in Sect.~\ref{SectionEigenspectrum}, this leads to a dramatic
difference in integrability properties of the two-loop dilatation operator in
QCD and SYM theories.

The two-particle kernel $\mathbb{V}_{\rm gauge}^{(1)}$ does not depend on the
renormalization scheme employed---it is the same within the regularization by
means of DREG or DRED provided that the perturbation series is in terms of
$\rm\overline{MS}-$coupling constant. Since it more natural to define the
coupling constant in regularization via DRED in the $\rm\overline{DR}-$scheme,
the two-particle dilatation operator acquires an extra term as explained below
in Sect.\ \ref{2pKernelFull}.

%%%%%%%%%%%%%%%%%%%%%%%%%%%%%%%%%%%%%%%%%%%%%%%%%%%%%%%%%%%%%%%%%%%%%
%            Figure
%%%%%%%%%%%%%%%%%%%%%%%%%%%%%%%%%%%%%%%%%%%%%%%%%%%%%%%%%%%%%%%%%%%%%
%%%%%%%%%%%%%%%%%%%%%%%%%%%%%%%%%%%%%%%%%%%%%%%%%%%%%%%%%%%%%%%%%%%%%

%%%%%%%%%%%%%%%%%%%%%%%%%%%%%%%%%%%%%%%%%%%%%%%%%%%%%%%%%%%%%%%%%%%%%
\subsubsection{Scalar contribution}
\label{Scalars}
%%%%%%%%%%%%%%%%%%%%%%%%%%%%%%%%%%%%%%%%%%%%%%%%%%%%%%%%%%%%%%%%%%%%%

The scalars contribute to the two-particle evolution kernel in the $\mathcal{N}
=2$ and $\mathcal{N}=4$ SYM theories. The corresponding Feynman diagrams are shown
in Fig.\ \ref{scalarvertex}. As before, the ladder diagram in Fig.\ \ref{scalarvertex}
(e) vanishes due to the maximal $R-$charge of the three-fermion operator. The color
factors of the diagrams in Fig.~\ref{scalarvertex} (a), (b), (c) and (f) coincide
with those shown in Fig.\ \ref{gluonthreeparticle} (c), (d), (e) and (b), respectively,
and can be read from the Table given above. In particular, the diagram in
Fig.~\ref{scalarvertex} (f) has a vanishing color factor and does not contribute. The
remaining diagrams take the form of the one-loop graph (see Fig.~\ref{oneloopkernel})
with the propagators and the vertices ``dressed'' by the scalar interaction. Therefore,
it is not surprising that their contribution is proportional to the one-loop
$f-$function \re{f-def} decorated by additional logarithms
\ba
\nonumber \mathbb{V}^{(1)}_{\rm scalar} (u_1, u_2 | v_1, v_2) \!\!\!&=&\!\!\!
\frac{n_s}{6}  {\delta (u_1 + u_2 - v_1 - v_2)}  \bigg\{\theta(v_1-u_1) \left(
\ln\frac{u_1}{v_1} - \frac{4}{3} \right) f \left(  {u_1} , {v_1} \right)
\\
&&\qquad\qquad\! + \theta(v_2-u_2)\left( \ln\frac{u_2}{v_2} - \frac{4}{3} \right)
f \left(  {u_2} ,  {v_2} \right) \bigg\} \, ,
\label{V-scalar}
\ea
with $n_s=2(\mathcal{N}-1)$ being the total number of scalars, Eq.~\re{n_s}.
The kernel \re{V-scalar} is the same in the $\rm\overline{MS}-$ and
$\rm\overline{DR}-$schemes.

Notice that the scalar contribution \re{V-scalar} is similar to the $V_\beta-$term,
Eq.~\re{V-beta}. In the gluon contribution, Eqs.~\re{Def-ERBL-tra} and
\re{Def-FurSubCon}, this term is accompanied by the prefactor $\sim
\beta_{{\scriptscriptstyle\rm QCD},0}=11N_c/3-2n_f/3$, which coincides upon
substitution \re{SYM-sub} with the one-loop beta-function in SYM theory modulo
the scalar contribution given by $-n_s/6$. Combining together \re{Def-FurSubCon}
and \re{V-scalar}, one finds that the logarithmically enhanced part of the scalar
contribution \re{V-scalar} is accompanied by the factor $+n_s/6$. Writing this
factor as  $-n_s/6+n_s/3$ and splitting the scalar contribution into two terms,
one finds that the first term forces the coefficient in front to the $V_\beta-$term
to be equal to the complete one-loop beta-function in the SYM. As we will show
in Sect.~\ref{ConformalConstraints}, the second term is an artefact of the
$\overline{\rm MS}-$renormalization scheme used in the evaluation of \re{V-scalar}.

%%%%%%%%%%%%%%%%%%%%%%%%%%%%%%%%%%%%%%%%%%%%%%%%%%%%%%%%%%%%%%%%%%%%%
\subsubsection{Two-particle kernel}
\label{2pKernelFull}
%%%%%%%%%%%%%%%%%%%%%%%%%%%%%%%%%%%%%%%%%%%%%%%%%%%%%%%%%%%%%%%%%%%%%

Let us combine together various two-loop contributions to the two-particle
kernel $\mathbb{V}^{(1)}_{12}$, Eq.~\re{TwoLoopThreeParticle}. They come
from three different sources: the gluon kernel \re{Def-FurSubCon}, scalar
kernel \re{V-scalar} and two-particle kernel induced by the subtraction
terms in \re{Dec-Vcon}
\be
\mathbb{V}^{(1)}_{12}=\mathbb{V}^{(1)}_{\rm gluon} (u_1, u_2 | v_1,
v_2)+\mathbb{V}^{(1)}_{\rm scalar} (u_1, u_2 | v_1, v_2)+\mathbb{V}^{(1)}_{\rm
sub} (u_1, u_2 | v_1, v_2)\,.
\label{V12-full}
\ee
To identify the last term one substitutes the three-particle kernel \re{Dec-Vcon}
into \re{TwoLoopThreeParticle} and selects the two-particle contribution
proportional to $\delta(u_3-v_3) = \delta(u_1+u_2-v_1-v_2)$. In this way one gets
\be
\mathbb{V}^{(1)}_{\rm sub} (u_1, u_2 | v_1, v_2) = \delta(u_1+u_2-v_1-v_2) \left[
W(u_1,u_2| v_1, v_2) + W(u_2,u_1| v_2, v_1) \right]
\label{sub-term}
\ee
where
\be
W(u_1,u_2| v_1, v_2) = \int_0^1 du_3'\, \mathbb{V}^{(1)}_{123} (u_1, 1-u_1-u_3',
u_3' | v_1, v_2, v_3)
\ee
with $u_2=v_1+v_2-u_1$ and $v_1+v_2+v_3=1$. The explicit form of
$\mathbb{V}^{(1)}_{\rm sub} (u_1, u_2 | v_1, v_2)$ can be found in
Appendix~\ref{DoublePlusAppendix}.

The two-particle kernel \re{V12-full}, contrary to its three-particle counterpart,
is not universal and depends on the gauge theory under consideration. Substituting
\re{Def-FurSubCon}, \re{V-scalar} and \re{sub-term} into \re{V12-full} one finds
after some algebra
\ba
\label{V2-final}
\mathbb{V}^{(1)}_{12} \!\!\!\!&=&\!\!\! \delta(u_1+u_2-v_1-v_2) \bigg[
\theta(\bar{u}_2) \theta (u_2 - v_1 - v_2) \mathcal{F}_1(\bit{u} | \bit{v})
\\
&&\qquad\qquad\qquad\qquad + \theta(u_1) \theta(v_1 - u_1) \mathcal{F}_2(\bit{u}
| \bit{v}) + \frac{5}{8} \theta(u_2) \theta(v_1 - u_2) \mathcal{F}_3 (\bit{u} |
\bit{v}) + \left\{ u_1 \rightleftarrows u_2 \atop v_1\rightleftarrows v_2
\right\} \bigg] \! , \nonumber
\ea
where $\bar u_j\equiv 1-u_j$ and we used the results \re{Def-Delf1} --
\re{Def-Delfa} from Appendix~\ref{DoublePlusAppendix}. The $\{\ldots\}-$term
ensures symmetry of the two-particle kernel under interchange of two particles.
The $\mathcal{F}-$functions which enter as coefficients in front of different
step-functions arise either from the subtraction terms alone ($\mathcal{F}_1$)
or the genuine two-particle irreducible kernel ($\mathcal{F}_3$) or the sum of
both ($\mathcal{F}_2$). The former being defined by the three-particle irreducible
kernel is given by a universal expression
\ba
\label{Def-F1TwoPar}
\mathcal{F}_1(\bit{u} | \bit{v}) = \frac{\bar{u}_2}{(\bar{v}_1 - v_2)(v_1 - u_1)}
\!\!\!&-&\!\!\! \ft12 L(- u_1, \bar{v}_1 - v_2, v_2 - u_2) - \ft12 L(- u_1,
\bar{v}_1 - v_2, v_1)
\\ \nonumber
&+&\!\!\! \ft12 L(u_2 - v_2, \bar{v}_2, \bar{v}_1 - v_2) + \ft12 L(u_2 - v_2,
\bar{v}_2, v_1) \, ,
\ea
with $L(u,v,w)= \frac{1}{2 w} \ln\frac{u}{v} \left[ 4 + \ln \frac{u}{v} \right]$.
The two remaining functions are defined as follows:

\medskip

\noindent $\bullet$ In SYM theory with $\mathcal{N}$ supercharges one has
\ba
\label{V2-SYM}
\mathcal{F}_2 (\bit{u} | \bit{v})\big|_{\rm \scriptscriptstyle SYM}
\!\!\!&=&\!\!\! \mathcal{F}_0(\bit{u} | \bit{v}) + \frac{1}{2} \left(
\frac{\beta_{\mathcal{N},0}}{N_c} + \mathcal{N} - 1 \right) \left[
\ln\frac{u_1}{v_1} + \frac53 \right] f(u_1,v_1) - (\mathcal{N} - 1) f(u_1,v_1)
\, , \qquad \\
\nonumber %\label{F2SUSY}
\mathcal{F}_3 (\bit{u} | \bit{v})\big|_{\rm
\scriptscriptstyle SYM} \!\!\!&=&\!\!\! 0 \, ,
\ea
where $\beta_{\mathcal{N},0}=N_c(4-\mathcal{N})$ is the one-loop beta-functions
in SYM, Eqs.~\re{beta-SYM}, and
\ba
\mathcal{F}_0(\bit{u} | \bit{v}) \!\!\!&=&\!\!\! \ft12 L(u_2 - v_2,
\bar{v}_2, \bar{v}_1 - v_2) + \ft12 L(u_2 - v_2, \bar{v}_2, v_1) - \ft12 L(u_2 -
v_2, v_1, \bar{v}_1 - v_2)
\\[2mm]
&& + \frac{1}{v_1} - \frac{1}{4v_1} \ln^2\left( 1 - \frac{u_1}{v_1} \right) +
\left[ \frac{2}{3} - \frac{\pi^2}6 - \frac{3}{2} \ln\frac{u_1}{v_1} + \frac{1}{4}
\ln^2\frac{u_1}{v_1} \right] f(u_1,v_1)
\nonumber\\
&&- \ft12 \left[f(u_1,v_1) - f(u_2,v_2)\right]
\ln\frac{u_1}{v_1} \ln \left(1 - \frac{u_1}{v_1} \right)
+
\frac{v_1 + v_2}{2 v_1 v_2} \ln\frac{u_1}{v_1 + v_2}\ln\frac{u_2}{v_1 + v_2}
\nonumber
\ea
with the $f-$function defined in \re{f-def}.

\medskip

\noindent $\bullet$ In QCD the same functions are given by
\ba
\mathcal{F}_2(\bit{u} | \bit{v})\big|_{\rm \scriptscriptstyle QCD}
\!\!\!&=&\!\!\! \mathcal{F}_0(\bit{u} | \bit{v}) + \frac{3}{8}
\beta_{{\scriptscriptstyle \rm QCD},0} \left[\ln\frac{u_1}{v_1} + \frac53 \right]
f(u_1,v_1) + \frac58 \, \mathcal{F}_{\rm\scriptscriptstyle QCD}(\bit{u} |
\bit{v})\,,
\label{V2-QCD}
\\ \nonumber
\mathcal{F}_3(\bit{u} | \bit{v})\big|_{\rm \scriptscriptstyle QCD}
\!\!\!&=&\!\!\! - \frac{u_1}{v_2 (v_1+v_2)} + \left[f(u_2,v_2)- f(u_1,v_1)
\right] \left[ 2{\rm Li}_2\left(1-\frac{u_2}{v_2}\right)+\ln^2
\frac{v_2}{v_1+v_2} \right]
\\
&+&\!\!\!
2 f(u_2,v_2) \left[{\rm Li}_2\left(\frac{v_1}{v_1+v_2}\right)
-
\ln\frac{u_2}{v_1+v_2} \ln\frac{v_2}{v_1+v_2} \right]
+
2 f(u_1,v_1) {\rm Li}_2\left(\frac{u_1}{v_1+v_2}\right)
\,, \nonumber
\ea
where $\beta_{{\scriptscriptstyle \rm QCD},0}={11} -\frac23 n_f$ is one-loop
beta-functions in QCD, Eq.~\re{beta-QCD}, and
\ba
\mathcal{F}_{\rm \scriptscriptstyle QCD} = \frac{4}{3} f(u_1,v_1) -
\frac{u_1}{v_1(v_1+v_2)} \!\!\!&+&\!\!\! 2 f(u_2,v_2) \ln\frac{u_2}{v_1+v_2}
\ln\frac{v_1}{v_1+v_2}
\\
&-&\!\!\!
2 f(u_1,v_1) \left[ {\rm Li}_2
\left(\frac{u_1}{v_1+v_2}\right) + {\rm Li}_2 \left(\frac{v_2}{v_1+v_2}\right)
\right] \, . \nonumber
\ea
The following comments are in order.

The step-function structure in front of $\mathcal{F}_1$ looks quite unusual and
even appears not to contribute to the moments \re{V-moments} which are defined
with the measure $[d u]_3$ possessing the support region $\{ 0 \leq u_i \leq 1 ,
\sum_i u_i = 1 \}$. If one uses the momentum-conserving condition from the
$\delta$-function accompanying $\mathcal{F}_1$ to express $\theta (u_2 - v_1 - v_2)$
as $\theta (- u_1)$ one would immediately come to this conclusion. This is not the
case however since this contribution emerged from the subtraction term in the
definition of the double plus-distribution in Eq.\ \re{Dec-Vcon}. As it is clear
from there, while the $\mathbb{V}_{123}-$kernel itself resides on the simplices
\re{simplex} and \re{u-simplex}, the subtraction terms ``live'' in the entire
domain of $u_i$. Their contribution to the two-particle subchannels was
reconstructed from the subtracted expression by ``integrating-in'' corresponding
$\delta-$functions and assuming that the support of the resulting subtracted
kernel coincides with the original unsubtracted one. Therefore, the inconsistency
which arises for the $\mathcal{F}_1-$term only is simply resolved by ignoring the
spectral constraint on the total sum of momentum fractions after integrating out
the momentum-conserving $\delta-$function in the measure $[d u]_3$. Namely,
\be
\int_0^1 [d u]_3 \delta(u_1 + u_2 - v_1 - v_2) \theta(\bar{u}_2) \theta (u_2 -
v_1 - v_2) \mathcal{F}_1 (\bit{u} | \bit{v}) = \int_{v_1 + v_2}^1 d u_2 \,
\mathcal{F}_1 (v_1 + v_2 - u_2, u_2, v_3 | v_1, v_2, v_3) \, ,
\ee
with $v_3 = 1 - v_1 - v_2$ on the right-hand side. Analogously, one treats other
terms with the $\mathcal{F}_1$ structure.

The expression \re{V2-final} was obtained within the $\rm
\overline{MS}-$renormalization scheme. Going over to the $\rm
\overline{DR}-$scheme, one finds that contribution of all diagrams except
the one containing gluon self-energy, Fig.~\ref{twoloopkernel} (f), remain
the same. For the latter diagram, the difference between the two schemes
arises at the level of $\mathcal{O}(\varepsilon)$ corrections to the gluon
self-energy (see Fig.~\ref{GluonPolarizationOneLoop}). These corrections
interfere with a double pole coming from the Feynman integral to produce a
nontrivial contribution to the two-particle evolution kernel which is
proportional to the one-loop evolution kernel. We will return to this issue
in Sect.~\ref{SchemeDependence}.

Comparing \re{V2-SYM} and \re{V2-QCD} one notices that the two-particle kernel in
SYM theories has a simpler form and, most importantly, it does not involve the
${\rm Li}_2-$function. This property is nontrivial since ${\rm Li}_2$ enters into
the expression for gluon kernel \re{Def-FurSubCon} through the abelian
contribution $\mathbb{V}^{(1)}_{\scriptscriptstyle\rm SV}$,
Eq.~\re{SelEneVer-tra}. A close examination reveals that in the expression for
the two-particle kernel \re{V12-full} the ${\rm Li}_2-$terms coming from gluon
kernel cancel against similar terms generated by subtraction term \re{sub-term}.
Similar cancellation takes place inside the QCD expression \re{V2-QCD} but in
that case ${\rm Li}_2-$term also comes from the nonplanar kernel
$\mathbb{V}_A^{(1)}$. In SYM theories this term does not contribute to the
three-particle operators since it is accompanied by the color factor $C_F-N_c/2$
which vanishes upon the substitution \re{LOsunstitutionColorCF}.

As was mentioned in Sect.~\ref{Lagrangians}, the conformal symmetry is broken in
gauge theories with nonvanishing beta-function. One observes that the kernels
\re{V2-SYM} and \re{V2-QCD} contain terms proportional to the one-loop
beta-function. In the light-like axial gauge, they come from the one-loop Feynman
diagram in which the internal gluon gets ``dressed'' by a one-loop self-energy
correction (see Fig.~\ref{twoloopkernel} (f)). The corresponding terms are
responsible for explicit breaking of conformal invariance of the evolution
kernels at two loops. Other more subtle symmetry breaking effects will be
addressed below in Sect.\ \ref{ConformalConstraints}.

In distinction with the three-particle irreducible kernel \re{Def-Vcon}, the
two-particle kernels \re{V2-SYM} and \re{V2-QCD} are singular for $u_j \to v_j$.
Carefully examining \re{V2-final} in this limit, one finds that the two-particle
kernel $\mathbb{V}^{(1)}_{12}$ develops simple poles $\sim 1/(u_j-v_j)$ and, as
a consequence, its moments are divergent. In the axial gauge with the principal
value prescription, these poles are regularized according to
\re{PrincipalValuePrescription} and lead to $\sim \ln \delta$ contribution to
the moments. These singular terms can be separated with the help of the
plus-distribution as in Eq.\ (\ref{SinglePlusDistr}). For the pair-wise
kernels $\mathbb{V}^{(1)}_{12}=\mathbb{V}^{(1)}(u_1, u_2 | v_1, v_2)$ it
reads
\be
[\mathbb{V}^{(1)}(u_1, u_2 | v_1, v_2)]_+ = \mathbb{V}^{(1)}(u_1, u_2 | v_1, v_2)
- \delta (u_1 - v_1) \delta (u_2 - v_2)
\dashint
d u_1 du_2\, \mathbb{V}^{(1)} (u_1, u_2 | v_1, v_2) \, ,
\label{V2-plus}
\ee
where the integral in the right-hand side is evaluated using the principal value
prescription. The two-particle kernel $[\mathbb{V}^{(1)}_{12}]_+$ defined in this
way has finite moments and satisfies the polynomiality condition.

In the SYM theories with $\mathcal{N}$ supercharges, the $\mathcal{N}-$dependence
enters into the two-particle kernel \re{V2-final} through the last two terms in
the right-hand side of \re{V2-SYM}. Taking into account that $\beta_{\mathcal{N},
0}=(4-\mathcal{N}) N_c$ one finds that the $\mathcal{N}-$dependence only resides
in the last term proportional to the one-loop $f-$function, Eq.~\re{f-def}.
Together with \re{OneLoopMomDOBL} this implies that the two-particle kernels in
SYM theories with different number of supercharges are related to each other in a
simple manner
\be
\mathbb{V}^{(1)}_{12}\bigg|_{\mathcal{N}} =
\mathbb{V}^{(1)}_{12}\bigg|_{\mathcal{N}=1} -
(\mathcal{N}-1)\mathbb{V}^{(0)}_{12}\,,
\label{V2-N}
\ee
where $\mathbb{V}^{(0)}_{12}=\mathbb{V}^{(0)} ( u_1, u_2 | v_1, v_2 )$ is the
one-loop expression, Eq.~\re{OneLoopMomDOBL}. This suggests that the two-loop
dilatation operators for different $\mathcal{N}$ should be related to each other
in a simple manner (see Eq.~\re{N-factorization} below).

%%%%%%%%%%%%%%%%%%%%%%%%%%%%%%%%%%%%%%%%%%%%%%%%%%%%%%%%%%%%%%%%%%%%%
\subsection{One-particle contributions}
%%%%%%%%%%%%%%%%%%%%%%%%%%%%%%%%%%%%%%%%%%%%%%%%%%%%%%%%%%%%%%%%%%%%%

%%%%%%%%%%%%%%%%%%%%%%%%%%%%%%%%%%%%%%%%%%%%%%%%%%%%%%%%%%%%%%%%%%%%%
%            Figure
%%%%%%%%%%%%%%%%%%%%%%%%%%%%%%%%%%%%%%%%%%%%%%%%%%%%%%%%%%%%%%%%%%%%%
\begin{figure*}[t]
\begin{center}
\mbox{
\begin{picture}(0,75)(195,0)
\put(0,-8){\insertfig{13}{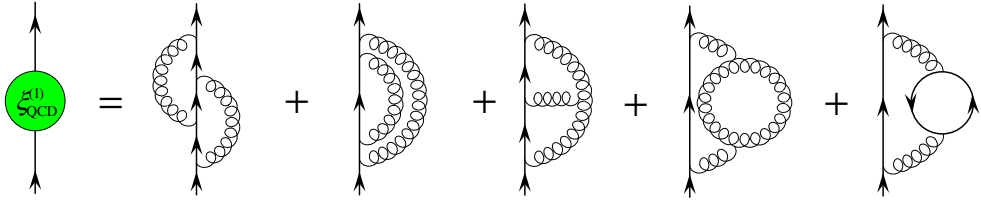}}
\end{picture}
}
\end{center}
\caption{\label{GauginoSEgluons2loop} Two-loop fermion wave functions
renormalization in QCD and $\mathcal{N} = 1$ SYM.}
\end{figure*}
%%%%%%%%%%%%%%%%%%%%%%%%%%%%%%%%%%%%%%%%%%%%%%%%%%%%%%%%%%%%%%%%%%%%%

\begin{figure*}[t]
\begin{center}
\mbox{
\begin{picture}(0,67)(245,0)
\put(-0.2,-8){\insertfig{17.3}{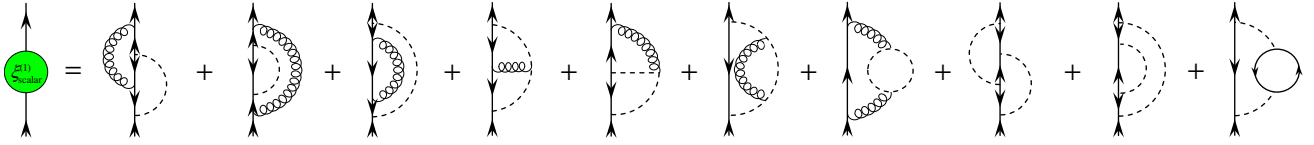}}
\end{picture}
}
\end{center}
\caption{\label{GauginoSEscalars2loop} Two-loop Feynman graphs for wave
function renormalization involving scalar fields propagating in internal lines.}
\end{figure*}

Finally let us consider single-particle irreducible contributions
$\mathbb{V}^{(1)}_i$ to the dilatation operator \re{TwoLoopThreeParticle}.
They come from the fermion wave function renormalization and are defined
by the Feynman diagrams shown in Figs.~\ref{GauginoSEgluons2loop} and
\ref{GauginoSEscalars2loop}.

The two-loop renormalization constants $\xi^{(1)}_i$ in Eq.\ (\ref{V-one-to-xi})
depend on the renormalization scheme used for their evaluation. In addition, in
the light-like axial gauge $A_+(x)=0$ supplemented with the principal value
prescription \re{PrincipalValuePrescription}, they depend on the ratio of the
light-cone momentum carried by the fermion and the regularization parameter
$\delta$
\be
\xi^{(1)}_i = \xi^{(1)} (v_i/\delta)
\, .
\label{sum-xi}
\ee
The result of the calculation in the $\rm \overline{MS}-$scheme of $\xi^{(1)}
(v/\delta)$ both in QCD \cite{Sar84} and SYM theories is
\ba
\xi^{(1)}_{\scriptscriptstyle\rm QCD}
(k_+/\delta)
\!\!\!&=&\!\!\!
\frac{N_c}{C_F}
\left(
\frac{17}{24} + \frac{11}{3} \zeta (2) - 3 \zeta (3)
+
\left( \frac{67}{9} - 2 \zeta (2) \right) \ln \frac{\delta}{k_+}
\right)
\nonumber\\
&+&\!\!\! \frac{3}{8} - 3 \zeta (2) + 6 \zeta (3)
-
\frac{n_f}{C_F}
\left(
\frac{1}{12} + \frac{2}{3} \zeta (2) + \frac{10}{9} \ln \frac{\delta}{k_+}
\right)
\, ,
\\
\xi^{(1)}_{\scriptscriptstyle \mathcal{N}}
(k_+/\delta) \!\!\!&=&\!\!\!
1 + 3 \zeta (3) + \left( \frac{19}{3} - 2 \zeta(2) \right)
\ln \frac{\delta}{k_+}
+
\frac{n_s^2}{8} - n_s \left( \frac{11}{12} + \ln \frac{\delta}{k_+} \right)
\, . \qquad
\ea
We recall that for three-quark operators in QCD the color factors take the values
$C_F=4/3$ and $N_c=3$.

In Eq.~\re{TwoLoopThreeParticle}, the self-energy correction \re{sum-xi} is
accompanied by the delta-functions $\delta(u_1-v_1)\delta(u_2-v_2)$. Similar
contribution also comes from the subtraction terms entering into the definition
the double-plus distribution for three-particle kernel \re{Dec-Vcon} and from the
plus distribution for two-particle kernel \re{V2-plus}. Combining them together
one gets
\be
\label{NLOGamma}
\Gamma^{(1)} = \frac{1}{3} \int [du]_3 \, \mathbb{V}^{(1)}_{123} (\bit{u} | \bit{v})
-
\frac{1}{3} \dashint du_1 du_2
\mathbb{V}^{(1)}_{12}(u_1,u_2|v_1,v_2)
-
\frac{1}{6} \left[\xi^{(1)} (v_1/\delta) + \xi^{(1)} (v_2/\delta)\right]
+
\ldots
\ee
where ellipsis denote the terms needed to restore the symmetry under cyclic
permutations of particles $(v_1,v_2,v_3)\to (v_2,v_3,v_1) \to (v_3,v_1,v_2)$. Two
comments follow. First, notice that the integrals of three- and two-particle
irreducible kernels in the right-hand side of \re{NLOGamma} contain a nontrivial
dependence on momentum fractions $v_i$. However, in the combination \re{NLOGamma}
this dependence entirely cancels as a manifestation of the restoration of the
polynomiality condition. Second, the contributions of the two-particle kernel and
self-energy corrections contain an explicit dependence on the axial gauge
regularization parameter $\delta$. The gauge invariance requires that the
divergences $\sim \ln\delta$ should cancel in their sum. Indeed, one can verify
that $\Gamma^{(1)}$ stays finite for $\delta\to 0$. It reads in the
$\overline{\rm MS}$ scheme for the QCD case with $N_c = 3$
\be
\Gamma^{(1)}_{\scriptscriptstyle\rm QCD}
=
\frac{27}{16} + \frac{13}{32} \beta_{{\scriptscriptstyle\rm QCD} ,0}
\, ,
\label{single-QCD}
\ee
and for the SYM case
\be
\Gamma^{(1)}_{\scriptscriptstyle\rm SYM}
= \frac{4}{3} - \frac{1}{12} n_s - \frac{1}{8} n_s^2
\, ,
\label{single-SYM}
\ee
with $n_s = 2(\mathcal{N}-1)$. By the construction, $3\Gamma^{(1)}$ defines the
two-loop correction to the anomalous dimension of the local three-fermion
operators, Eqs.~\re{B0} and \re{GammaPertExpansion}, in the ${\rm
\overline{MS}}-$scheme. The one-loop correction is given by \re{Gamma0}.

%%%%%%%%%%%%%%%%%%%%%%%%%%%%%%%%%%%%%%%%%%%%%%%%%%%%%%%%%%%%%%%%%%%%%
%            Figure
%%%%%%%%%%%%%%%%%%%%%%%%%%%%%%%%%%%%%%%%%%%%%%%%%%%%%%%%%%%%%%%%%%%%%
%%%%%%%%%%%%%%%%%%%%%%%%%%%%%%%%%%%%%%%%%%%%%%%%%%%%%%%%%%%%%%%%%%%%%

%%%%%%%%%%%%%%%%%%%%%%%%%%%%%%%%%%%%%%%%%%%%%%%%%%%%%%%%%%%%%%%%%%%%%
\subsection{Two-loop result}
\label{TwoLoopResultTotal}
%%%%%%%%%%%%%%%%%%%%%%%%%%%%%%%%%%%%%%%%%%%%%%%%%%%%%%%%%%%%%%%%%%%%%

Summarizing the calculation, the two-loop correction to the dilatation operator
\re{TwoLoopThreeParticle} reads
\be
\mathbb{V}^{(1)} = \left[\mathbb{V}^{(1)}_{123}\right]_{++} +
\left[\mathbb{V}^{(1)}_{231}\right]_{++} +
\left[\mathbb{V}^{(1)}_{312}\right]_{++} + \left[\mathbb{V}^{(1)}_{12}\right]_+ +
\left[\mathbb{V}^{(1)}_{23}\right]_+ + \left[\mathbb{V}^{(1)}_{31}\right]_+
- 3 \Gamma^{(1)} \delta (u_1 - v_1) \delta (u_2 - v_2) \, ,
\label{V-two-loop}
\ee
where each term in the right-hand side has finite moments and satisfies the
polynomiality condition. Here the three-particle kernel $\mathbb{V}^{(1)}_{123}$
is defined in \re{Def-Vcon}, the two-particle kernel $\mathbb{V}^{(1)}_{12}$ is
given by \re{V2-final} whereas the remaining kernels can be obtained from these
two with the help of \re{V-symmetry}.  The `++'-- and `+'--distributions were
introduced in \re{Dec-Vcon} and \re{V2-plus}, respectively. The single-particle
contribution is specified in \re{single-QCD} and \re{single-SYM}.

%%%%%%%%%%%%%%%%%%%%%%%%%%%%%%%%%%%%%%%%%%%%%%%%%%%%%%%%%%%%%%%%%%%%%
\subsubsection{Scheme dependence}
\label{SchemeDependence}
%%%%%%%%%%%%%%%%%%%%%%%%%%%%%%%%%%%%%%%%%%%%%%%%%%%%%%%%%%%%%%%%%%%%%

The above mentioned results were obtained within the $\rm
\overline{MS}-$renormalization scheme. As was already mentioned, the two-particle
and single-particle contributions have different form in the $\rm \overline{MS}-$
and $\rm \overline{DR}-$schemes. In QCD both schemes are equally suitable while
in the SYM theory the $\rm \overline{DR}-$scheme is advantageous as it preserves
supersymmetry. Therefore, as a final step we have to transform \re{V-two-loop} to
the supersymmetry preserving dimensional reduction scheme. This scheme
transformation is achieved via a finite renormalization of the nonlocal
light-cone operators \re{O-1/2} renormalized according to the conventional
$\overline{\rm MS}$ scheme. For nonlocal operator of arbitrary length $L$ it
reads
\begin{equation}
\mathbb{O}_{\overline{\scriptscriptstyle\rm DR}}(z_1,\ldots,z_L) = {w} (g^2)\,
\mathbb{O}_{\overline{\scriptscriptstyle\rm MS}}(z_1,\ldots,z_L)  \, ,
\end{equation}
where $w(g^2)$ is, in general, a complicated integral operator acting on
$z-$coordinates. Substituting this relation into the evolution equation \re{RG}
one finds that the evolution kernels in the two schemes are related as
\begin{equation}
\mathbb{H}_{\overline{\scriptscriptstyle\rm DR}} (g^2)
=
w (g^2) \, \mathbb{H}_{\overline{\scriptscriptstyle\rm MS}} (g^2) \, w^{-1} (g^2)
-
\beta (g^2) \frac{\partial  w (g^2)}{\partial g^2} \, w^{-1} (g^2) \, .
\label{w-matrix}
\end{equation}
Notice that in the conformal limit, for $\beta (g^2)=0$, the two evolution
kernels are related to each other through similarity transformation and,
therefore, have the same eigenvalues. The latter determine the anomalous
dimension of Wilson operators in the underlying gauge theory.

Equation \re{V-two-loop} defines the two-loop correction to
$\mathbb{H}_{\overline{\scriptscriptstyle\rm MS}} (g^2)$ in the momentum
representation. In order to transform it into the ${\overline{\rm DR}}-$scheme
with the help of \re{w-matrix}, it suffices to determine $w(g^2)$ to one loop. To
this accuracy, $w$ arises due to the fact that the one-loop diagrams shown in
Fig.~\ref{oneloopkernel} have different finite parts when renormalized in the
${\overline{\rm DR}}-$ and $\overline{\rm MS}-$ schemes. It is easy to see that
for maximal-helicity operators, $w$ only receives contribution from the gluon
self-energy, Fig.\ \ref{oneloopkernel} (b), while the diagrams (a) and (c) are
the same in both schemes to the $\sim \varepsilon^0$ accuracy. Therefore, $w$
is given by the unit operator times a factor. The latter can be easily computed
with the result
\begin{equation}
w (g^2) = 1 - L \frac{N_c}{4} \frac{g^2}{8 \pi^2}\,.
\end{equation}
Yet in the above scheme transformation, the gauge coupling is still defined in
the $\overline{\rm MS}$ scheme. Thus, one has to further renormalize the coupling
according to the DRED conventions, this yields its finite renormalization to
two-loop accuracy \cite{SchSasKor87}
\begin{equation}
g^2_{\overline{\scriptscriptstyle\rm MS}} = g^2_{\overline{\scriptscriptstyle\rm
DR}} \left( 1 - \frac{N_c}{6} \frac{g^2_{\overline{\scriptscriptstyle\rm DR}}}{8
\pi^2} \right) \, .
\end{equation}
Thus, for the evolution kernel in the $\overline{\rm MS}-$scheme given by
\be
\mathbb{H}_{\overline{\scriptscriptstyle\rm MS}}=
\lambda_{\overline{\scriptscriptstyle\rm MS}}
\mathbb{H}^{(0)}+\lambda_{\overline{\scriptscriptstyle\rm MS}}^2
\mathbb{H}^{(1)}+\mathcal{O}(\lambda^3)\,,
\ee
the same kernel in the $\overline{\rm DR}-$scheme takes the form
\be\label{equi}
\mathbb{H}_{\overline{\scriptscriptstyle\rm DR}}
=\left(1-\frac{\lambda_{\overline{\scriptscriptstyle\rm
DR}}}6\right)\left[\lambda_{\overline{\scriptscriptstyle\rm DR}}
\mathbb{H}^{(0)}+\lambda_{\overline{\scriptscriptstyle\rm DR}}^2
\mathbb{H}^{(1)}\right]
-
\frac{L}{4}\frac{\beta_0}{N_c}\lambda_{\overline{\scriptscriptstyle\rm
DR}}^2+\mathcal{O}(\lambda^3)\,,
\ee
or equivalently
\be
\mathbb{H}_{\overline{\scriptscriptstyle\rm DR}}
=
\lambda_{\overline{\scriptscriptstyle\rm DR}} \mathbb{H}^{(0)}
+
\lambda_{\overline{\scriptscriptstyle\rm DR}}^2
\left(\mathbb{H}^{(1)}-\frac16
\mathbb{H}^{(0)}-\frac{L}{4}\frac{\beta_0}{N_c}\right)+\mathcal{O}(\lambda^3) \,.
\label{H-DR}
\ee
According to \re{equi}, the two-loop dilatation operators in the two schemes
differ by an overall normalization only and, therefore, they share the same
symmetry properties (if any). This property is a unique feature of the dilatation
operator in the maximal helicity sector and it holds in SYM theory independently
of the number of supercharges $\mathcal{N}$.

At $L=3$, one finds from \re{H-DR} that the two-loop evolution kernel for
three-particle operators in SYM theory is given in the $\overline{\rm DR}-$scheme
in the momentum representation by
\be
\mathbb{V}_{\overline{\scriptscriptstyle\rm DR}}^{(1)} = \mathbb{V}^{(1)}-\frac16
\mathbb{V}^{(0)} -\frac{3}{4}(4-\mathcal{N})
\II
\ee
with $\mathbb{V}^{(0)}$ and $\mathbb{V}^{(1)}$ defined in Eqs.~\re{PairWiseLOkernel}
and \re{V-two-loop}, respectively, while $\II$ is the identity operator $\II =
\delta (u_1 - v_1) \delta (u_2 - v_2)$.

%%%%%%%%%%%%%%%%%%%%%%%%%%%%%%%%%%%%%%%%%%%%%%%%%%%%%%%%%%%%%%%%%%%%%
\subsubsection{Limiting cases}
\label{Tw2LimitingCases}
%%%%%%%%%%%%%%%%%%%%%%%%%%%%%%%%%%%%%%%%%%%%%%%%%%%%%%%%%%%%%%%%%%%%%

The two-loop evolution kernel $\mathbb{V}^{(1)}$ is given by a rather complicated
expression. There are however two limiting cases in which its form simplifies
significantly. Let us examine $\mathbb{V}^{(1)}$ for  $u_j\to v_j$ (with
$j=1,2,3$). As was explained in Sect.~\ref{OneLoopDilOper}, this corresponds to
the limit when gluons exchanged between the fermions in the Feynman diagrams
shown in Fig.~\ref{twoloopkernel} carry vanishing light-cone momenta $u_j - v_j
\to 0$. We argued in Sect.~\ref{2pKernelFull}, that for $u_j\to v_j$ the dominant
contribution to $\mathbb{V}^{(1)}$ comes from the two-particle evolution kernels
$[\mathbb{V}^{(1)}_{jk}]_+$ which develop poles $\sim 1/(u_j-v_j)$. It is
straightforward to verify using \re{V2-final} that these poles are generated by
terms in the expression for $\mathbb{V}^{(1)}_{jk}$ involving the one-loop
$f-$functions, Eq.~\re{f-def},
\be
\mathbb{V}^{(1)}_{12} \stackrel{u_j\to v_j}{\sim} \delta(u_1+u_2-v_1-v_2)
\left[ {\theta(v_1-u_1)}f(u_1,v_1) +  {\theta(v_2-u_2)}f(u_2,v_2)\right]
\, . \label{IR-asym}
\ee
One recognizes the expression in the right-hand side as the one-loop kernel
$\mathbb{V}^{(0)}_{12}$ defined in \re{OneLoopMomDOBL}. Substituting \re{IR-asym}
into \re{V-two-loop} one finds that for $u_j-v_j\to 0$ the full two-loop
evolution kernel is proportional to the one-loop contribution,
Eq.~\re{PairWiseLOkernel},
\be
\mathbb{V}(\bit{u}|\bit{v}) =
\lambda\mathbb{V}^{(0)}(\bit{u}|\bit{v})+\lambda^2\mathbb{V}^{(1)}(\bit{u}|\bit{v})
+\mathcal{O}(\lambda^3) \stackrel{u_j\to v_j}{=} \frac12 \Gamma_{\rm
cusp}(\lambda)\cdot \mathbb{V}^{(0)}(\bit{u}|\bit{v})+ \ldots \,,
\label{V-IR-asym}
\ee
where ellipsis denote terms subleading as $u_j-v_j\to 0$. One can show that
\re{V-IR-asym} holds to all orders of perturbation theory~\cite{Kor89} with
$\Gamma_{\rm cusp}(\lambda)$ being the universal cusp anomalous
dimension~\cite{KorRad87} related to a cusp anomaly of Wilson loops \cite{Pol80}.
The two-loop expression for $\Gamma_{\rm cusp}(\lambda)$ can be obtained from
\re{V2-final}. In SYM theory, in the $\overline{\rm DR}-$scheme one finds
\cite{BelGorKor03}
\be
\Gamma_{\rm cusp,\overline{\rm DR}}(\lambda) = 2 \lambda + \lambda^2 \left[
2(4-\mathcal{N})-\frac{\pi^2}3\right]
\label{cusp-SYM}
\ee
with the coupling constant $\lambda=g^2_{\scriptscriptstyle \overline{\rm
DR}}N_c/(8\pi^2)$.

The second limiting case corresponds to the selection of terms inside
$\mathbb{V}(\bit{u}|\bit{v})$ proportional to the one-loop beta functions. These
terms are gauge invariant and originate in the axial gauge from a special subset
of Feynman diagrams involving gluon propagator dressed by one-loop corrections
(see Fig.~\ref{twoloopkernel} (f)). In QCD one can automatically select the required
terms by going over to the limit of large $\beta_0$, that is $g^2 \beta_0=\rm fixed$
for $\beta_0\to\infty$. Since $\beta_0=11-2n_f/3$, the large $\beta_0-$limit can be
formally achieved in QCD by continuing $\mathbb{V}(\bit{u}|\bit{v})$ to the
number of quark flavors $n_f \to -\infty$. Similar procedure can not be performed
in the SYM theory since the number of gaugino equals the number of supercharges,
Eq.~\re{n_s}. According to \re{V2-final}, the $\beta_0-$term only comes from
two-particle kernels and has the same form in QCD and SYM theories
\be
\mathbb{V}^{(1)}_{12} = \frac{\beta_{0}}{2C_R}
\delta(u_1+u_2-v_1-v_2)\left[\theta(u_1) \theta(v_1 - u_1)
\left( \ln\frac{u_1}{v_1} + \frac53 \right)
f(u_1,v_1) + \left\{ u_1 \rightleftarrows u_2 \atop v_1\rightleftarrows v_2
\right\}\right]
+
\ldots
\label{V12-beta}
\ee
where $C_R=N_c$ in SYM theories and $C_R=4/3$ in QCD. A unique feature of
$\beta_0$ enhanced terms is that they can be resummed to all loops. To $n^{\rm
th}$ order of perturbation theory, these terms take the form $\lambda (\lambda
\beta_0)^n$ and come from the one-loop Feynman diagrams in which the bare gluon
propagator is dressed by ``renormalon'' chain involving $n$ one-loop polarization
operators (see Fig.~\ref{GluonPolarizationOneLoop}). These diagrams contribute to
the two-particle kernels and lead to the following all-loop expression for the
evolution kernel in the large $\beta_0-$limit
\be
\mathbb{V}(\bit{u}|\bit{v}) \stackrel{\beta_0\to \infty}{=} \mathbb{V}_{12} +
\mathbb{V}_{23} + \mathbb{V}_{31}
\ee
with
\ba
\label{V12-large-beta}
\mathbb{V}_{12} \!\!\!&=&\!\!\! \lambda \mathbb{V}^{(0)}_{12} + \lambda^2
\mathbb{V}^{(1)}_{12}+ \lambda^3 \mathbb{V}^{(2)}_{12}+ \ldots = \lambda
\varphi\left( \frac{\beta_{0} g^2}{16\pi^2}\right)\delta(u_1+u_2-v_1-v_2)
\\
&\times&\!\!\! \left[\theta(u_1) \theta(v_1 - u_1)
\left(\frac{u_1}{v_1}\right)^{\frac{\beta_{0} g^2}{16\pi^2}} f(u_1,v_1) + \left\{
u_1 \rightleftarrows u_2 \atop v_1\rightleftarrows v_2 \right\}\right] \, ,
\nonumber
\ea
where $\lambda = g^2 C_R/(8\pi^2)$ and the notation was introduced for the
function
\be
\varphi(x) = \frac{(1+x)\Gamma(4+2x)}{6\Gamma(1-x)\Gamma^3(2+x)}
\,.
\ee
It is easy to verify that the first two terms of the expansion of
\re{V12-large-beta} in powers of $\lambda$ are in agreement with \re{V12-beta}.
Comparing \re{V12-large-beta} with the one-loop expression for the evolution kernel
\re{OneLoopMomDOBL} one notices that higher order corrections ``renormalize'' the
conformal spin $j$ of the fermion fields and induce the additional normalization
factor \cite{BelMul97,Mik00}
\be
\label{KernelLargeBetaRenormConfSpin}
\mathbb{V}(\bit{u}|\bit{v})
\stackrel{\beta_0\to \infty}{=}
\lambda
\,
\varphi\left( \frac{\beta_{0} g^2}{16\pi^2}\right)
\mathbb{V}^{(0)}(\bit{u}|\bit{v})\bigg|_{j=1+ \frac{g^2\beta_{0}}{32\pi^2}}
\, .
\ee
This relation holds both in QCD and SYM theories.

%%%%%%%%%%%%%%%%%%%%%%%%%%%%%%%%%%%%%%%%%%%%%%%%%%%%%%%%%%%%%%%%%%%%%
\subsubsection{$\mathcal{N}-$dependence}
\label{NdepSection}
%%%%%%%%%%%%%%%%%%%%%%%%%%%%%%%%%%%%%%%%%%%%%%%%%%%%%%%%%%%%%%%%%%%%%

To one-loop order, the kernel $\mathbb{V}^{(0)}$ depends on $\mathcal{N}$ through the
normalization constant $\Gamma_{\scriptscriptstyle \rm SYM}^{(0)}$, Eq.~\re{Gamma0}.
To two-loop order, in Eq.~\re{V-two-loop}, the $\mathcal{N}-$dependence resides in the
two-particle kernels $\mathbb{V}_{jk}^{(1)}$ and the normalization constant $\xi^{(1)}$.
Taking into account \re{V2-N}, one finds
\be
\mathbb{V}^{(1)}_{\mathcal{N}} = \mathbb{V}^{(1)}_{\mathcal{N}=1} -
(\mathcal{N}-1)\left[\mathbb{V}^{(0)}_{\mathcal{N}=1}+ {\rm const}\right]\,,
\ee
where the subscript refers to the number of supercharges. This leads to the
following remarkable relation between the two-loop dilatation operators in SYM
theories with different number of supercharges
\be\label{N-factorization}
\mathbb{H}_{\mathcal{N}}(\lambda) = \big[1-(\mathcal{N}-1)\lambda \big]\cdot
\mathbb{H}_{\mathcal{N}=1}(\lambda)
+ {\rm const}
+ \mathcal{O}( \lambda^3 )
\,.
\ee
We conclude that to two-loop accuracy the spectrum of anomalous dimensions in the
sector of maximal helicity operators looks alike in {\sl all} SYM theories
including the maximally supersymmetric $\mathcal{N}=4$ theory. If the dilatation
operators in these theories have different properties, the difference could only
appear starting from three loops. Eq.~\re{N-factorization} is one of the main
results of this paper.

%%%%%%%%%%%%%%%%%%%%%%%%%%%%%%%%%%%%%%%%%%%%%%%%%%%%%%%%%%%%%%%%%%%%%
\section{Eigenspectrum of the dilatation operator}
\label{SectionEigenspectrum}
%%%%%%%%%%%%%%%%%%%%%%%%%%%%%%%%%%%%%%%%%%%%%%%%%%%%%%%%%%%%%%%%%%%%%

In the previous section, we have calculated the two-loop dilatation operator
in the momentum representation. Let us now turn to the analysis of its symmetry
properties. Our strategy will be to solve the spectral problem for the two-loop
evolution kernel $\mathbb{V}(\bit{u}|\bit{v})$, Eq.~\re{spectral}, and, then,
look for the hidden symmetry of its spectrum.

As we already observed in Sect.~\ref{MixingMatrix}, the dilatation operator is
invariant under translations and dilatations along the light-ray, Eqs.~\re{trans}
and \re{dilat}, respectively. The latter are the light-cone projections of the
Poincar\'e transformations in the underlying gauge theory. In the momentum
representation, this symmetry leads to scaling property of the evolution kernel
\re{rescale} which in its turn allows one to identify its eigenstates as
homogeneous polynomials, Eq.~\re{homo}. It is well-known that classical Yang-Mills
theory is invariant under an even larger set of space-time transformations---the
$SO(4,2)$ conformal group. The conformal symmetry imposes additional constraints
both on the properties of the evolution kernels and their eigenspectra. In
particular, it allows one to resolve the mixing between the Wilson operators
involving total derivatives \cite{BraKorMul03,BelRad05}.

On quantum level, the conformal symmetry becomes anomalous for $\beta(g^2) \neq 0$.
As we will explain later in this section, the conformal anomaly affects the dilatation
operator starting from two loops only and, therefore, the one-loop dilatation operator,
Eqs.~\re{LOkernel} and \re{PairWiseLOkernel}, inherits the conformal symmetry of the
classical theory. We will demonstrate that the conformal symmetry allows one to
diagonalize the two-particle evolution kernels in the coordinate and momentum
representations, Eqs.~\re{LightConeKernel} and \re{OneLoopMomDOBL}, respectively, but
it is not sufficient to diagonalize the three-particle dilatation operators.

Before we turn to the three-particle operators, let us examine the spectrum of
the dilatation operators for the twist-two operators, Eqs.~\re{O2-QCD} and
\re{O2-SYM}. The reason for doing this is twofold. Firstly, the evolution kernel
for twist-two operators contributes to the two-particle kernel $\mathbb{V}_{12}$
in the expression for the evolution kernel \re{V-two-loop}. Secondly, this offers
a convenient framework for discussing the conformal symmetry breaking.

%%%%%%%%%%%%%%%%%%%%%%%%%%%%%%%%%%%%%%%%%%%%%%%%%%%%%%%%%%%%%%%%%%%%%
\subsection{Twist-two operators}
\label{EigenspectrumTw2Section}
%%%%%%%%%%%%%%%%%%%%%%%%%%%%%%%%%%%%%%%%%%%%%%%%%%%%%%%%%%%%%%%%%%%%%

The generating function for twist-two operators in QCD and SYM theories is defined
in \re{O2-QCD} and \re{O2-SYM}, respectively. Similar to \re{O_q}, the twist-two
operators are uniquely defined by polynomials $P(u_1,u_2)$ which are eigenstates of
the evolution kernel in the momentum representation, $\mathbb{V} (u_1,u_2|v_1,v_2)$.
In QCD the two-loop evolution kernel $\mathbb{V}_{\scriptscriptstyle\rm QCD}(u_1,
u_2 | v_1, v_2)$ was defined in \re{V-QCD}. In SYM theories the evolution kernel
$\mathbb{V}_{\scriptscriptstyle\rm SYM}(u_1, u_2 | v_1, v_2)$ can be obtained
from the QCD expression by properly adjusting the color factors and adding the
contribution of scalars determined in Sect.~\ref{Scalars}.

%%%%%%%%%%%%%%%%%%%%%%%%%%%%%%%%%%%%%%%%%%%%%%%%%%%%%%%%%%%%%%%%%%%%%
\subsubsection{Eigenspectrum at one-loop}
%%%%%%%%%%%%%%%%%%%%%%%%%%%%%%%%%%%%%%%%%%%%%%%%%%%%%%%%%%%%%%%%%%%%%

To one-loop order, the evolution kernel for twist-two maximal-helicity operators
in QCD and SYM are given by the same universal expression \re{BLqcd}
\be
\mathbb{V}^{(0)}_{\rm tw-2} (\mbox{\boldmath $u$}| \mbox{\boldmath $v$} ) =
2\left[ \mathbb{V}^{(0)} ( u_1, u_2 | v_1, v_2 ) \right]_+ - 2\Gamma^{(0)}
\delta(u_1-v_1)\delta(u_2-v_2) \, ,
\ee
with $\Gamma^{(0)}$ defined in \re{Gamma0} and $\mathbb{V}^{(0)}$ being the
two-particle evolution kernel, Eq.~\re{OneLoopMomDO}. Solving the spectral
problem for this operator
\be\label{V2-eig}
\int [du]_2\left[ \mathbb{V}^{(0)} ( u_1, u_2 | v_1, v_2 ) \right]_+ P^{(\ell)}_n
(u_1,u_2) = - \gamma(n) P^{(\ell)}_n (v_1, v_2) \, ,
\ee
with $[du]_2=du_1 du_2 \delta(u_1+u_2-v_1-v_2)$, it is straightforward to verify
that its eigenstates are given by Gegenbauer polynomials \cite{BroLep79,EfrRad79}
\be\label{Gegenbauer}
P^{(\ell)}_n (u_1,u_2) = (u_1+u_2)^{n+\ell}
\textrm{C}_n^{3/2}\left(\frac{u_1-u_2}{u_1+u_2} \right)\,,
\ee
with $n$ and $\ell$ nonnegative integers. The corresponding eigenvalues do not
depend on $\ell$ and are given by the Euler digamma function
\be
\label{Tw2Eigenvalue}
\gamma(n) = 2 [ \psi (n + 2) - \psi (2)]\,.
\ee
The eigenstates of the two-particle evolution kernel, Eq.~\re{Gegenbauer}, define
the local twist-two operators $\mathcal{O}_{nl}(0)$ with $l\ge n$
\be\label{conf-op}
\mathcal{O}_{nl}(0)=P^{(l-n)}_n (\partial_1, \partial_2)\, \mathbb{O}(z_1,z_2)
\big|_{z_1=z_2=0}\,,
\ee
where $\mathbb{O}(z_1,z_2)$ is given by \re{O2-QCD} and \re{O2-SYM}. As follows from
\re{Gegenbauer}, the operators $\mathcal{O}_{nl}(0)$ with $l > n$ can be obtained
from the one with $l=n$ by applying the total derivatives, $\mathcal{O}_{nl}(0)=
\partial_+^{l-n} \mathcal{O}_{nn}(0)$.

The operators \re{conf-op} have an autonomous scale dependence to one-loop order
\be
\mu\frac{d}{d\mu} {\mathcal{O}}_{nl}(0) = -
\frac{g^2}{8\pi^2}\gamma_n^{(0)} {\mathcal{O}}_{nl}(0) + \mathcal{O}(g^4) \, ,
\ee
with the anomalous dimension given in QCD and in the SYM theory with
$\mathcal{N}$ supercharges by
\ba \label{gamma-0-QCD}
\gamma_n^{(0)}\big|_{\scriptscriptstyle\rm QCD}
\!\!\!&=&\!\!\! C_F \left[ 4 \psi (n + 2) - 4\psi (2) +1
\right],
\\ \label{gamma-0-SYM}
\gamma_n^{(0)}\big|_{\scriptscriptstyle\rm SYM}
\!\!\!&=&\!\!\! N_c \left[ 4 \psi (n + 2) - 4\psi (2) +
{\mathcal{N}}  \right].
\ea
Although these expressions define the anomalous dimensions for integer $n\ge 0$,
they can be analytically continued to the complex $n-$plane. In particular, for
$n=-1$ one finds in the SYM case that the anomalous dimension coincides with the
one-loop beta-function, Eq.~\re{beta-SYM},
\be
\gamma_{n=-1}^{(0)} =  - \beta_{\mathcal{N}, 0}
=
N_c \left[ \mathcal{N}-4 \right]
\ee
As we will argue in Sect.~\ref{TwistTwoTwoLoopADs}, this leading order property
is not accidental and can be extended to all loops.

%%%%%%%%%%%%%%%%%%%%%%%%%%%%%%%%%%%%%%%%%%%%%%%%%%%%%%%%%%%%%%%%%%%%%
\subsubsection{Conformal symmetry}
%%%%%%%%%%%%%%%%%%%%%%%%%%%%%%%%%%%%%%%%%%%%%%%%%%%%%%%%%%%%%%%%%%%%%

As was already mentioned, the explicit form of the one-loop eigenstates
\re{Gegenbauer} is uniquely determined by the conformal symmetry. We remind
that the nonlocal light-cone operators $\mathbb{O}(z_1,z_2)$ are built from
elementary fields $X(zn_\mu)$ located on the light ray, defined by the null
vector $n_\mu$. For such operators the full $SO(4,2)$ conformal group consisting
of Lorentz rotations $\mathbb{M}_{\mu\nu}$, translations $\mathbb{P}_\mu$,
dilatation $\mathbb{D}$ and special conformal boosts $\mathbb{K}_\mu$ is
effectively reduced to its ``collinear'' $SL(2)$ subgroup which acts
nontrivially on functions of the light-cone coordinate $z = x \cdot n^\ast$,
defined by means of a tangent vector $n^\ast_\mu$ to the light cone such that
$n \cdot n^\ast = 1$. The generators of the collinear subgroup are
\begin{equation}
\mathbb{L}^- = - i \mathbb{P}_+ \, , \qquad \mathbb{L}^+ = \ft{i}{2} \mathbb{K}_-
\, , \qquad \mathbb{L}^0 = \frac{i}{2} ( \mathbb{M}_{-+} + \mathbb{D} ) \, ,
\end{equation}
with the standard commutation relations
\begin{equation}
[\mathbb{L}^0 , \mathbb{L}^{\pm}] = \pm \mathbb{L}^{\pm} \, , \qquad
[\mathbb{L}^+ , \mathbb{L}^-] = \mathbb{L}^0 \, .
\end{equation}
These quantum field operators can be expressed in terms of the energy-momentum
tensor in the underlying gauge theory. Still, for their action on the elementary
field operators $X(z_k)\equiv X(z_k n_\mu)$ they can be represented by
differential operators acting on the light-cone coordinate
\ba
{} [\mathbb{L}^- , X(z_k)] &\equiv& L_k^- X(z_k) = - \partial_{z_k} X(z_k)  \, ,
\quad
\nonumber \\
{} [\mathbb{L}^+ , X(z_k)] &\equiv& L_k^+ X(z_k) = (z_k^2
\partial_{z_k} + 2 z_k j ) X (z_k) \, , \quad
\label{L-generators}
\\
{} [\mathbb{L}^0 , X (z_k)] &\equiv& L_k^0 X(z_k) = (z_k
\partial_{z_k} + j ) X (z_k) \, ,
\nonumber
\ea
where the notation was introduced for the differential operators $L_k^\alpha$
(with $\alpha=\pm,\,0$) acting on the light-cone coordinate $z_k$. Here $j$ is
the conformal spin of the field constructed from its scaling dimension $d$ and
the projection of the spin onto the light-cone $s$
\be\label{conf-spin}
j = \frac12(d + s)\,.
\ee
For gaugino/quark operators defined in \re{1/2-SYM} and \re{1/2-QCD}, these
parameters admit the values at the classical level $d=3/2$ and $s=1/2$ leading
to $j=1$. In quantum theory, the scaling dimension $d$ acquires an anomalous
contribution while $s$ is protected to all orders of the perturbation theory.

The field $X(z_k)$ defines a representation of the $SL(2;\mathbb{R})$ group which
we shall denote as $\mathcal{V}_j$. The representation space $\mathcal{V}_j$
possesses the highest weight $X(0)$, such that $L^+ X(0)=0$ and $L^0 X(0)=j X(0)$
while the descendant states are given by its derivative $(L^-)^k X(0) = (-1)^k
\partial_+^k X(0)$. The nonlocal operators $\mathbb{O}(z_1,z_2)$ belong to the
tensor product of two $SL(2;\mathbb{R})$ modules which can be decomposed over the
irreducible components as
\be\label{tensor_prod}
\mathcal{V}_{j} \otimes \mathcal{V}_{j} = \sum_{n \ge 0} \mathcal{V}_{2j+ n}\,.
\ee
The representation space of $\mathcal{V}_{2j+ n}$ is spanned by local twist-two
Wilson operators $\mathcal{O}_{nl}(0)$. Among these operators one can distinguish
the highest weight $l=n$ satisfying the relations \cite{Mak80,Ohr81}
\be\label{hw}
[\mathbb{L}^+ , \mathcal{O}_{nn}(0)] = 0\,,\qquad [\mathbb{L}^0 ,
\mathcal{O}_{nn}(0)] = (2j+n)\mathcal{O}_{n0}(0)
\ee
and its descendants with $l>n$
\be\label{desc}
\mathcal{O}_{nl}(0) = [[\mathcal{O}_{nn}(0),\mathbb{L}^-],\mathbb{L}^-]\ldots
\mathbb{L}^-]] = (\partial_+)^{l-n} \mathcal{O}_{nn}(0)\,.
\ee
The operators $\mathcal{O}_{nl}(0)$ carry the conformal spin $2j+n$ and have the
canonical dimension $l+3$.

It is straightforward to verify that the twist-two operators defined in
\re{conf-op} and \re{Gegenbauer} satisfy the relations \re{hw} and \re{desc} and,
therefore, they belong to the $SL(2;\mathbb{R})$ module $\mathcal{V}_{2j+ n}$.
Then, the conformal invariance implies that the Wilson operators belonging to
different $SL(2)$ modules do not mix with each other and, in addition, the
operators belonging to the same module have the same anomalous dimension.

Let us define the representation of the conformal group generators for the
nonlocal light-cone operators $\mathbb{O} (z_1,z_2)$ similar to \re{L-generators}.
Evaluating $[\mathbb{L}^\alpha , \mathbb{O} (z_1,z_2)]$ (with $\alpha=\pm,\, 0$)
and making use of \re{L-generators}, it is easy to see that to the lowest order
of perturbation theory the generators $\mathbb{L}^\alpha$ are given by the sum of
differential operators $L_1^\alpha+L_2^\alpha$ defined in \re{L-generators}. In
two of these operators, $\mathbb{L}^-$ and $\mathbb{L}^0$, we immediately
recognize (up to an additive constant) the differential operators which were
found to commute with the dilatation operator, see Eqs.\ (\ref{dilat}) and
(\ref{trans}). The conformal symmetry implies that the one-loop dilatation
operator also commutes with the conformal boost $\mathbb{L}^+$ leading to
\be
{}[L_1^\alpha+L_2^\alpha, \mathbb{H}^{\scriptscriptstyle (0)}_{12}] = 0\,,
\label{conf-inv}
\ee
with $\alpha=\pm,\, 0$. Notice that in distinction with (\ref{dilat}) and
(\ref{trans}) this property is lost starting from two loops due to the conformal
anomaly. We remind that the evolution kernel $\mathbb{V}^{(0)} ( u_1, u_2 | v_1,
v_2 )$ defines the operator $ \mathbb{H}^{\scriptscriptstyle  (0)}_{12}$,
Eq.~\re{LightConeKernel}, in the momentum representation.

It follows from \re{conf-inv} that the two-particle evolution kernel
$\mathbb{H}^{\scriptscriptstyle (0)}_{12}$ is a function of the quadratic Casimir
operators defined on the tensor product \re{tensor_prod}
\ba\label{Casimir}
L_{12}^2 &=& (L_{12}^0)^2 + \frac12 \left( L_{12}^+ L_{12}^- + L_{12}^-
L_{12}^+\right) \\ \nonumber &=& -(z_1-z_2)^{2(1-j)} \partial_{z_1}\partial_{z_2}
(z_1-z_2)^{2j}= J_{12} (J_{12}-1)
\ea
with $L_{12}^\alpha=(L_1+L_2)^\alpha$ and $J_{12}$ being the two-particle
conformal spin. In the coordinate representation, the $SL(2)$ module $V_{2j+n}$
is spanned by homogeneous polynomials
\begin{equation}
\label{SL2Module} \Psi_n^{(0)} (z_1, z_2) = (z_1 - z_2)^n \, , \qquad
\Psi_n^{(\ell)} (z_1, z_2) = (L_{12}^+)^\ell \Psi_n^{(0)} (z_1, z_2) \, .
\end{equation}
with $\Psi_n^{(0)}$ the lowest weight, $L_{12}^-\Psi_n^{(0)}=0$, and
$\Psi_n^{(\ell)}$ its descendant. These polynomials are in the one-to-one
correspondence with similar polynomials in the momentum representation,
Eq.~\re{Gegenbauer}, and satisfy the orthogonality condition similar to
\re{ortho}. The eigenstates \re{SL2Module} diagonalize the Casimir operator
\re{Casimir} and, as a consequence, the Hamiltonian \re{LightConeKernel} has the
same eigenvalues on these eigenfunctions independent of $\ell$,
\ba
\mathbb{H}^{\scriptscriptstyle (0)}_{12} \cdot \Psi^{(\ell)}_n(z_1, z_2) &=& 2[
\psi (n + 2j) - \psi (2j) ]\Psi^{(\ell)}_n(z_1, z_2)\,, \label{1-loop-AD}
\\ \nonumber
L_{12}^2 \,\Psi^{(\ell)}_n(z_1, z_2)&=&(n+2j)(n+2j-1)\Psi^{(\ell)}_n(z_1, z_2)
\ea
with $\psi(x)=d\ln\Gamma(x)/dx$. This relation is a counter-part of \re{V2-eig}
in the momentum representation. Together with $J_{12} \cdot \Psi^{(\ell)}_n= (n +
2j) \Psi^{(\ell)}_n$ it allows one to write down the two-particle kernel in the
$SL(2)$ invariant form
\be\label{H12=psi}
\mathbb{H}^{\scriptscriptstyle (0)}_{12} = 2[ \psi (J_{12}) - \psi (2j) ]\,.
\ee
Here $j=1$ is the conformal spin of the quark/gaugino fields to the lowest order
of perturbation theory.

%%%%%%%%%%%%%%%%%%%%%%%%%%%%%%%%%%%%%%%%%%%%%%%%%%%%%%%%%%%%%%%%%%%%%
\subsubsection{Conformal symmetry breaking}
\label{ConfSymmBreakSection}
%%%%%%%%%%%%%%%%%%%%%%%%%%%%%%%%%%%%%%%%%%%%%%%%%%%%%%%%%%%%%%%%%%%%%

Let us now examine the two-loop corrections to the dilatation operator for the
twist-two operators. As was explained in Sect.~\ref{SchemeDependence}, starting
from two loops the dilatation operator depends on the renormalization scheme.
Throughout the paper we employ the $\overline{\rm MS}-$scheme and its spin-off
the $\overline{\rm DR}-$scheme. Another subtle issue emerging at two loops is the
breaking of conformal symmetry in the dilatation operator within these two
schemes even for $\beta(g^2)=0$.

It is a matter of traditional folklore that in a generic gauge theory the
conformal symmetry is broken if the beta-function is nonvanishing. This is a
direct consequence of the dimensional transmutation phenomenon which generates
an intrinsic mass scale in gauge theory modifying the scaling behavior of
correlation functions. The nonzero beta-function induces an anomaly in the
trace of the energy-momentum tensor. The latter is then generates in turn
anomalous dimensions of composite operators when their product is renormalized
in the conformal Ward identities, as we will demonstrate below. However, even
in cases when the four-dimensional beta function is zero to all orders of
perturbation theory, there is yet another source of the conformal symmetry
breaking related to the choice of the regularization procedure. We remind that
in the dimensional regularization with $D=4-2\varepsilon$ the beta-function
acquires an additional contribution $\beta_{\varepsilon}(g^2)\sim - 2\varepsilon$,
Eq.~\re{beta-epsilon}, and, therefore, the conformal symmetry is violated for
$\varepsilon\neq 0$. According to \re{O-renorm} and \re{H from Z}, the dilatation
operator is related to the residue of a simple pole $1/\varepsilon$ in the
expression for renormalized light-cone operator $\mathbb{O}(z_i)$. Subtracting
divergences and sending $\varepsilon\to 0$ afterwards, one generates symmetry
breaking contribution to the dilatation operator coming from terms $\sim
\beta_{\varepsilon}(g^2)/ \varepsilon$. This source of the symmetry breaking
is a peculiar feature of dimensional regularization rather than intrinsic
property of the dilatation operator. In other words, in gauge theories with
vanishing beta-function the conformal symmetry breaking terms can be removed
by performing a scheme transformation of the dilatation operator and by going
over to the so-called conformal scheme. Obviously, this transformation does
not affect the eigenvalues of the dilatation operator but it does change the
form of the corresponding eigenstates.

The conformal operators $\mathbb{O}_{nl}(0)$ define the basis of twist-two
operators in gauge theory. To one-loop order these operators have an autonomous
scale dependence but they mix starting from two loops
\be
\left(
\mu \frac{\partial}{\partial \mu} + \beta(g^2) \frac{\partial}{\partial g^2}
\right)
\mathcal{O}_{nl}(0)
=
- \sum_{k = 0}^n {\gamma}_{nk} (g^2) \,
\mathcal{O}_{kl}(0) \, .
\ee
The mixing matrix is given by the perturbative expansion
\be\label{gamma-dec}
{\gamma}_{nk} (g^2) = \frac{g^2}{8\pi^2}\gamma^{(0)}_n \delta_{nk} +
\left(\frac{g^2}{8\pi^2} \right)^2{\gamma}_{nk}^{(1)} + \mathcal{O}(g^6)
\ee
and its form to higher orders is constrained by the Lorentz symmetry. To see this,
one considers the operator $\mathcal{O}_{nn}(z n_\mu)$ having the canonical
dimension $n+3$ and carrying the maximal possible conformal spin. It could mix
with operators $\mathcal{O}_{mn}(z n_\mu)$ of the same canonical dimension but
smaller conformal spin $m<n$. Notice that these operators necessarily involve
$(n-m)$ total derivatives. Then, expanding the operators around the origin $z=0$,
one finds that $\mathcal{O}_{nl}(0)=\partial_+^{l-n}\mathcal{O}_{nn}(0)$ could
only mix with the operators $\mathcal{O}_{kl}(0)$ with $k \le n$. This means that
the matrix of anomalous dimensions ${\gamma}_{nk} (g^2)$ has a triangular form,
$\gamma_{nk} = 0$ for $n < k$, or explicitly
\begin{equation}
\label{DiagonalNonDiagonalADs} \bit{\gamma}^{(1)} = \left(
\begin{array}{ccc}
\gamma_{00}^{(1)} & & \bit{0} \\
 \vdots           & \ddots &  \\
\gamma_{n0}^{(1)}           &    \cdots     & \gamma_{nn}^{(1)}
\end{array}
\right) \equiv \gamma_n^{(1)} \delta_{nk} + \gamma_{nk}^{\scriptscriptstyle\rm
ND} \theta_{nk} \, ,
\end{equation}
with the diagonal entries $\gamma_n^{(1)} \equiv \gamma_{nn}^{(1)}$ and
non-diagonal elements accompanied by the step function $\theta_{nk} = \{ 1 ,
n > k ; 0 , n \leq k \}$. The eigenvalues of the mixing matrix
\re{DiagonalNonDiagonalADs} are only determined by the diagonal matrix
elements $\gamma_n^{(1)}$ while the off-diagonal matrix elements only
affect the form of the corresponding eigenstates.

A convenient approach to study manifestation of the conformal symmetry breaking
in the mixing matrix \re{DiagonalNonDiagonalADs} is through the conformal Ward
identities. Below we briefly review this formalism, for a detailed discussion
see \cite{BelMul98} and reviews \cite{BraKorMul03,BelRad05}. Let us examine
the variation of the correlation function $\langle \mathcal{O}_{nl}(0) \ X (z_1)
\dots X (z_L) \rangle$ under transformations generated by the $SL(2)$ generator
$\mathbb{L}^0$
\begin{eqnarray}\label{CWI}
&&\sum_k^L \left( j + z_k  \partial_k \right) \langle \mathcal{O}_{nl}(0) \ X
(z_1) \dots X (z_L) \rangle \equiv - \langle \mathcal{O}_{nl}(0) \ \delta_0
\left( X (z_1) \dots X (z_L) \right) \rangle
\\
&&\qquad\qquad\qquad\qquad = \langle \delta_0 \mathcal{O}_{nl}(0)\, X (z_1) \dots
X (z_L) \rangle + \langle  i \delta_0 S_{\scriptscriptstyle \rm
YM}\mathcal{O}_{nl}(0)  X (z_1) \dots X (z_L) \rangle \, ,\nonumber
\end{eqnarray}
with $\delta_0 X (z_k) = -[\mathbb{L}^0,X(z_k)]$ defined in \re{L-generators}.
The last term in the right-hand side is due to nonvanishing variation of the
regularized Yang-Mills action. It is generated by a nonvanishing trace of the
energy-momentum tensor in the $D$-dimensional gauge theory $\delta_0
S_{\scriptscriptstyle \rm YM} = - i\int d^D z \, \Theta_{\mu\mu} (z)$. The
product of the conformal operator and the trace anomaly requires an additional
renormalization. This produces an anomalous contribution proportional to the
anomalous dimension of the conformal operator $\mathcal{O}_{nl}$
\begin{equation}
\label{DilatationConfOperRenorm}
\mathcal{O}_{nl}(0) \delta_0 S_{\scriptscriptstyle \rm YM}
=
\frac{i}{2}
\gamma_{nk} (g^2) \mathcal{O}_{k l}(0) + \dots
\, ,
\end{equation}
where ellipsis denote ``regular'' terms. Here and in what follows the summation
over the repeated index $0\le k \le n$ is implied. Combining together \re{CWI}
and \re{DilatationConfOperRenorm} one obtains that the conformal operator is
transformed as
\begin{equation}\label{CWI0}
{}[ \mathcal{O}_{nl} (0) , \mathbb{L}^0]
=
- \frac{1}{2}
\left[
\left( l + 4j \right) \delta_{nk} + \gamma_{nk} (g^2)
\right] \mathcal{O}_{kl} (0)
-
\frac{i}{2}
\left[
\mathcal{O}_{nl} (0) \int d^4 z \, {\mit\Theta}_{\mu\mu} (z)
\right]_{\scriptscriptstyle\rm R}
+
\dots \, ,
\end{equation}
where $j$ is the conformal spin of the constituent field and the subscript `R'
indicates that the product of two operators is renormalized. In a similar
manner, one can examine the variation of correlation functions under
transformations generated by the $SL(2)$ generator $\mathbb{L}^+$. They lead
to the following commutation relations
\begin{equation}\label{CWI+}
{}[\mathcal{O}_{nl} (0) , \mathbb{L}^+]
=
\frac{i}{2}
\left[ a_{nk} (l) + \gamma^c_{nk} (l; g^2) \right] \mathcal{O}_{k l - 1}
-
\frac{i}{2}
\left[ \mathcal{O}_{nl} (0) \int d^4 z \, 2
z^- {\mit\Theta}_{\mu\mu} (z)
\right]_{\scriptscriptstyle\rm R}
+
\dots \, ,
\end{equation}
where $a_{nk} (l) \equiv a (n,l) \delta_{nk} = 2(n - l)(n + l + 4 j - 1)
\delta_{nk}$ is a diagonal matrix and the matrix $\gamma^c_{nk} (l; g^2)$ is
the so-called special conformal anomaly. This matrix has a well-defined
perturbative expansion in powers of the gauge coupling and similar to the
scale anomaly, Eq.~\re{DilatationConfOperRenorm}, it arises from the
renormalization of the product of the conformal operator and the variation
of the Yang-Mills action,
\begin{equation}
\mathcal{O}_{n l}(0)  \delta_+ S_{\scriptscriptstyle \rm YM}
=
\frac{1}{2}
\gamma^c_{nk} (l; g^2) \mathcal{O}_{k l - 1}
+
\dots \, .
\end{equation}
The one-loop expression for $\gamma^c_{nk}$ has been found in Ref.~\cite{BelMul98}.

We remind that the $SL(2)$ generators satisfy the commutation relation
$[\mathbb{L}^0 , \mathbb{L}^{+}] = \mathbb{L}^{+}$. Evaluating the
commutators with $\mathcal{O}_{n l}(0)$ on both sides of this relation
and making use of \re{CWI0} and \re{CWI+} one obtains the consistency condition
\begin{equation}
\label{CommutatorConstraint} \left[ \bit{\gamma} (g^2) , \bit{a} + \bit{\gamma}^c
(g^2)
+
\beta (g^2) \bit{b} \right] = 0 \, ,
\end{equation}
which relates to each other the two anomaly matrices. Here, the $\bit{a}-$matrix
was defined in \re{CWI+} while the matrix $\bit{b}$ possesses the following
off-diagonal elements $b_{nk} = - 2 (2 k + 3)\theta_{nk}$. In distinction with
$\gamma_{nk} (g^2)$, the matrix elements $\gamma^c_{nk} (l; g^2)$ depend on the
scaling dimension of the conformal operator $l$. This dependence can be fixed from
the commutation relations $[\mathbb{L}^0 , \mathbb{L}^{-}] = \mathbb{L}^{-}$
leading to
\be
\gamma^c_{nk} (l+1; g^2)-\gamma^c_{nk} (l; g^2)=-2\gamma_{nk} (g^2)\,.
\ee

Since the one-loop correction to the matrix of anomalous dimension is diagonal,
Eq.~\re{gamma-dec}, the relation \re{CommutatorConstraint} is satisfied
automatically to the lowest order of perturbation theory. Beyond leading
order one finds that the special conformal anomaly has off-diagonal elements,
$\gamma^c_{nk} (l; g^2) \neq 0$ for $n > k$, and as a consequence the anomalous
dimension matrix $\gamma_{nk} (g^2)$ also acquires nonzero off-diagonal elements,
even in theory with vanishing four-dimensional beta function. The reason for this
apparently counterintuitive result is due to the fact that the beta-function is
different from zero in $D=4-2\varepsilon$ dimensions, the anomaly propagates
into the finite result for the off-diagonal matrix elements of $\gamma_{nk}
(g^2)$.

Expanding the left-hand side of (\ref{CommutatorConstraint}) in powers of $g^2$
one observes that (\ref{CommutatorConstraint}) mixes orders of perturbative
expansion. Obviously, this relation does not provide any information on the
diagonal elements of the anomalous dimension matrix $\gamma_{nn}$ but it does
constrain the off-diagonal elements $\gamma_{nk}$ with $n > k$. In two-loop
approximation, taking into account the explicit form of the one-loop special
conformal anomaly matrix $\gamma^c_{nk} (l; g^2)$, one finds the off-diagonal
elements of the anomalous dimension matrix as~\cite{BelMul98}
\begin{equation}
\label{ADinBetaZero} \left. \gamma_{nk}^{(1)} \right|_{n > k} = - \left(
\gamma_{n}^{(0)} - \gamma_{k}^{(0)} \right) \left\{ d_{nk}
(\gamma_{k}^{(0)}+\beta_0) + g_{nk} \right\} \, ,
\end{equation}
where the coefficients $d_{nk}$ are defined in terms of the derivative of the
Gegenbauer polynomials over its index as
\begin{equation}\label{index-der}
\left. \frac{d}{d \nu}  C_n^\nu (x)\right|_{\nu = 3/2} = - 2 \sum_{k = 0}^n
d_{nk} C_k^{3/2} (x) \, ,\qquad  d_{nk} =   - \frac{2k + 3}{(n - k)(n + k +
3)}\,,
\end{equation}
while the explicit form of $g_{nk}-$coefficients is not relevant for our purposes and
can be found in \cite{BelMul98}. As we show in the next section, Eq.~\re{ADinBetaZero}
allows one to restore the contribution of scalars to the two-loop evolution kernels
in SYM without actual loop calculations.

As we just explained, the conformal symmetry is broken in dimensional regularization
even when the four-dimensional $\beta$-function vanishes identically. In the latter
case, the conformal symmetry can be restored however by performing a finite
renormalization of the conformal operators, Eq.~\re{conf-op}
\begin{equation}
\label{MStoCS} \mathcal{O}_{nl}(0) = {\cal B}_{nk}(g^2) \widehat
{\mathcal{O}}_{kl}(0) \, .
\end{equation}
The ${\cal B}$-matrix absorbs symmetry breaking effects and rotates the conformal
operators to a new basis, in which the operators $\widehat {\mathcal{O}}_{kl}(0)$
do not mix and have an autonomous scale dependence (for $\beta(g^2)=0$)
\begin{equation}
\label{RGeqOhat} \mu\frac{d}{d\mu} \widehat {\mathcal{O}}_{nl}(0) = -
\gamma_n(g^2) \widehat {\mathcal{O}}_{nl}(0) \, .
\end{equation}
The anomalous dimensions $\gamma_n(g^2)$ coincide with the diagonal matrix elements
of the mixing matrix \re{gamma-dec}, $\gamma_{n} \equiv \gamma_{nn}(g^2)$. The
transformation \re{MStoCS} defines a new scheme known as the conformal subtraction
scheme, or simply $\overline{\rm CS}$. By construction, the rotation matrix ${\cal B}$
diagonalizes the conformal anomaly matrix
\begin{equation}\label{B-gamma}
\left({\cal B}^{- 1}\,\bit{\gamma}(g^2)\, {\cal B} \right)_{nk} = \gamma_n(g^2)
\delta_{nk} \, .
\end{equation}
Wherefrom one can find that $\mathcal{B}^{-1}_{nk} = \delta_{nk} + \gamma^c_{nk}
(k; g^2)/a (n,k)$. Then, it follows from \re{CommutatorConstraint} that, in the
conformal limit $\beta(g^2)=0$, the ${\cal B}-$matrix also diagonalizes the
special conformal anomaly matrix
\be\label{B-gamma-c}
\left({\cal B}^{- 1}[\bit{a} + \bit{\gamma}^c (g^2)]\,{\cal B}\right)_{nk}
=
2(n-l)(n+l+4j-1+\gamma_n(g^2))\delta_{nk}\,.
\ee
In this limit, the energy-momentum tensor has vanishing trace,
$\Theta_{\mu\mu}=0$, and, as a consequence, Eqs.~\re{CWI0} and \re{CWI+} simplify
significantly. Combining them with \re{MStoCS}, \re{B-gamma} and \re{B-gamma-c}
we conclude that the operators $\widehat {\mathcal{O}}_{nl}(0)$ (with $l\ge n$)
define a representation of the collinear $SL(2)$ group. The highest weight of
this representation, $\widehat {\mathcal{O}}_{nn}(0)$, satisfies the relations
\ba\label{rest}
[\mathbb{L}^+ , \widehat{\mathcal{O}}_{nn}(0)] =
0\,,\qquad{}[\mathbb{L}^0,\widehat{\mathcal{O}}_{nn} (0)]
\!\!\!&=&\!\!\!
(n+2j+\gamma_n(g^2))\widehat{\mathcal{O}}_{nn} (0)\,,
\ea
which should be compared with \re{hw}. One observes that, in the conformal limit,
higher order corrections renormalize the conformal spin of the $SL(2)$ module,
to which the conformal operators $\widehat {\mathcal{O}}_{nl}(0)$ belong to, by
an amount proportional to their anomalous dimension. This property alone fixes
the explicit form of the operators $\widehat {\mathcal{O}}_{nl}(0)$. As before,
the conformal symmetry constrains the form of twist-two operators but it does not
allow us to determine the explicit expressions for their anomalous dimension.

%%%%%%%%%%%%%%%%%%%%%%%%%%%%%%%%%%%%%%%%%%%%%%%%%%%%%%%%%%%%%%%%%%%%%
\subsubsection{Contribution of scalars from one-loop special conformal anomaly}
\label{ConformalConstraints}
%%%%%%%%%%%%%%%%%%%%%%%%%%%%%%%%%%%%%%%%%%%%%%%%%%%%%%%%%%%%%%%%%%%%%

Let us give a brief demonstration of the application of the conformal
constraint \re{ADinBetaZero} at two-loop order and how it compares with
the conventional diagonalization procedure of the evolution kernels deduced
in the previous sections. In QCD, the twist-two two-loop evolution kernel
in the momentum representation was given in Eq.\ \re{V-QCD}. In SYM case,
the two-loop evolution kernel receives an additional contribution from
scalars determined by the Feynman diagrams shown in Fig.~\ref{scalarvertex}.
As was already explained in Sect.~\ref{TwoLoopTwoParticleContrSection},
the scalar contribution to the twist-two evolution kernel coincides,
modulo the overall color factor, with the two-particle kernel \re{V-scalar}
contributing to the three-fermion evolution kernel.

In the basis of conformal operators ${\mathcal{O}}_{nl}(0)$, Eqs.~\re{conf-op}
and \re{Gegenbauer}, the scalar kernel \re{V-scalar} is represented by the mixing
matrix $\gamma_{nk}^{(1)\rm sc}$. Similar to \re{V2-eig}, it can be found as
\be\label{gamma-sc}
 \int [du]_2\,
2\mathbb{V}^{(1)}_{\rm\scriptstyle scalar} (u_1, u_2 | v_1,v_2)
\textrm{C}_n^{3/2} \left( \frac{u_1 - u_2}{u_1 + u_2} \right) = - \sum_{k=0}^n
\left({\gamma_{nk}^{(1)\rm sc}}/{N_c^2}\right)\,
\textrm{C}_k^{3/2} \left( \frac{v_1 - v_2}{v_1 + v_2} \right) \, ,
\ee
with the measure $[du]_2$ defined after Eq.\ \re{V2-eig}. Here, the factor $2$ in
the left-hand side accounts for the difference in color factors of singlet and
octet kernels, cf.\ Table in Sect.~\ref{TwoLoopGluonContributionSection}. As
expected, the mixing matrix has a triangular form, $\gamma_{nk}^{(1)\rm sc}=0$ for
$k>n$. Making use of \re{V2-eig}, it is easy to see that the off-diagonal matrix
elements $\gamma_{nk}^{(1)\rm sc}$ with $k < n$ are generated by the
$\ln(u/v)-$enhanced terms in the right-hand side of \re{V-scalar}.

Let us demonstrate that the contribution of scalars to the two-loop mixing
matrix $\gamma_{nk}^{(1)\rm sc}$ is fully consistent with the conformal symmetry
constraints \re{CommutatorConstraint} and \re{ADinBetaZero}. One notices that
the scalars contribute to the right-hand side of \re{ADinBetaZero} through the
one-loop anomalous dimension, Eq.~\re{gamma-0-SYM}, and the one-loop beta-function
\be
\gamma_n^{(0)}=\frac12 n_s N_c + \ldots
\, , \qquad
\beta_0 = - \frac16 n_s N_c + \ldots\,,
\ee
where $n_s=2(\mathcal{N} - 1)$ and ellipsis denote the remaining,
$n_s-$independent terms. Substituting these expressions into \re{ADinBetaZero},
we can predict the contribution of scalars to the off-diagonal part of the
two-loop mixing matrix
\begin{equation}
 \gamma^{(1)\rm sc}_{nk} \bigg|_{n > k} = - \left(
\gamma_n^{(0)} - \gamma_k^{(0)} \right) d_{nk} \left( - \frac{n_s}{6} +
\frac{n_s}{2} \right) N_c \, ,
\end{equation}
where $\gamma_n^{(0)}$ is given by \re{gamma-0-SYM} and the coefficients $d_{nk}$
were defined in \re{index-der}. A straightforward calculation of the integral in
the left-hand side of \re{gamma-sc} leads to the same expression.

%%%%%%%%%%%%%%%%%%%%%%%%%%%%%%%%%%%%%%%%%%%%%%%%%%%%%%%%%%%%%%%%%%%%%
\subsubsection{Anomalous dimension of twist-two operators}
\label{TwistTwoTwoLoopADs}
%%%%%%%%%%%%%%%%%%%%%%%%%%%%%%%%%%%%%%%%%%%%%%%%%%%%%%%%%%%%%%%%%%%%%

Let us finally turn to the calculation of the eigenvalues of the two-loop
mixing matrices in QCD and SYM theories. We remind that the mixing matrix has
a triangular form \re{DiagonalNonDiagonalADs} and, therefore, its eigenvalues
coincide with the diagonal matrix elements, $\gamma_n(g^2)=\gamma_{nn}(g^2)$.
The off-diagonal matrix elements correspond to the admixture of operators with
total derivatives and they do not affect $\gamma_n(g^2)$. One can easily get
rid of this redundant mixing by neglecting the operators with total derivatives.
This can be rigorously done by going over from the operators to their forward
matrix elements over some reference momentum state, the so-called forward limit
\be
\mathcal{O}_{nn} (0) \to \langle p | \mathcal{O}_{nn} (0) | p \rangle\,.
\ee
In distinction with $\mathcal{O}_{nn} (0)$ its forward matrix elements have an
autonomous scale dependence since the operators with total derivatives do not
contribute $\langle p | \mathcal{O}_{nn} (0) | p \rangle$. In the momentum
representation, the forward limit amounts to setting the total momentum equal to
zero, $\sum_k u_k=\sum_k v_k=0$.

To two-loop order, the eigenvalue of the mixing matrix can be written as
\be\label{gamma-general}
\gamma_n(g^2)=\frac{g^2}{8 \pi^2} \gamma_n^{(0)} + \left( \frac{g^2}{8 \pi^2}
\right)^2 \gamma_n^{(1)} + \dots \, .
\ee
Here $\gamma_n^{(0)}$ is the same in the $\overline{\rm MS}-$ and $\overline{\rm
DR}-$schemes while $\gamma_n^{(1)}$ is scheme-dependent. In QCD, the eigenvalue
of the two-loop evolution kernel \re{Def-ERBL-tra} can be extracted from the
available expression for the anomalous dimension of the quark transversity operator
in the $\overline{\rm MS}$ scheme \cite{Vog97}
\begin{eqnarray}\label{QCD-1-loop}
\gamma_n^{(0)\phantom{,{\overline{\scriptscriptstyle\rm MS}}}}\!\!\!&=&\!\!\! C_F
\left\{ 4 S_1(n) - 3 \right\}
\\
\gamma^{(1),{\overline{\scriptscriptstyle\rm MS}}}_{n} \!\!\!&=&\!\!\! C_F^2
\left\{ - \frac{3}{4} + 6 S_2(n) + 16 S_{- 2, 1}(n) - 8 S_3(n) - 8 S_{- 3}(n) - 8
S_1(n) ( S_2(n) + 2 S_{- 2}(n) ) \right\}
\nonumber\\
&+&\!\!\!
C_F N_
c\left\{ - \frac{17}{12} + \frac{134}{9} S_1(n)
- \frac{22}{3} S_2(n) - 8 S_{- 2, 1}(n) + 4 S_3(n) + 4 S_{- 3}(n) + 8 S_1(n) S_{-2} (n)
\right\}
\nonumber\\ &+&\!\!\!
C_F n_f \left\{ \frac{1}{6} - \frac{20}{9} S_1(n) +
\frac{4}{3} S_2(n) \right\} \, , \label{QCD-2-loop}
\end{eqnarray}
for odd $n$. It is expressed in terms of the harmonic sums
\begin{equation}\label{sums}
S_{\pm \ell}(n) = \sum_{k = 1}^{n + 1} \frac{(\pm 1)^k}{k^\ell} \, , \qquad
S_{\pm \ell,m}(n) = \sum_{k = 1}^{n + 1} \frac{(\pm 1)^k}{k^\ell} S_m (k-1) \, .
\end{equation}
The one-loop result \re{QCD-1-loop} coincides with \re{gamma-0-QCD}. The two-loop
result can be reproduced by calculating the mixing matrix in the basis of
conformal operators, $\gamma_{nk}$, and, then, finding its eigenvalues.

To obtain the eigenvalues of the two-loop mixing matrix in SYM case, we have to
``supersymmetrize'' the QCD expression \re{QCD-1-loop} and \re{QCD-2-loop} by
adjusting the values of the color factors, then adding the contribution of
scalars present in SYM theories with extended supersymmetry and, finally,
transforming the entire result to the DRED scheme. The first of these steps is
trivially accomplished via the identification
\begin{equation}
\label{Supersymmetrization} C_F \to N_c \, , \qquad n_f \to  N_c\,\mathcal{N} \,.
\end{equation}
The anomalous dimension obtained via this procedure will be denoted
$\gamma^{{\scriptscriptstyle\rm sQCD}, {\overline{\scriptscriptstyle\rm
MS}}}_{n}$. We remind that in SYM with $\mathcal{N}$ supercharges, the gaugino
fields belong to the adjoint representation and their total number is related to
the number of scalars as $ n_f = 1 + {n_s}/{2} = \mathcal{N}$. So we set it to
this value in what follows.

Let us now add the contribution of scalars. To one- and  two-loop orders it comes
from diagrams shown in Fig.\ \ref{oneloopkernel} (c) and Figs.\ \ref{scalarvertex},
\ref{GauginoSEscalars2loop}, respectively. These diagrams have been calculated in
the previous section as a part of the evolution kernel for three-gaugino operators.
For the twist-two operators, the only difference is that the pair of gaugino
fields carries zero color charge which affects the overall color factors. Using
our findings from the previous section, we get for the contribution of scalars
in the $\overline{\rm MS}$ scheme
\begin{equation}
\gamma_{n}^{{\rm sc}, \overline{\scriptscriptstyle\rm MS}}
(g^2) = \frac{g^2N_c}{8 \pi^2} \frac{n_s}{2} + \left( \frac{g^2 N_c}{8 \pi^2}
\right)^2 \left\{ - \frac{n_s^2}{4} + n_s \left( \frac{7}{4} - \frac{8}{9} S_1(n)
- \frac{2}{3} S_2(n) \right) \right\} \, .
\end{equation}
Combining this expression together with the QCD result ``supersymmetrized'' via
(\ref{Supersymmetrization}), we get the anomalous dimension of the twist-two
gaugino operator \re{O2-SYM} in the $\overline{\rm MS}-$scheme
\begin{equation}\label{SYM-MS}
\gamma^{{\overline{\scriptscriptstyle\rm MS}}}_n (g^2) =
\gamma^{{\scriptscriptstyle\rm sQCD}, {\overline{\scriptscriptstyle\rm MS}}}_n
(g^2) + \gamma^{{\rm sc}, \overline{\scriptscriptstyle\rm MS}}_n (g^2) \, .
\end{equation}
As a final step we have to transform this result to the supersymmetry preserving
dimensional reduction scheme. This scheme transformation was described in detail
in Sect.~5.3.1 for a general $L-$particle operator. Applying the formulas
\re{H-DR} for $L=2$ we obtain the anomalous dimension in the dimensional
reduction scheme as
\begin{equation}\label{SYM-DR}
\gamma^{{\overline{\scriptscriptstyle\rm DR}}}_n
(g^2_{\overline{\scriptscriptstyle\rm DR}}) =
\gamma^{{\overline{\scriptscriptstyle\rm MS}}}_n \left(
g^2_{\overline{\scriptscriptstyle\rm DR}} - \frac{N_c}{6}
\frac{g^4_{\overline{\scriptscriptstyle\rm DR}}}{8 \pi^2} \right) - \frac{\beta_0
N_c}{2} \left( \frac{g^2_{\overline{\scriptscriptstyle\rm DR}}}{8 \pi^2}
\right)^2  + \mathcal{O}(g^6)\, .
\end{equation}
Combining together \re{SYM-MS} and \re{SYM-DR}, we find the one- and two-loop
corrections to the anomalous dimension \re{gamma-general} in the SYM theory as
functions of the number of supercharges $\mathcal{N}$
\begin{eqnarray}
\gamma_n^{(0)\phantom{,\overline{\scriptscriptstyle\rm DR}}} \!\!\!&=&\!\!\! N_c
\big\{ \mathcal{N}- 4 + 4 S_1(n)  \big\}
\, , \nonumber \\[2mm]
\gamma_n^{(1),\overline{\scriptscriptstyle\rm DR}} \!\!\!&=&\!\!\! N_c^2 \left\{
-(\mathcal{N}-4)(\mathcal{N}-2) - 4 \left[ \mathcal{N} - 4 + 2 S_2(n) + 2 S_{-
2}(n) \right] S_1(n) \nonumber
\right. \\[2mm] && \left. \qquad
- 4 S_3(n) - 4 S_{- 3}(n) + 8 S_{- 2, 1}(n) \right\} \, . \label{uni}
\end{eqnarray}
Here, the $\mathcal{N}-$dependence accounts for contribution from
$n_s=2(\mathcal{N}-1)$ scalar and $n_f=\mathcal{N}$ gaugino fields.
The one-loop result matches \re{gamma-0-SYM}.

The following comments are in order.

In SYM theory, the maximal helicity twist-two operator \re{O2-SYM} is a
component of a supermultiplet of twist-two operators. Supersymmetry implies
that all components of the supermultiplet have the same anomalous dimension,
Eqs.~\re{gamma-general} and \re{uni}. In the maximally supersymmetric
$\mathcal{N}=4$ SYM theory, all twist-two operators belong to a single
supermultiplet and, as a consequence, their anomalous dimension is given
by the so-called universal anomalous dimension.%
\footnote{In SYM theory with $\mathcal{N}<4$ supercharges there are two
different supermultiplets of twist-two operators and, as a consequence,
there are two universal anomalous dimensions~\cite{BelDerKorMan04}. The
maximal helicity operators belong to one of the supermultiplets.} One can
verify that for $\mathcal{N}=4$ our result for the anomalous dimension,
Eqs.~\re{gamma-general} and \re{uni}, agrees with the one of Refs.\
\cite{KotLip02,DolOsb05}.

The expression for the two-loop anomalous dimension, Eqs.~\re{gamma-general} and
\re{uni}, can be analytically continued from nonnegative integer $n$ to the
entire complex $n-$plane. The value of the anomalous dimension for $n=-1$ is of a
special interest. Formally, it corresponds to an anomalous dimension of a
nonlocal operator $\lambda^{\{A}\partial_+^{-1} \lambda^{B\}}$ with canonical
dimension $2$. However, supersymmetry implies that the same quantity determines
the anomalous dimension of a local operator, given by the product of two complex
holomorphic scalar fields. The latter operator is protected by supersymmetry and
its anomalous dimension is given to all orders by the beta-function of the
corresponding SYM model~\cite{BlaLemMagSorTanVenVil00}. This leads to the
following relation
\begin{equation}\label{minus-one}
\gamma_{n = -1}^{\overline{\scriptscriptstyle\rm DR}}
(g^2_{\overline{\scriptscriptstyle\rm DR}})
=
\beta_{\scriptscriptstyle\rm SYM} (g^2_{\overline{\scriptscriptstyle\rm DR}}) \, .
\end{equation}
Indeed, one verifies using \re{sums} that for $n=-1$ all harmonic sums vanish in
the left-hand side of \re{uni} and the relation \re{minus-one} holds true to two
loops.

We argued in Sect.~\ref{NdepSection}, that the $\mathcal{N}-$dependence of
the evolution kernel for three-gaugino operators can be factored out into
c-valued normalization factors. One can verify that the same relation
\re{N-factorization} fulfilled by the anomalous dimension of the twist-two
operators, Eqs.~\re{gamma-general} and \re{uni}. Moreover, taking into
account \re{minus-one} one can write this relation in a concise form as
\be
\label{tw-2-relation}
\left[\gamma_{n}(g^2) - \beta_{\scriptscriptstyle\rm SYM}(g^2)\right]_{\mathcal{N}}
=\left(1-(\mathcal{N}-1)\frac{g^2 N_c}{8\pi^2}\right)
\left[\gamma_{n}(g^2) - \beta_{\scriptscriptstyle\rm SYM}(g^2)\right]_{\mathcal{N}=1}
+
\mathcal{O}(g^6)
\, ,
\ee
where we introduced the subscript to indicate the value of $\mathcal{N}=0,1,2,4$.
We conclude that the anomalous dimensions of twist-two operators \re{O2-SYM} in
{\sl all} SYM theories are related to each other to two-loop accuracy as in
\re{tw-2-relation}. Taking into account that $\beta_{\scriptscriptstyle\rm
SYM}(g^2)=0$ for $\mathcal{N}=4$, the same relation can be written as
\be
\left[\gamma_{n}(g^2) - \beta_{\scriptscriptstyle\rm SYM}(g^2)\right]_{\mathcal{N}}
=
\left(1-(\mathcal{N}- 4)\frac{g^2 N_c}{8\pi^2}\right)
{}[ \gamma_{n}(g^2) ]_{\mathcal{N}=4} + \mathcal{O}(g^6) \,.
\ee

Let us examine the asymptotics of the anomalous dimension at large $n$. For $n\gg
1$ the leading contribution to \re{uni} comes from terms involving $S_1(n)\sim
\ln n$. In this way one finds that the anomalous dimension scales logarithmically
at large $n$
\be
\gamma_{n}(g^2) =  4 \left(\frac{g^2 N_c}{8\pi^2} - \left(\frac{g^2
N_c}{8\pi^2}\right)^2 \left[ \mathcal{N}-4+\frac{\pi^2}6\right] \right) \ln n +
\mathcal{O}(n^0)\,.
\ee
Comparing this relation with \re{cusp-SYM} one observes that the prefactor
coincides with the cusp anomalous dimension leading to
\be\label{tw-2-cusp}
\gamma_{n}(g^2) = 2 \Gamma_{\rm cusp}(\lambda) \ln n + \mathcal{O}(n^0)\,,
\ee
with the coupling $\lambda=g^2 N_c/(8\pi^2)$ defined in \re{CouplingConstant}.
One can show that this relation holds true to all loops with $\Gamma_{\rm
cusp}(\lambda)$ modified by higher order corrections~\cite{Kor89}. Substituting
\re{tw-2-cusp} into \re{tw-2-relation} one finds that as far as the
$\mathcal{N}-$dependence is concerned the cusp anomaly satisfies the same
relation, Eq.~\re{tw-2-relation}.

%%%%%%%%%%%%%%%%%%%%%%%%%%%%%%%%%%%%%%%%%%%%%%%%%%%%%%%%%%%%%%%%%%%%%
\subsection{Three-particle operators}
\label{3Pspectrum}
%%%%%%%%%%%%%%%%%%%%%%%%%%%%%%%%%%%%%%%%%%%%%%%%%%%%%%%%%%%%%%%%%%%%%

Let us now turn to diagonalization of the evolution kernel for the three-fermion
operators in QCD and SYM theories. As for two particle operators, we start in
Sect.\ \ref{SubSubSec-EigSpeLO} with finding the eigenspectrum of the dilatation
operator to one-loop order.  This allows us to elucidate in Sect.\
\ref{CriteriumSection} the symmetries of the all-order dilatation operator and
their manifestation in the eigenspectrum. Then in Sect.\ \ref{NumericsSection} we
numerically study  the eigenspectrum of the dilatation operator
(\ref{V-two-loop}) at two loops and consider afterwards the limits of large spin
and $\beta_0\to\infty$.

%%%%%%%%%%%%%%%%%%%%%%%%%%%%%%%%%%%%%%%%%%%%%%%%%%%%%%%%%%%%%%%%%%%%%
\subsubsection{Eigenspectrum to one-loop}
\label{SubSubSec-EigSpeLO}
%%%%%%%%%%%%%%%%%%%%%%%%%%%%%%%%%%%%%%%%%%%%%%%%%%%%%%%%%%%%%%%%%%%%%

The one-loop dilatation operator has been determined in
Sect.~\ref{OneLoopDilOper}. We demonstrated there that it is given in QCD
and SYM theories by the same universal expression. In the coordinate
representation, the eigenproblem for the one-loop dilatation operator reads
\be\label{H0-coor} \mathbb{H}^{\scriptscriptstyle
(0)}_{\phantom{i}} \Psi_{\bit{{\scriptstyle q}}}(z_1,z_2,z_3)
=\gamma_{\bit{{\scriptstyle q}}}^{\scriptscriptstyle
(0)}\Psi_{\bit{{\scriptstyle q}}}(z_1,z_2,z_3)
\ee
with the kernel $\mathbb{H}^{\scriptscriptstyle (0)}_{\phantom{i}}$
\re{LOkernel} built from the pairwise interactions \re{LightConeKernel}
and $\bit{q}$ enumerating the solutions. In the momentum representation,
the same relation looks like
\be
\int [du]_3 \mathbb{V}^{\scriptscriptstyle (0)} (\bit{u}| \bit{v} )
\,P_{\bit{{\scriptstyle q}}}(u_1,u_2,u_3) = - \gamma^{\scriptscriptstyle
(0)}_{\bit{{\scriptstyle q}}} \, P_{\bit{{\scriptstyle q}}}(v_1,v_2,v_3) \,,
\ee
with the evolution kernel $\mathbb{V}^{\scriptscriptstyle (0)} (\bit{u}|
\bit{v})$ given by \re{PairWiseLOkernel} and the integration measure $[du]_3$
defined in Eq.~\re{3-measure}. The advantage of the coordinate representation is
that the conformal symmetry is manifest (see Eq.~\re{H12=psi}). On the other
hand, the momentum representation is more convenient for actual evaluation of the
eigenvalues $\gamma^{\scriptscriptstyle (0)}_{\bit{{\scriptstyle q}}}$ since it
allows one to eliminate the contribution of operators with total derivatives by
going over to the forward limit $\sum_k u_k = \sum_k v_k = 0$ (see
Appendix~\ref{ForwardLimitAppendix}). The eigenstates in the two representations,
$\Psi_{\bit{{\scriptstyle q}}}(z_i)$ and $P_{\bit{{\scriptstyle q}}}(v_i)$ are
homogeneous polynomials in light-cone coordinates and momentum, respectively.
They are related to each other through the orthogonality condition \re{ortho}. We
remind that $P_{\bit{{\scriptstyle q}}} (v_i)$ determines the explicit form of
local Wilson operator \re{O_q} while $\Psi_{\bit{{\scriptstyle q}}}(z_i)$ defines
the coefficient function accompanying this operator in the operator product
expansion \re{B-expansion}.

The dilatation operator $\mathbb{H}^{\scriptscriptstyle (0)}_{\phantom{i}}$,
Eq.~\re{LOkernel}, is invariant under the cyclic permutation of particles,
$\mathbb{P}$, and the interchange of any their pair, say $\mathbb{P}_{12}$,
\be\label{P-symmetry}
[\mathbb{P}, \mathbb{H}] = [\mathbb{P}_{12}, \mathbb{H}] = 0 \, ,
\ee
with the corresponding discrete operators defined as
\be\label{P-definition}
\mathbb{P}\, \Psi_{\bit{\scriptstyle q}}
(z_1, z_2, z_3) = \Psi_{\bit{\scriptstyle q}} (z_2, z_3, z_1) \, ,
\qquad \mathbb{P}_{12} \Psi_{\bit{\scriptstyle q}} (z_1, z_2, z_3)
= \Psi_{\bit{\scriptstyle q}} (z_2, z_1, z_3) \, .
\ee
Since these two operators do not commute, $[ \mathbb{P}, \mathbb{P}_{12}] \neq
0$, the eigenfunctions $\Psi_{\bit{\scriptstyle q}}(z_i)$ can diagonalize
either $\mathbb{P}$ or $\mathbb{P}_{12}$, but not both simultaneously. If
$\Psi_{\bit{\scriptstyle q}}(z_i)$ diagonalizes $\mathbb{P}$, then the
eigenstates of \re{H0-coor} can be classified according to their quasimomentum
$\theta_{\bit{\scriptstyle q}}$,
\be
\mathbb{P}
\,\Psi_{\bit{\scriptstyle q}} (z_1, z_2, z_3) = {\rm e}^{i
\theta_{\bit{\scriptscriptstyle q}}} \Psi_{\bit{\scriptstyle q}}
(z_1, z_2, z_3) \, ,
\ee
which takes three different values satisfying $ {\rm e}^{3i
\theta_{\bit{\scriptscriptstyle q}}} =1$ %(exponential of the cubic roots of unity)
\be
\theta_{\bit{\scriptstyle q}} = 0, \pm \frac{2
\pi}{3} \, .
\ee
Then, the eigenfunctions of \re{H0-coor} with definite parity are given by
\be \label{DefParityStates}
\Psi_{\bit{\scriptstyle q}}^\pm (\bit{z}) = \ft1{\sqrt{2}} (1 \pm
\mathbb{P}_{12}) \Psi_{\bit{\scriptstyle q}} (\bit{z}) \, , \qquad
\mathbb{P}_{12} \Psi_{\bit{\scriptstyle q}}^\pm (\bit{z}) = \pm
\Psi_{\bit{\scriptstyle q}}^\pm (\bit{z}) \, .
\ee
Invariance of the dilatation operator under the discrete transformations
\re{P-symmetry} immediately implies that its spectrum is double degenerate except
for the eigenstates with zero quasimomentum $\theta_{\bit{\scriptstyle
q}}=0$~\cite{BraDerMan98}. To see this one applies the eigenstate
$\Psi_{\bit{{\scriptstyle q}}} (z_1,z_2,z_3)$ to both sides of the relation
$\mathbb{P}\, \mathbb{P}_{12}\, \mathbb{P} = \mathbb{P}_{12}$ which follows from
the definition \re{P-definition}. In this way, one obtains
\be
\mathbb{P}
\left( \mathbb{P}_{12} \Psi_{\bit{{\scriptstyle q}}}(z_i) \right)
=
\e^{-i \theta_{\bit{{\scriptstyle q}}}}
\mathbb{P}_{12} \Psi_{\bit{{\scriptstyle q}}}(z_i)
\, ,
\ee
and, therefore, $\mathbb{P}_{12}\Psi_{\bit{{\scriptstyle q}}}(z_i)$ defines
yet another eigenstate with the same ``energy'' $\gamma^{\scriptscriptstyle
(0)}_{\bit{{\scriptstyle q}}}$ unless $\theta_{\bit{{\scriptstyle q}}} = 0$,
leading to
\be\label{degeneracy}
[\mathbb{H}^{\scriptscriptstyle (0)} -
\gamma_{\bit{\scriptstyle q}} (\lambda)] \Psi_{\bit{\scriptstyle
q}}^\pm(z_i) = 0 \, .
\ee
As we will see in a moment, integrability extends this property to the
eigenstates with $\theta_{\bit{{\scriptstyle q}}} = 0$. We would
like to stress that the degeneracy of the eigenstates with
$\theta_{\bit{\scriptstyle q}}\neq 0$ follows from the symmetry of
the dilatation operator under discrete transformations \re{P-definition}
and, therefore, it holds to all loops.

The additional restrictions on the eigenspectrum of the dilatation operator
are imposed by the properties of the nonlocal light-cone operators \re{O3-SYM}
and \re{O3-QCD}. Since both operators are invariant under cyclic permutations
of fields, the same property should hold true for the polynomials
$\Psi_{\bit{\scriptstyle q}}^\pm(z_i)$ entering \re{B-expansion}. Therefore,
their quasimomentum ought to be equal to zero, $\theta_{\bit{\scriptstyle q}}
= 0$. One can ``revive'' the remaining eigenstates with
$\theta_{\bit{\scriptstyle q}}\neq 0$ by assigning an additional ``flavor''
index to fermions. In QCD this procedure has a direct physical meaning (there
are six different species of quarks) while in the SYM case it would violate
supersymmetry by breaking a balance between bosonic and fermionic degrees of
freedom.

The one-loop dilatation operator $\mathbb{H}^{\scriptscriptstyle
(0)}_{\phantom{i}}$ is invariant under the conformal
transformations, $[L_1^\alpha+L_2^\alpha+L_3^\alpha,
\mathbb{H}^{\scriptscriptstyle (0)}] = 0$, with the $SL(2)$
generators $L_k^\alpha$ defined in \re{L-generators}. This allows
one to classify its eigenstates $\Psi_{\bit{{\scriptstyle
q}}}(z_i)$ according to irreducible components in the tensor
product of three $SL(2)$ modules, $\mathcal{V}_{j} \otimes
\mathcal{V}_{j} \otimes \mathcal{V}_{j}$. Applying the relation
\re{tensor_prod} twice, one finds that $\Psi_{\bit{{\scriptstyle
q}}}(z_i)$ carries the total three-particle conformal spin equal
to $J_{123}={3j + N}$ with $j=1$ and integer $N\ge 0$. This is the
eigenvalue of the three-particle quadratic Casimir
\be
\label{ThreePartConfSpinChange}
q_2^{(0)} = L_{12}^2 + L_{23}^2 + L_{31}^2 - 3 j (j - 1) \, ,
\ee
such that
\be
q_2^{(0)}\, \Psi_{\bit{{\scriptstyle q}}}(z_i) = J_{123} (J_{123} - 1)
\Psi_{\bit{{\scriptstyle q}}}(z_i) \, ,
\ee
with $J_{123}={3 + N}$. On the other hand, the two-particle conformal spin in any
subchannel, say $(12)$, is not fixed anymore $2j\le J_{12}\le 2j+N$. This means
that, in distinction with the $L=2$ particle operators discussed in the previous
section (see Eq.~\re{SL2Module}), the conformal invariance alone does not permit
us to determine the eigenstates of the three-particle dilatation operator. Still,
it allows us to write a general expression for the eigenstates
$\Psi_{\bit{{\scriptstyle q}}}(z_i)$ as a sum over the states possessing definite
conformal spins $J_{123}=3j+N$ and $J_{12}=2j+n$ and reduce the eigenproblem
\re{H0-coor} to finding the corresponding expansion coefficients.

The conformal symmetry also dictates that the eigenstates
$\Psi_{\bit{{\scriptstyle q}}}(z_i)$ have to be orthogonal to each other with
respect to the $SL(2)$ scalar product (for a review see
Refs.~\cite{BraKorMul03,BelDerKorMan05}). In the momentum representation, the same
condition looks like
\be
\vev{\Psi_{\bit{{\scriptstyle q}}}|\Psi_{\bit{{\scriptstyle q'}}}}
=
\int_0^1 [d u]_3 \, u_1 u_2 u_3
P_{\bit{{\scriptstyle q}}}(u_i) P_{\bit{{\scriptstyle q'}}}(u_i)
=
\delta_{\bit{{\scriptstyle q}}\bit{{\scriptstyle q'}}}
\,. \label{scalar_product}
\ee
This
relation is a counter-part of the orthogonality condition for the Gegenbauer
polynomials defining the conformal polynomials for $L=2$ operators,
Eq.~\re{Gegenbauer}. For the eigenstates of definite parity, the
orthogonality condition looks like
\be
\label{parity-norm}
\vev{\Psi^\pm_{\bit{{\scriptstyle
q}}}|\Psi^\pm_{\bit{{\scriptstyle q'}}}}=
\delta_{\bit{{\scriptstyle q}}\bit{{\scriptstyle q'}}}\,,\qquad
\vev{\Psi^\pm_{\bit{{\scriptstyle
q}}}|\Psi^\mp_{\bit{{\scriptstyle q'}}}}=0\,.
\ee
The dilatation
operator is a self-adjoint operator with respect to the scalar
product \re{scalar_product} and, as a consequence, its eigenvalues
$\gamma_{\bit{{\scriptstyle q}}}^{(0)}$ are real.

The eigenproblem for the one-loop dilatation operator \re{H0-coor} can
be solved exactly thanks to the existence of the hidden, $SL(2)$
invariant conserved charge $q_3^{(0)}$
\be\label{q-H-com}
{}[
\mathbb{H}^{(0)} , q_3^{(0)} ] = {}[
L_1^\alpha+L_2^\alpha+L_3^\alpha, q_3^{(0)} ] = 0\,,
\ee
where we introduced a superscript to indicate that this charge could
acquire higher order perturbative corrections in coupling constant. The
charge $q_3^{(0)}$ is defined as
\be
\label{Q3chargeLO}
q_3^{(0)} = \frac{i}2 [ L_{12}^2, L_{23}^2] = \frac{i}2 [ L_{31}^2, L_{12}^2]\,,
\ee with the two-particle Casimir operators $L_{jk}^2$ given by \re{Casimir}. As
a result, the Schr\"odinger equation \re{H0-coor} turns out to be completely
integrable and the one-loop dilatation operator $\mathbb{H}^{(0)}$ can be mapped
into the Hamiltonian of a completely integrable Heisenberg magnet of length $L=3$
and spins being the generators of the conformal $SL(2)$ group (for a review see
Ref.~\cite{BelBraGorKor04}).

According to \re{q-H-com}, the eigenstates of the dilatation
operator $\mathbb{H}^{(0)}$ can be chosen to diagonalize the
conserved charge
\be\label{q3-eigen}
q_3^{(0)}\Psi_{\bit{\scriptstyle q}} (z_1, z_2, z_3) = q
\,\Psi_{\bit{\scriptstyle q}} (z_1, z_2, z_3)\,.
\ee
Together with
the total conformal spin $J_{123}=3j+N$, the charge $q$ define the
total set of quantum numbers parameterizing the solutions to the
spectral problem \re{H0-coor}, $\bit{q}=(N,q)$. A distinguished
feature of $q_3^{(0)}$ is that it is invariant under cyclic
permutations and changes the sign under permutation of any pair of
fields
\be \label{DiscreteSymmetriesQ3} {}[ \mathbb{P},q_3^{(0)}]
= \{ \mathbb{P}_{12}, q_3^{(0)}\} = 0\,.
\ee
As far as the
dependence of $\mathbb{H}^{(0)}$ on the conserved charge is
concerned, one finds using \re{P-symmetry}
\be\label{H-parity}
\mathbb{H}^{(0)}(q_3^{(0)}) =
\mathbb{P}_{12}\mathbb{H}^{(0)}(q_3^{(0)})\mathbb{P}_{12} =
\mathbb{H}^{(0)}(\mathbb{P}_{12}\,q_3^{(0)}\mathbb{P}_{12}) =
\mathbb{H}^{(0)}(-q_3^{(0)}) \, ,
\ee
that is, the eigenstates carrying opposite values of the conserved
charge have the same energy
\be\label{q-parity} \Psi_{-\bit{\scriptstyle q}} (z_i)
=\mathbb{P}_{12}\,\Psi_{\bit{\scriptstyle q}}(z_i)\,,\qquad
\gamma_{\bit{\scriptstyle q}}^{(0)} = \gamma_{-\bit{\scriptstyle
q}}^{(0)}\,,
\ee
where $\bit{-q}=(N,-q)$. For $q=0$ one finds from
\re{q-parity} that $\Psi_{\bit{\scriptstyle q}=0}(z_i)$ has the
parity $+1$ and, therefore, $\Psi^-_{\bit{{\scriptstyle
q}}}(z_i)=0$. For $q\neq 0$, it follows from \re{q3-eigen} and
\re{q-parity} that the operator $q_3^{(0)}$ maps into each other
``degenerate'' eigenstates with positive and negative parity
\be\label{q3-map}
q_3^{(0)}\Psi^\pm_{\bit{\scriptstyle q}} (z_1,
z_2, z_3) = q \,\Psi^\mp_{\bit{\scriptstyle q}} (z_1, z_2, z_3)\,.
\ee
We already observed that the symmetry of the dilatation operator under
permutations of fields leads to the degeneracy of eigenstates with nonvanishing
quasimomentum. According to \re{q-parity}, integrability extends this property to
the eigenstates with $\theta_{\bit{\scriptstyle q}} =0$ and $q\neq 0$. The
eigenstate with $q=0$ is the only one which is not paired. It has the
quasimomentum $\theta_{\bit{\scriptstyle q}=0}=0$ and only appears for even $N$.

The exact solution to the eigenproblem is given by the Bethe
Ansatz technique. For given $N$, the eigenspectrum is determined
by a set of $N$ real numbers $\lambda_1,\ldots,\lambda_N$, the
so-called Bethe roots, satisfying the system of transcendental
equations
\be\label{ABA1}
\left(\frac{\lambda_n+i}{\lambda_n-i}\right)^3 =
\prod_{k=1,k\neq n}^N
\frac{\lambda_n-\lambda_k-i}{\lambda_n-\lambda_k+i}\,,\qquad
(n=1,\ldots,N)\,.
\ee
The corresponding exact values of the ``energy'', quasimomentum and the conserved
charge are given, respectively, by
\ba\label{ABA2}
\gamma^{\scriptscriptstyle (0)}_{\bit{\scriptstyle q}} =  \sum_{k=1}^N
\frac2{\lambda_k^2+1}\,, \qquad \e^{i\theta_{\bit{\scriptstyle q}}} =
\prod_{k=1}^N \frac{\lambda_k-i}{\lambda_k+i}\,, \qquad q = - 2 \Im
\prod_{k=1}^N\left(1- \frac{i}{\lambda_k}\right)\,.
\ea
We remind that cyclic symmetry of the three-particle states imposes the
selection rule $\theta_{\bit{\scriptstyle q}}=0$. The results of the
calculation of the one-loop anomalous dimension
$\gamma^{\scriptscriptstyle (0)}_{\bit{\scriptstyle q}}$ for
$0 \le N \le 20$ is shown in Fig.~\ref{2LoopSpectrumFig}. Although
$\gamma^{\scriptscriptstyle (0)}_{\bit{\scriptstyle q}}$ can not
be written in a closed form, one can work out asymptotic
expressions which approximate the exact result with a good
accuracy \cite{Kor95,BraDerMan98,Bel99,DerKorMan00}.

%%%%%%%%%%%%%%%%%%%%%%%%%%%%%%%%%%%%%%%%%%%%%%%%%%%%%%%%%%%%%%%%%%%%%
\subsubsection{Criterium of higher loop integrability}
\label{CriteriumSection}
%%%%%%%%%%%%%%%%%%%%%%%%%%%%%%%%%%%%%%%%%%%%%%%%%%%%%%%%%%%%%%%%%%%%%

Integrability played a vital role in the diagonalization of the
one-loop dilatation operator. It appeared atop of the conformal
symmetry and simplified significantly the solution of the spectral
problem \re{H0-coor} but its origin remained unclear. The question
arises whether this symmetry will survive to higher loops in QCD
and SYM theories. If integrability were ultimately tied to the
conformal symmetry then one should expect, based on the discussion
in Sect.~\ref{ConfSymmBreakSection}, that both symmetries are
broken starting from two loops in gauge theory with
$\beta(g^2)\neq 0$. As we will argue below this does not happen
and integrability survives the conformal symmetry breaking.

To two-loop order, the dilatation operator takes the form
$\mathbb{H} = \lambda \mathbb{H}^{(0)} + \lambda^2
\mathbb{H}^{(1)} + \mathcal{O} (\lambda^3) $, Eq.~\re{H-dec} with
the two-loop correction given in the momentum representation by
\re{V-two-loop}. The eigenspectrum of $\mathbb{H}^{(0)}$ was
determined in the previous section. Thinking about $\mathbb{H}$ as
a quantum mechanical Hamiltonian, one can apply the perturbation
theory to calculate the corrections to the eigenspectrum of the
two-loop dilatation operator induced by the ``perturbation''
$\mathbb{H}^{(1)}$. In spite of the fact that $\mathbb{H}$ does
not respect the conformal symmetry, the eigenfunctions of the
one-loop dilatation operator provide a complete set of functions
endowed with the scalar product \re{scalar_product}. In other
words, the eigenstates of the all-loop dilatation operator
$\mathbb{H}$ can be decomposed over the leading order eigenstates
$\Psi_{\bit{\scriptstyle q}} (z_i)$. Since $\mathbb{H}$ has the
same symmetry under cyclic permutations of particles as
$\mathbb{H}^{(0)}$, this decomposition involves the states with
the same quasimomentum $\theta_{\bit{\scriptstyle q}}$. As before,
the eigenstates of $\mathbb{H}$ with $\theta_{\bit{\scriptstyle
q}}\neq 0$ are double-degenerate while for
$\theta_{\bit{\scriptstyle q}}=0$ the degeneracy can be lifted in
general.

Applying the conventional perturbation theory we find that the
eigenstate of $\mathbb{H}$ is given by
\be\label{Psi1-naive}
\Psi^{(1)}_{\bit{\scriptstyle q}}(z_i) = \Psi_{\bit{\scriptstyle
q}} (z_i) + \lambda \sum_{\bit{\scriptstyle q'}\neq
\bit{\scriptstyle q}} \frac {\vev{\Psi_{\bit{\scriptstyle q'}}|
\mathbb{H}^{(1)}|\Psi_{\bit{\scriptstyle
q}}}}{\gamma_{\bit{{\scriptstyle q}}}^{\scriptscriptstyle
(0)}-\gamma_{\bit{{\scriptstyle q'}}}^{\scriptscriptstyle (0)}}
\Psi_{\bit{\scriptstyle q'}} (z_i)+ \mathcal{O}(\lambda^2)\,.
\ee
However, due to degeneracy of the spectrum \re{q-parity}, the
expansion is not well-defined for $\bit{q'}=-\bit{q}$. We recall
that the paired eigenstates have opposite parity,
$\Psi^\pm_{\bit{\scriptstyle q}} (z_i)$, Eq.~\re{DefParityStates}.
The perturbation $\mathbb{H}^{(1)}$ preserves the parity,
$[\mathbb{H}^{(1)}, \mathbb{P}_{12}] =0$, and, therefore, the
eigenstates with positive and negative parity get decoupled from
each other
\be \vev{\Psi^\pm_{\bit{\scriptstyle q'}}|
\mathbb{H}^{(1)}|\Psi^\mp_{\bit{\scriptstyle q}}} = 0\,.
\ee
Then, in the sector with a definite parity, the energy levels of
$\mathbb{H}^{(0)}$ are {\sl not} degenerate and the conventional
formulas are at work. In particular, the two-loop correction to
the eigenvalue of the dilatation operator with positive/negative
parity is given by
\be \gamma^{(1),\pm}_{\bit{\scriptstyle q}} =
\vev{\Psi^\pm_{\bit{\scriptstyle
q}}|\mathbb{H}^{(1)}|\Psi^\pm_{\bit{\scriptstyle q}}}\,,
\ee
with the normalization condition \re{parity-norm}. Substituting
$\Psi^\pm_{\bit{ \scriptstyle q}}(z_i)$ by their expressions
\re{DefParityStates} the matrix element can be evaluated as
\be
\vev{\Psi_{\bit{\scriptstyle q}}^{ \pm } |\mathbb{H}^{(1)}|
\Psi_{\bit{\scriptstyle q}}^{ \pm }} =
\vev{\Psi_{\bit{\scriptstyle q}} |\mathbb{H}^{(1)}|
\Psi_{\bit{\scriptstyle q}}} \pm \vev{\Psi_{\bit{\scriptstyle q}}
|\mathbb{H}^{(1)}| \Psi_{-{\bit{\scriptstyle q}}}}\,,
\ee
where we took into account that $\vev{\Psi_{-{\bit{\scriptstyle q}}}
|\mathbb{H}^{(1)}|\Psi_{-{\bit{\scriptstyle q}}}}=
\vev{\Psi_{\bit{\scriptstyle q}}
|\mathbb{P}_{12}\mathbb{H}^{(1)}\mathbb{P}_{12}|\Psi_{\bit{\scriptstyle
q}}} =\vev{\Psi_{\bit{\scriptstyle
q}}|\mathbb{H}^{(1)}|\Psi_{\bit{\scriptstyle q}}}$. Thus, the
two-loop correction to the dilatation operator lifts the leading
order degeneracy, $\gamma^{(0),\pm}_{\bit{\scriptstyle
q}}=\gamma^{(0)}_{\bit{\scriptstyle q}}$, Eq.~\re{degeneracy},
\be\label{lift}
\gamma^{(1),+}_{\bit{\scriptstyle
q}}-\gamma^{(1),-}_{\bit{\scriptstyle q}} =
2\vev{\Psi_{\bit{\scriptstyle
q}}|\mathbb{H}^{(1)}|\Psi_{-{\bit{\scriptstyle q}}}} \,.
\ee
We expect that $\gamma^{(1),+}_{\bit{\scriptstyle q}}
= \gamma^{(1),-}_{\bit{\scriptstyle q}}$ for the eigenstates with
$\theta_{\bit{\scriptstyle q}}\neq 0$. Indeed, examining the identity
$\vev{\Psi_{\bit{\scriptstyle q}}|[\mathbb{P},
\mathbb{H}^{(1)}]|\Psi_{-{\bit{\scriptstyle q}}}}=0$ it is easy to see that the
matrix element in the right-hand side of \re{lift} vanishes for
$\theta_{\bit{\scriptstyle q}}\neq 0$. Thus, the two-loop correction to the
dilatation operator could only affect the pairing of the eigenstates with the
quasimomentum $\theta_{\bit{\scriptstyle q}} = 0$.

The necessary and sufficient condition for pairing to persist to two loops is
(for $q\neq 0$)
\be\label{zero1}
\vev{\Psi_{\bit{\scriptstyle
q}}|\mathbb{H}^{(1)}|\Psi_{-{\bit{\scriptstyle q}}}} = 0\,.
\ee
It turns out that this relation implies the existence of the
conserved charge
\be\label{q3-dec}
q_3 = q_3^{(0)} + \lambda
q_3^{(1)} + \mathcal{O}(\lambda^2)\,,
\ee
which has the same discrete symmetry as the leading order charge,
$[ \mathbb{P}, q_3]
= \{ \mathbb{P}_{12}, q_3\} = 0$ and commutes with the dilatation
operator $[q_3,\mathbb{H}]=0$. When expanded in powers of
$\lambda$, the last relation reads
\be\label{q3-2-loop}
[q_3^{(0)},\mathbb{H}^{(1)}] + [q_3^{(1)},\mathbb{H}^{(0)}] = 0\,.
\ee
Going over to the matrix elements one obtains
\be\label{mat-elem}
(q - q') \vev{\Psi_{\bit{\scriptstyle
q}}|\mathbb{H}^{(1)}|\Psi_{{\bit{\scriptstyle q}}'}} =
(\gamma_{\bit{\scriptstyle q}}^{(0)}-\gamma_{{\bit{\scriptstyle
q}}'}^{(0)}) \vev{\Psi_{\bit{\scriptstyle
q}}|q_3^{(1)}|\Psi_{{\bit{\scriptstyle q}}'}}\,.
\ee
One puts $q'=-q$, applies \re{q-parity} and arrives at \re{zero1}.
Eq.~\re{mat-elem} can be used to determine the matrix elements
$\vev{\Psi_{\bit{\scriptstyle q}}|q_3^{(1)}
|\Psi_{{\bit{\scriptstyle q}}'}}$ for $q\neq \pm q'$. For $q=\pm
q'$ both sides of the relation \re{mat-elem} vanish
simultaneously. The reason for this is that \re{q3-2-loop} defines
the charge $q_3^{(1)}$ up to an addition of an operator commuting
with the one-loop dilatation operator. This ambiguity can be fixed
by imposing additional conditions on $q_3^{(1)}$. For instance, we
will show below that $q_3^{(1)}$ can be defined in such a way that
the two-loop charge \re{q3-dec} has the same eigenvalues as the
one-loop charge while its eigenstates are modified by
$\sim\lambda$ corrections.

Let us examine expressions for the eigenstates of two-loop
dilatation operator
\be\label{Psi-1} \Psi_{\bit{\scriptstyle
q}}^{(1),\pm}(z_i) = \Psi_{\bit{\scriptstyle q}}^{\pm}(z_i) +
\lambda \sum_{q'\neq q} \frac{\vev{\Psi_{{\bit{\scriptstyle
q}}'}^{\pm} |\,\mathbb{H}^{(1)} | \Psi_{\bit{\scriptstyle
q}}^{\pm}} }{ \gamma_{\bit{\scriptstyle q}}^{(0)} -
\gamma_{{\bit{\scriptstyle q}}'}^{(0)}} \Psi_{{\bit{\scriptstyle
q}}'}^{\pm}(z_i) \equiv (\II + \lambda \mathbb{Z}
)\,\Psi_{\bit{\scriptstyle q}}^{\pm}(z_i) \, ,
\ee
where in distinction with \re{Psi1-naive} the sum in the right-hand
side is well-defined for all $q'$. Here the notation was introduced for
the operator $\mathbb{Z}$
\be\label{Z-def} \mathbb{Z}
\ket{\Psi_{\bit{\scriptstyle q}}^{\pm}} = \sum_{q'\neq q}
\frac{\vev{\Psi_{{\bit{\scriptstyle q}}'}^{\pm}|\,\mathbb{H}^{(1)}
|\Psi_{\bit{\scriptstyle q}}^{\pm}} }{ \gamma_{\bit{\scriptstyle
q}}^{(0)}-\gamma_{{\bit{\scriptstyle q}}'}^{(0)}}
\ket{\Psi_{{\bit{\scriptstyle q}}'}^{\pm}} \, .
\ee
It is easy to verify that this operator is antihermitian, $(\mathbb{Z})^\dagger =
- \mathbb{Z}$.

The operator $(\II + \lambda \mathbb{Z})$ has a simple physical meaning---it
rotates the one-loop eigenstates into the two-loop ones. Let us perform the
following unitary transformation of the two-loop Hamiltonian
\be\label{tilde-H} \widetilde{\mathbb{H}} =
(\II - \lambda \mathbb{Z})\mathbb{H}(\II + \lambda \mathbb{Z}) =
\mathbb{H}^{(0)} + \lambda\left(\mathbb{H}^{(1)}+
[\mathbb{H}^{(0)}, \mathbb{Z}]\right) + \mathcal{O}(\lambda^2)\,.
\ee
By construction, $\widetilde{\mathbb{H}}$ has the same
eigenvalues as the two-loop Hamiltonian $\mathbb{H}$ whereas the
corresponding eigenstates coincide with the one-loop eigenstates
$\widetilde{\mathbb{H}}\,\ket{\Psi_{\bit{\scriptstyle q}}^{\pm}} =
\gamma_{\bit{\scriptstyle q}}^{\pm}\ket{\Psi_{\bit{\scriptstyle
q}}^{\pm}}$ with $\gamma_{\bit{\scriptstyle q}}^{\pm} =
\gamma_{\bit{\scriptstyle q}}^{(0)}+\lambda
\gamma_{\bit{\scriptstyle q}}^{(1),\pm}+\mathcal{O}(\lambda^2)$.
Combining this relation together with \re{q-parity} one obtains
\be
(\widetilde{\mathbb{H}}\, q_3^{(0)}
-
q_3^{(0)}\,\widetilde{\mathbb{H}})\ket{\Psi_{\bit{\scriptstyle q}}^{\mp}}
=
q (\gamma_{\bit{\scriptstyle q}}^\pm
-
\gamma_{\bit{\scriptstyle q}}^\mp)
\ket{\Psi_{\bit{\scriptstyle q}}^{\pm}}
=
\left[\lambda q (\gamma_{\bit{\scriptstyle q}}^{(1),\pm}
-
\gamma_{\bit{\scriptstyle q}}^{(1),\mp})
+
\mathcal{O}(\lambda^2) \right]
\ket{\Psi_{\bit{\scriptstyle q}}^{\pm}}\,.
\ee
If the degeneracy of eigenvalues survives to two loops, $q
(\gamma_{\bit{\scriptstyle q}}^{(1),\pm}-\gamma_{\bit{\scriptstyle
q}}^{(1),\mp}) = 0$, then the operators $\widetilde{\mathbb{H}}$
and $q_3^{(0)}$ commute leading to
\be
[\widetilde{\mathbb{H}},q_3^{(0)}]=[{\mathbb{H}},(1+\lambda\mathbb{Z})q_3^{(0)}
(1-\lambda\mathbb{Z})]=0+\mathcal{O}(\lambda^2)\,.
\ee
We deduce from this relation that the pairing of eigenvalues allows us to
construct the operator satisfying \re{q3-2-loop} and \re{q3-dec}
\be
\label{q3-rot}
q_3 = q_3^{(0)} - \lambda [q_3^{(0)}, \mathbb{Z} ]
+\mathcal{O}(\lambda^2)\,.
\ee
Taking into account \re{Z-def} and \re{q3-map}, it is straightforward to verify
that the operator $q_3^{(1)} = - [q_3^{(0)}, \mathbb{Z} ]$ satisfies \re{mat-elem}
and has the same discrete symmetries, Eq.~\re{DiscreteSymmetriesQ3}, as the leading
term $q_3^{(0)}$. By construction, the operator $q_3$ commutes (up to higher order
corrections) with the two-loop Hamiltonian. Since $q_3$ is obtained from $q_3^{(0)}$
by the same unitary transformation as in \re{tilde-H}, its eigenvalues do not receive
perturbative corrections while the eigenstates are given by \re{Psi-1}
\be
\label{q3-gauge}
q_3\ket{\Psi^{(1)}_{\bit{\scriptstyle q}}} =
\left[q+\mathcal{O}(\lambda^2)\right]\ket{\Psi^{(1)}_{\bit{\scriptstyle
q}}}\,.
\ee
We recall that the definition of $q_3$ is ambiguous since one can fulfill
\re{q3-2-loop} by adding to $q_3^{(1)}$ an arbitrary function of $q_3^{(0)}$.
Eq.~\re{q3-gauge} corresponds to a particular choice of the `gauge' in which
eigenvalues of $q_3$ do not receive radiative corrections to two loops.

In an analogous manner one finds the higher order analogue $q_2$ of the quadratic
conformal Casimir $q_2^{(0)}$, Eq.~\re{ThreePartConfSpinChange}. It is determined
by the same rotation matrix $\mathbb{Z}$ such that $q_2 = q_2^{(0)} - \lambda
[q_2^{(0)}, \mathbb{Z} ] + \mathcal{O}(\lambda^2)$. Notice however that the
charge $q_2$, having an opposite permutation symmetry compared to $q_3$, does not
require the pairing of eigenvalues with opposite parity for its commutativity
with the Hamiltonian $\mathbb{H}$ to hold. As a consequence of our construction,
we found a unitary transformation which gives the two-loop Hamiltonian
$\widetilde{\mathbb{H}}$ which can be diagonalized simultaneously with mutually
commuting leading order charges $q_2^{(0)}$ and $q_3^{(0)}$.

We would like to stress that existence of $q_3$ is tied solely to degeneracy of
the eigenvalues of the two-loop Hamiltonian $\mathbb{H}$. The same relation works
in the opposite direction. If the Hamiltonian $\mathbb{H}$ possesses the conserved
charge $q_3$ which is odd under permutation of particles then the degeneracy of the
eigenspectrum follows in the same manner as at the leading order, Eq.~\re{H-parity}.

The commutativity of the rotated two-loop Hamiltonian $\widetilde{\mathbb{H}}$
with the quadratic Casimir $q_2^{(0)}$ does not imply the restoration of the
conformal symmetry. The persistence of the latter would require that the local
Wilson operators $\mathcal{O}_{\bit{\scriptstyle q}} (0)$, Eq.~\re{B-expansion},
corresponding to the eigenstates \re{Psi-1} have to satisfy the relations similar
to \re{rest}, that is $[\mathcal{O}_{\bit{\scriptstyle q}} (0),\mathbb{L}^+]=0$
and $[\mathcal{O}_{\bit{\scriptstyle q}} (0),\mathbb{L}^0]\sim
\mathcal{O}_{\bit{\scriptstyle q}} (0)$. Notice that $\mathbb{H}$ defines
radiative corrections to $\mathbb{L}^0$ and the second relation is indeed
satisfied in the basis \re{Psi-1}. The question thus boils down to the study of
$\mathbb{L}^+$. A simple consideration demonstrates that the higher order
corrections in theories with nonvanishing $\beta-$function lead to
$[\mathcal{O}_{\bit{\scriptstyle q}} (0),\mathbb{L}^+]\sim
\beta(g^2)$ (see, e.g., \cite{BraKorMul03,BelRad05}). For the sake of the argument let us
discuss the well-studied twist-two sector, which was explored in detail in
Sect.~\ref{ConfSymmBreakSection}. As we have shown there, for $\beta = 0$, one
can bring the Ward identities for both $\mathbb{L}^0$, Eq.~\re{CWI0}, and
$\mathbb{L}_+$, Eq.~\re{CWI+}, to the needed form \re{rest} by going to the
$\overline{\rm CS}$ scheme with the transformation matrix $\mathcal{B}$,
\re{B-gamma} and \re{B-gamma-c}, respectively. Thus, the rotated operator
$\widehat{\mathcal{O}}_{nn}$ verifies \re{rest} and defines the lowest weight
vector of the irreducible representation of the conformal group. However, in case
when the $\beta-$function is nonzero, while the $\mathbb{L}^0$ operator can still
be diagonalized with the rotation matrix $\mathcal{B}$, the commutator of the
conformal boost $\mathbb{L}^+$ with the rotated Wilson operator
$\widehat{\mathcal{O}}_{nn}$ is proportional to $\beta(g^2)$ and, therefore,
$\widehat{\mathcal{O}}_{nn}$ does not represent the lowest weight. The same
phenomenon re-occurs for three-particle operators -- while one is able to
diagonalize $\mathbb{L}^0$, with $\mathbb{Z}$, it is not the case for
$\mathbb{L}^+$ unless $\beta = 0$.

%%%%%%%%%%%%%%%%%%%%%%%%%%%%%%%%%%%%%%%%%%%%%%%%%%%%%%%%%%%%%%%%%%%%%
\subsubsection{Numerical diagonalization of the two-loop dilatation operator}
\label{NumericsSection}
%%%%%%%%%%%%%%%%%%%%%%%%%%%%%%%%%%%%%%%%%%%%%%%%%%%%%%%%%%%%%%%%%%%%%

We argued in the previous section that the degeneracy of the eigenvalues of
the two-loop dilatation operator could serve as a criterium for two-loop
integrability. Let us now use the expressions for the two-loop dilatation
operators obtained in Sect.~\ref{DilOpTwoLoop} to evaluate their eigenspectrum
and verify whether the pairing of the eigenvalues present at one-loop in QCD
and SYM theories survives to two loops.

We recall that the two-loop evolution kernels have been evaluated
in the momentum representation. In this representation the
eigenvalue problem looks like
\be
\label{OffForwardEigenProblem}
\int [du]_3 \mathbb{V} ( \bit{u}| \bit{v}) \,
P_{\bit{\scriptstyle q}}(\bit{u}) = - \gamma_{\bit{\scriptstyle q}}
(\lambda)\,P_{\bit{\scriptstyle q}} (\bit{v}) \, ,
\ee
where the integration measure was defined in Eq.\ \re{3-measure}, while
the integration region over the $u-$variables is determined by the support
properties of the evolution kernel $\mathbb{V} ( \bit{u}| \bit{v})$. Notice
that the eigenvalues $\gamma_{\bit{\scriptstyle q}} (\lambda)$ do not depend
on the ``external'' momenta $\bit{v} = (v_1, v_2, v_3)$. One can explore this
fact to simplify the actual calculation of the anomalous dimensions. Indeed,
the calculation of the two-loop evolution kernel has been performed in
Sect.~\ref{DilOpTwoLoop} in the kinematics $0\le v_1,v_2,v_3 \le 1$ and
$v_1+v_2+v_3=1$. In this case, the phase space for integrated $u-$momenta is
the one shown in Fig.~\ref{region}.

As explained in Sec.\ \ref{ConfSymmBreakSection} for two-particle operators,
the conformal symmetry is broken beyond leading order in gauge coupling and
so the eigenfunctions, depending on the quantum number $N$, of the dilatation
operator are changed by $\mathcal{O} (\lambda)$ corrections. This symmetry
breaking is induced by the renormalization prescription of the composite
operators. In fact the breakdown of special conformal symmetry arises
already at one-loop level and this determines via the conformal constraint
(\ref{CommutatorConstraint}) the eigenfunctions of the two-loop dilatation
operator. Hence, one can get rid of this conformal symmetry breaking at this
order by means of a finite renormalization of the operator to one-loop accuracy.
It can be shown in general that the known special conformal anomaly in the
two-particle subchannels to one-loop accuracy together with the conformal
constraint, similar to Eq.\ (\ref{CommutatorConstraint}), is consistent with
our result (\ref{V-two-loop}).

For the $\beta(g)=0$ case, the finite renormalization alluded to above yields
the anomalous dimension matrix which is identical to the one found by means of
taking the forward limit. Therefore, the simplest way to calculate the
eigenvalues of the two-loop evolution kernel within the conformal subtraction
scheme is to take the forward limit of the kernel, i.e., impose the following
condition on the light-cone particle momenta
\be
\label{ForwardThreeParticle}
u_1 + u_2 + u_3 = v_1 + v_2 + v_3 = 0
\, .
\ee
As was already explained in
Sect.~\ref{TwistTwoTwoLoopADs}, the forward limit effectively
corresponds to taking forward matrix elements of three-particle
operators. This does not change the eigenspectrum of the
dilatation operator but drastically simplifies the numerical
calculation of moments since the mixing with the operators
containing total derivatives is automatically removed.

In the forward limit, the evolution kernel looks differently as compared to
the one in the off-forward kinematics $\sum_n v_n=1$. It turns out that the
two kernels are related to each other through the following limiting procedure
\cite{MulDitRobGeyHor98}. Let us set
\begin{eqnarray}
u_i = \frac{x_i}{\tau} \, , \qquad v_i = \frac{y_i}{\tau} \, ,
\qquad x_1 + x_2 + x_3 = y_1 + y_2 + y_3 = \tau
\end{eqnarray}
and take the limit $\tau\to 0$. Then, the kernel in the forward
limit reads
\begin{eqnarray}
\label{Def-ForKer} \mathbb{P} (\bit{x}| \bit{y}) = {\rm LIM} \,
\mathbb{V} (\bit{u}| \bit{v}) \equiv \lim_{\tau\to 0}
\frac{1}{\tau^2} \mathbb{V} \left( \frac{x_1}{\tau},
\frac{\tau-x_1-x_3}{\tau}, \frac{x_3}{\tau} \right| \left.
\frac{y_1}{\tau}, \frac{\tau-y_1-y_3}{\tau}, \frac{y_3}{\tau}
\right) \, ,
\end{eqnarray}
where $\bit{x}$ denotes now the variables $x_1$ and $x_3$ (with
$x_2=-x_1-x_3$) and similarly for $\bit{y}$. Before the limit
(\ref{Def-ForKer}) can be performed, the support of the kernel
must be extended. This extension procedure is described in
Appendix~\ref{ForwardLimitAppendix}. Performing the limiting
procedure \re{Def-ForKer} in both sides of the evolution equation
\re{OffForwardEigenProblem} one obtains the eigenproblem in the
forward limit in the following form
\begin{eqnarray}
\int d\mbox{\boldmath $x$}\, \mathbb{P} (\bit{x}| \bit{y})
C_{\bit{\scriptstyle q}} (x_1,x_3) = - \gamma_{\bit{\scriptstyle
q}}(\lambda)\, C_{\bit{\scriptstyle q}} (y_1,y_3) \, ,
\end{eqnarray}
with $d\mbox{\boldmath $x$} = dx_1 dx_3$. Here the eigenvalues
$\gamma_{\bit{\scriptstyle q}}(\lambda)$ are the same as in
\re{OffForwardEigenProblem} while the eigenfunctions are related
as \be C_{\bit{\scriptstyle q}} (y_1,y_3)=\lim_{\tau\to 0} \tau^N
\,P_{\bit{\scriptstyle q}} \left( \frac{y_1}{\tau},
\frac{\tau-y_1-y_3}{\tau}, \frac{y_3}{\tau} \right)=
P_{\bit{\scriptstyle q}} \left( {y_1}, {-y_1-y_3}, {y_3} \right).
\ee In the last relation we used the fact that the eigenstates of
the evolution kernel are homogenous polynomials in momentum
fractions of degree $N\ge 0$. Obviously, the polynomials
$C_{\bit{\scriptstyle q}} (y_1,y_3)$ have the same property. They
are uniquely specified by the set of coefficients in the expansion
of $C_{\bit{\scriptstyle q}} (y_1,y_3)$ over the basis  $y_1^k
y_3^{N-k}$ (with $k=0,\ldots,N$). The total number of these
coefficients is $N+1$ and it matches the total number of
three-particle operators with $N$ derivatives having nonvanishing
matrix elements in the forward limit, i.e., free from total
derivatives. In the basis $y_1^k y_3^{N-k}$  the forward evolution
kernel $\mathbb{P} (\bit{x}| \bit{y})$ is represented by a finite
dimensional mixing matrix
\begin{eqnarray}
\int d\mbox{\boldmath $x$} \, \mathbb{P} (\bit{x} | \bit{y}) \,
x_1^{N-n} x_3^n = \sum_{m=0}^N \Lambda_{nk}(\lambda) \, y_1^{N-k}
y_3^k \, ,
\end{eqnarray}
whose entries depend both on $N$ and the gauge coupling $\lambda$. Then, to find
the spectrum of the anomalous dimensions it suffices to solve the characteristic
equation for the mixing matrix for each value of $N\ge 0$ \be \det \left(\Lambda
- \gamma_{\bit{\scriptstyle q}} (\lambda) \right) = 0\,. \ee Evaluation of the
two-loop mixing matrix in the forward limit is a straightforward but tedious
exercise. For given $N$, the mixing matrix has dimension $N+1$ and, therefore,
the characteristic equation has $N+1$ solutions for $\gamma_{\bit{\scriptstyle
q}}$. All eigenvalues of the mixing matrix are real. This property is a
consequence of the conformal invariance of the dilatation operator to one-loop
order. In addition, some of the eigenstates of the mixing matrix are degenerate.
To one-loop order all eigenstates except the one with $q=0$ are paired and the
eigenspectrum contains $[1+N/2]$ different energy levels. To two-loop order, the
pairing can be only lifted for the eigenstates with vanishing quasimomentum. We
remind that the cyclic symmetry of Wilson operators selects the latter
eigenstates.

%%%%%%%%%%%%%%%%%%%%%%%%%%%%%%%%%%%%%%%%%%%%%%%%%%%%%%%%%%%%%%%%%%%%%
%            Figure
%%%%%%%%%%%%%%%%%%%%%%%%%%%%%%%%%%%%%%%%%%%%%%%%%%%%%%%%%%%%%%%%%%%%%
\begin{figure*}[t]
\begin{center}
\mbox{
\begin{picture}(0,210)(225,0)
\put(0,0){\insertfig{7}{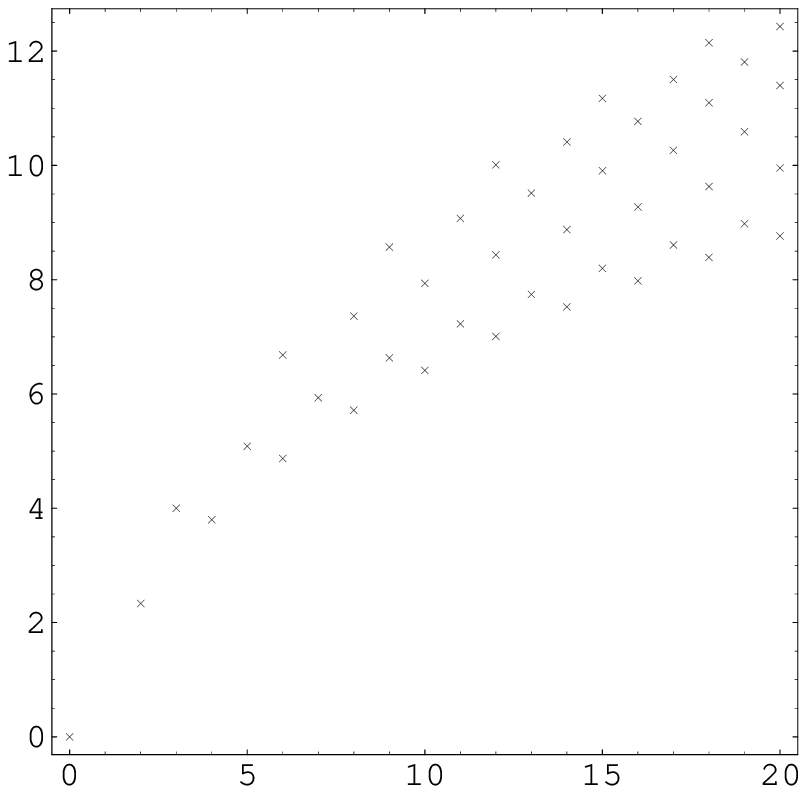}} \put(260,0){\insertfig{7}{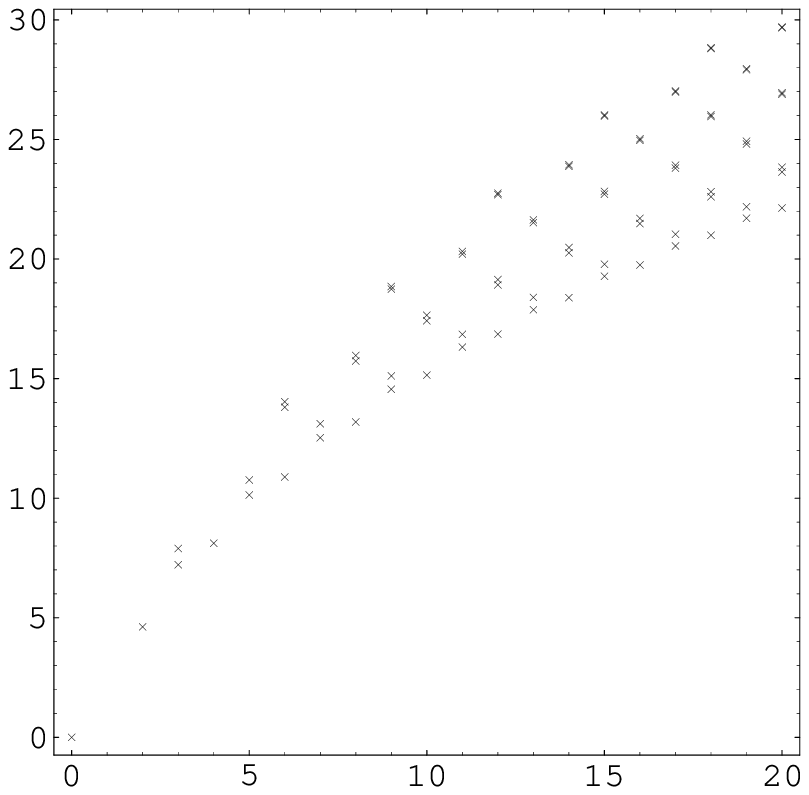}}
\put(-20,170){$\Delta\gamma^{(0)}$} \put(165,-10){$N$} \put(30,175){(a)}
\put(240,170){$\Delta\gamma^{(1)}$} \put(425,-10){$N$} \put(295,175){(b)}
\end{picture}
}
\end{center}
\caption{\label{2LoopSpectrumFig} The spectrum at one-loop order (a) and two-loop
order (b) in QCD in the $\overline{\rm MS}$ scheme. The eigenvalues with
$\theta_{\bit{\scriptstyle q}} \neq 0$ are removed from the spectra for better
comparison with SYM cases where only zero quasimomentum states are physical.}
\end{figure*}
%%%%%%%%%%%%%%%%%%%%%%%%%%%%%%%%%%%%%%%%%%%%%%%%%%%%%%%%%%%%%%%%%%%%%

To begin with, one performs numerical diagonalization of the one-loop mixing
matrix in QCD and SYM theories (see Fig.~\ref{2LoopSpectrumFig} (a)). In both
cases, the one-loop dilatation operator involves the same two-particle kernel
and, therefore, the one-loop mixing matrices are related to each other. One
verifies that, in agreement with the exact results based on the Algebraic Bethe
Ansatz, Eqs.~\re{ABA2} and \re{ABA1}, the one-loop eigenvalues coincide with the
energy levels of the Heisenberg spin magnet. For given total conformal spin
$J_{123}=3j+N$, there are $N+1$ eigenstates. For $N$ odd, all energy levels are
double degenerate while for $N$ even this is true for all levels except the
lowest lying one. The latter eigenstate carries the conserved charge $q=0$ and
carries the vanishing quasimomentum $\theta_{\bit{\scriptstyle q}}=0$. For
eigenstates with quasimomentum $\theta_{\bit{\scriptstyle q}}=\pm \frac23\pi$ the
degeneracy is a consequence of the symmetry of the evolution kernel under
discrete symmetries. The double degeneracy of eig\-en\-values with zero
quasimomentum implies integrability of the one-loop Schr\"odinger equation
\re{H0-coor}. For a given $N$, the total number $m$ of the eigenstates with
$\theta_{\bit{\scriptstyle q}} = 0$ is
\be
m = \frac{1}{3} \left( N - 1\right) + \frac23  \mbox{mod} (N - 1, 3) \, .
\ee
Among them, as we said above, there is one eigenstate with $q = 0$ for even
$N = 2 k$ with $k = 0,1,2, \dots$, while the rest are paired.

\begin{figure}[p]
\begin{center}
\mbox{
\begin{picture}(0,440)(230,0)
\put(11,240){\insertfig{7.1}{loqm0}}
\put(260,240){\insertfig{7.1}{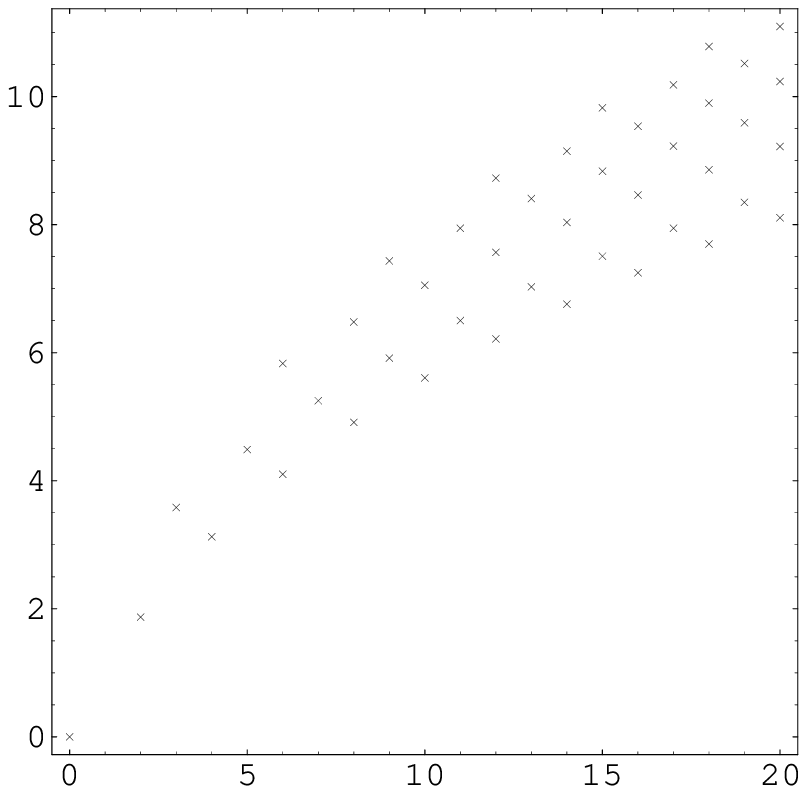}}
\put(0,0){\insertfig{7.5}{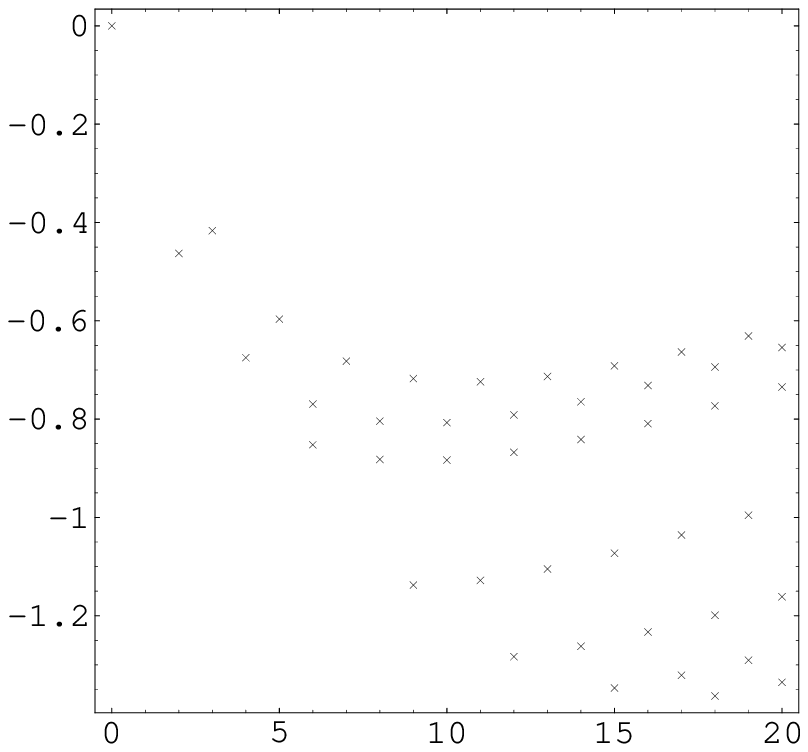}}
\put(255,3){\insertfig{7.3}{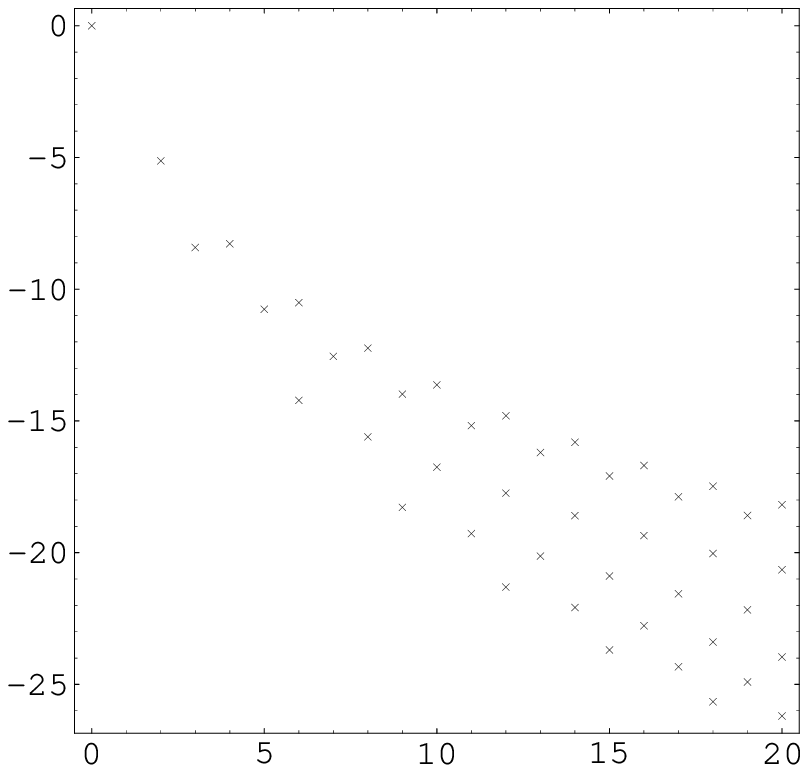}}
\put(-15,405){$\Delta\gamma^{(0)}$}
\put(180,230){$N$}
\put(50,415){(a)}
\put(235,405){$\Delta\gamma^{(1)}$}
\put(430,230){$N$}
\put(300,415){(b)}
\put(-15,180){$\Delta\gamma^{(1)}$}
\put(180,-5){$N$}
\put(50,180){(c)}
\put(235,180){$\Delta\gamma^{(1)}$}
\put(430,-5){$N$}
\put(300,180){(d)}
\end{picture}
}
\end{center}
\caption{\label{2LoopSpectrumSUSY} The spectrum at one-loop order
(a) and two-loop order in SYM theories in the $\overline{\rm DR}$
scheme for $\mathcal{N} = 1$ (b), $\mathcal{N} = 2$ (c) and
$\mathcal{N} = 4$ (d). Only eigenvalues with zero quasimomentum
are displayed.
}
\end{figure}

To two-loop order, the dilatation operators in QCD and SYM theories are given
by two different expressions and, therefore, the corresponding two-loop mixing
matrices are not related to each other in a simple manner. Diagonalizing the
latter in the forward limit, we observe the following remarkable phenomenon
(see Fig.~\ref{2LoopSpectrumFig} (b) and Figs.~\ref{2LoopSpectrumSUSY} (b,c,d)):
\begin{itemize}
\item In QCD, for the baryon operator $\mathbb{O}_{\rm fun}(z_1,z_2,z_3)$, the
pairing is lifted for the eigenvalues with zero quasimomentum but indeed it
persists for the eigenvalues with $\theta_{\bit{\scriptstyle q}} = \pm \frac23\pi$.
\item In SYM theories with arbitrary number of supercharges $\mathcal{N}$, the
one-loop pairing of eigenvalues carries on to two loops for the operator
$\mathbb{O}_{\rm adj}(z_1,z_2,z_3)$;
\end{itemize}
We recall that degeneracy of the eigenstates with nonzero
quasimomentum is not related to integrability.

Presenting the two-loop expressions for the eigenvalues it proves convenient to
introduce
\be
\Delta \gamma^\pm_{\bit{\scriptstyle q}} (\lambda) \equiv
\gamma_{\bit{\scriptstyle q}}^\pm (\lambda) - 3\Gamma (\lambda) \, ,
\ee
where $3\Gamma (\lambda)$ is the anomalous dimension of the local three-fermion
operator without derivatives, $N=0$. The anomalous dimension defined in such a
way satisfies the normalization condition $\Delta \gamma^\pm_{\bit{\scriptstyle
q}} (\lambda)=0$ for $N=0$. We recall that the anomalous dimension $\Gamma
(\lambda)$ has the perturbative expansion \re{GammaPertExpansion} with one- and
two-loop corrections in the $\overline{\rm MS}$ scheme defined in \re{Gamma0},
\re{single-QCD} and \re{single-SYM}. For SYM theories, we use the $\overline{\rm
DR}$ scheme with $\Gamma^{(1), \overline{\rm DR}}_{\scriptscriptstyle\rm SYM}$
related to \re{single-SYM} as
\be
\label{Gamma-2-loop}
\Gamma^{(1), \overline{\scriptscriptstyle\rm DR}}_{\scriptscriptstyle\rm SYM}
=
\Gamma^{(1)}_{\scriptscriptstyle\rm SYM} - \frac{1}{6}
\Gamma^{(0)}_{\scriptscriptstyle\rm SYM} + \frac{1}{4} (4 - \mathcal{N})
=
2 - \frac{1}{2} \mathcal{N} \left( \mathcal{N} - 1 \right)
\, .
\ee
The exact spectrum of eigenvalues $\Delta \gamma^\pm_{\bit{\scriptstyle q}}
(\lambda)$ up to conformal spin $N = 20$ is displayed in Figs.\
\ref{2LoopSpectrumFig} and \ref{2LoopSpectrumSUSY}. They clearly demonstrate the
lifting of the degeneracy in QCD for the baryon operator (\ref{O3-QCD}) compared
to $\mathcal{N}$-extended SYM where it persists for all eigenstates.

The first non-trial example of the pairing phenomenon arises for $N = 3$ and
demonstrates the main trend, recurring for higher $N^{\rm th}$. For $N=3$ there
are four different eigenstates. Two of them have the quasimomentum
$\theta_{\bit{\scriptstyle q}}=0$ while for the remaining two possess the
quasimomentum equal $\theta_{\bit{\scriptstyle q}}=\pm 2\pi/3$. As was already
explained, the latter eigenstates are degenerate to two loops but the
corresponding Wilson operators are not cyclically symmetric. The pair of the
eigenstates with $\theta_{\bit{\scriptstyle q}}=0$ is degenerate in SYM theories
for arbitrary $\mathcal{N}$ while in QCD the degeneracy is lifted to two loops.

%%%%%%%%%%%%%%%%%%%%%%%%%%%%%%%%%%%%%%%%%%%%%%%%%%%%%%%%%%%%%%%%%%%%%
%            Figure
%%%%%%%%%%%%%%%%%%%%%%%%%%%%%%%%%%%%%%%%%%%%%%%%%%%%%%%%%%%%%%%%%%%%%
%%%%%%%%%%%%%%%%%%%%%%%%%%%%%%%%%%%%%%%%%%%%%%%%%%%%%%%%%%%%%%%%%%%%%

To illustrate the phenomenon, we first present the explicit expressions for the
anomalous dimensions for $N=3$ in the $\mathcal{N}=1$ SYM theory in the
$\overline{\rm DR}$ scheme%
\footnote{In the $\overline{\rm MS}$ scheme these anomalous dimensions coincide
with Eq.\ (17) of Ref.~\cite{BelKorMul04}. Though, the anomalous dimensions
$\Delta \gamma_{\rm II}^\pm$ are not ``physical'' for cyclically symmetric
operators, we present them to demonstrate that the degeneracy is always present
for eigenvalues with nonzero quasimomentum.}
\ba \Delta \gamma_{\rm I}^\pm (\lambda)
\!\!\!&=&\!\!\!
\lambda \, 4  + \lambda^2
\left[
- \frac{11}{3} + \frac{29}{24} \frac{\beta_0}{C_R}
\right]
\, ,
\label{pair1}
\\
\Delta \gamma_{\rm II}^\pm (\lambda)
\!\!\!&=&\!\!\!
\lambda \frac{13}{4}
+
\lambda^2
\left[-\frac{465}{128} + \frac{199}{192}\frac{\beta_0}{C_R}
\right]
\, , \nonumber
\ea
where ${\beta_0}/{C_R} = 6$, and in QCD in the $\overline{\rm MS}$ scheme
\ba
\Delta \gamma_{\rm I}^+ (\lambda)
\!\!\!&=&\!\!\!
\lambda \, 4 +
\lambda^2 \left[ -\frac{101}{12} +  \frac{29}{24}\frac{\beta_0}{C_R} \right],
\nonumber \\
\Delta \gamma_{\rm I}^- (\lambda)
\!\!\!&=&\!\!\!
\lambda \, 4 +
\lambda^2 \left[ -\frac{291}{32} + \frac{29}{24}\frac{\beta_0}{C_R}
\right],
\label{pair2} \\
\Delta \gamma_{\rm II}^\pm (\lambda)
\!\!\!&=&\!\!\!
\lambda \frac{13}{4}
+
\lambda^2 \left[ -\frac{721}{96} + \frac{199}{192}\frac{\beta_0}{C_R}
\right],
\nonumber
\ea
with ${\beta_0}/{C_R}=\ft{33}2-N_f$. Here $\Delta \gamma_{\rm I}^\pm (\lambda)$
and $\Delta \gamma_{\rm II}^\pm (\lambda)$ correspond  to the eigenstates with
$\theta_{\bit{\scriptstyle q}} = 0$ and $\theta_{\bit{\scriptstyle q}} = \pm 2\pi/3$,
respectively. Thus, for quarks in the fundamental representation, the two
eigenstates with $\theta_{\bit{ \scriptstyle q}} = 0$ have different eigenvalues
starting from two loops, $\Delta \gamma_{\rm I}^+ \neq \Delta \gamma_{\rm I}^- $.
We remind that in QCD the three-quark baryon operator only exists for $N_c = 3$, so
that large-$N_c$ counting is unavailable and the non-planar diagram in Fig.\
\ref{twoloopkernel} (b) contributes and partially leads to breaking of integrability
in QCD at two loops. One immediately sees from the above equations that though the
degeneracy is generally lifted in QCD, the $\beta_0-$terms of the would-be-paired
eigenvalues do coincide. In SYM they do not break integrability either. The fact that
$\beta_0-$terms preserve degeneracy of the eigenstates implies that the two-loop
integrability is not related to conformal symmetry. We shall return to this issue
in Sect.\ \ref{LargeBetaLimitSect}.

As a next step, we examine the expression for the anomalous dimension of paired
eigenstates with vanishing quasimomentum, Eq.~\re{pair1}, in SYM theory with
$\mathcal{N}$ supercharges in the $\overline{\rm DR}$ scheme
\be\label{gamma-step}
\Delta \gamma_{N=3}^\pm (\lambda) = \lambda \, 4  + \lambda^2 \left[
\frac{91}{12}-4\mathcal{N}\right]\,.
\ee
Starting from two loops it depends linearly on the number of supercharges
$\mathcal{N}$ and satisfies \re{N-factorization}. It is easy to see that the two
loop correction to $\Delta \gamma_{\rm I}^\pm (\lambda)$ is positive for
$\mathcal{N}=1$ and negative for $\mathcal{N}=2$ and $\mathcal{N}=4$. As can be
seen from Fig.~\ref{2LoopSpectrumSUSY} the same is true for higher values of
conformal spin $N$.

The two-loop correction to \re{gamma-step} receives contributions from both two-
and three-particle kernels, Eq.~\re{V-two-loop}, and the question arises whether
each of these contributions preserves integrability separately. To this end, one
multiplies three-particle kernels in \re{V-two-loop} by some factor $\xi$ and
examines the $\xi-$dependence of the resulting expressions for the anomalous
dimension. In this way, one finds
\ba
\Delta \gamma_{\rm I}^+ (\lambda)
\!\!\!&=&\!\!\!
\lambda \, 4  + \lambda^2 \left[ \frac5{72}
\xi+\frac{541}{72}-4\mathcal{N}\right]
\, ,
\nonumber\\
\Delta \gamma_{\rm I}^- (\lambda)
\!\!\!&=&\!\!\!
\lambda\, 4  + \lambda^2 \left[ \frac1{18}
\xi+\frac{271}{36}-4\mathcal{N}\right]
\,.
\ea
For $\xi=1$ one recovers \re{gamma-step} while for $\xi\neq 1$ the degeneracy is
lifted. Thus, the two-loop integrability in SYM theories only arises in the sum
of two- and three-particle contributions.%
\footnote{One may also wonder whether integrability would persist if one merely
iterated the twist-two two-loop kernel $\mathbb{H}_{L=2}$ in all pairwise
subchannels. A straightforward analysis demonstrates that the answer to this
question is negative and the degeneracy of eigenvalues with zero quasimomentum is
lifted in this case.}

Let us present the exact two-loop eigenspectrum for a few lowest anomalous
dimensions with $\theta_{\bit{\scriptstyle q}} =0$ in SYM theories with
$\mathcal{N}$ supercharges. For odd conformal spin $N$ all eigenvalues with
zero quasimomentum are paired and we denote the corresponding two-loop anomalous
dimension matrix as $\Delta \gamma^{\pm}_{N}$. For even $N$ there is an additional
unpaired eigenstate with $q=0$ whose energy we denote as $\Delta \gamma_{N}$. Then
one finds
\ba
\Delta \gamma_{N=4} (\lambda)
\!\!\!&=&\!\!\!
\lambda\frac{19}{5}
+
\lambda^2\left[\frac{15581}{2250}-\frac{19}5\mathcal{N}\right]
\, ,
\nonumber \\
\Delta \gamma_{N=5}^\pm (\lambda)
\!\!\!&=&\!\!\!
\lambda\frac{61}{12}
+
\lambda^2 \left[\frac{55123}{5760}-\frac{61}{12}\mathcal{N}\right]
\, ,
\nonumber \\
\Delta \gamma_{N=6}^\pm (\lambda)
\!\!\!&=&\!\!\!
\lambda \frac{401}{60}
+
\lambda^2 \left[\frac{600697}{48000}-\frac{401}{60}\mathcal{N}\right]
\, ,
\nonumber \\
\Delta \gamma_{N=6} (\lambda)
\!\!\!&=&\!\!\!
\lambda\frac{341}{70}
+
\lambda^2 \left[ \frac{55402939}{6174000}-\frac{341}{70}\mathcal{N}\right]
\label{exam}
\, .
\ea

Notice the change of the sign of two-loop correction to the anomalous dimensions
as one increases the amount of supersymmetry present in the model. While for QCD
and $\mathcal{N} = 1$ SYM, the eigenvalues are strictly positive, the trend
starts to change for $\mathcal{N} = 2$ SYM, where the low-spin anomalous
dimensions are negative and decrease and then for higher $N$ the behavior starts
to change and the spectrum recovers to positive values at higher conformal spins.
For $\mathcal{N} = 4$ SYM, the spectrum is entirely negative. This change in the
 tendency can be traced back to the linear in $\mathcal{N}$ term in the right-hand
 side of \re{exam} which in its turn comes from \re{N-factorization}. It provides
 a large negative contribution to the anomalous dimensions for $\mathcal{N} = 2$
 and $\mathcal{N} =  4$, which competes with the positive one from other terms.

%%%%%%%%%%%%%%%%%%%%%%%%%%%%%%%%%%%%%%%%%%%%%%%%%%%%%%%%%%%%%%%%%%%%%
\subsubsection{Large-spin limit}
%%%%%%%%%%%%%%%%%%%%%%%%%%%%%%%%%%%%%%%%%%%%%%%%%%%%%%%%%%%%%%%%%%%%%

One can see from Figs.~\ref{2LoopSpectrumFig} and \ref{2LoopSpectrumSUSY} that
the anomalous dimensions of the three-particle operators occupy a band
\be
\gamma_{\rm min} (\lambda)
\leq
\gamma_{\bit{\scriptstyle q}} (\lambda)
\leq
\gamma_{\rm max} (\lambda)
\, .
\ee
Let us demonstrate that both boundaries scale as $\sim  \ln N$ at large conformal
spins $N \gg 1$ with certain computable overall coefficients.

It is well known that the asymptotics of the anomalous dimensions at large
conformal spins is governed by the terms in the evolution kernel which are
singular for $v_j \to u_j$. This correspond to the situation when the wavelength
of exchanged gluons is large. The behavior of the two-loop evolution kernels
for $v_j \to u_j$ was scrutinized in Sects.\ \ref{Tw2LimitingCases} and
\ref{TwistTwoTwoLoopADs}. We have found in section \ref{ThreeParticleIrrContribution}
that the three-particle irreducible part of the three-particle kernel is not
singular in this limit and, therefore, the leading asymptotic behavior of the
anomalous dimensions for $N \gg 1$ emerges entirely from the two-particle
irreducible kernels \re{IR-asym}. According to \re{V-IR-asym}, the all-loop
evolution kernel $\mathbb{H}(\lambda)$ is given for $v_j \to u_j$ by a remarkably
simple expression
\be
\label{H-large spin}
\mathbb{H}(\lambda) =\frac{1}{2} \Gamma_{\rm cusp} (\lambda)\,\mathbb{H}^{(0)}+
\ldots
\ee
where $\mathbb{H}^{(0)}$ is the one-loop evolution kernel and ellipses denote
terms subleading for $v_j \to u_j$. We would like to stress that this relation
captures the leading $\sim \ln N-$terms in the anomalous dimension and neglects
subleading $\sim N^0$ corrections.

The fact that $\mathbb{H}(\lambda)$ reduces to the one-loop evolution kernel
accompanied by a multiplicative factor of $\Gamma_{\rm cusp}(\lambda)$ implies
that the all-loop dilatation operator is integrable in the large-spin limit both
in QCD and SYM theories. The underlying integrable structure is the one
corresponding to the {\sl classical} Heisenberg magnet \cite{BelGorKor03} with
the Hamiltonian which can be cast into the form
\be
\label{ClassHamiltonian}
\mathbb{H} (\lambda) = \Gamma_{\rm cusp} (\lambda) \ln q_{3, \rm cl}^{(0)}+
O(N^0) \, ,
\ee
with $q_{3, \rm cl}^{(0)}$ being the classical counterpart of the quantum
integral of motion $q_{3}^{(0)}$. This leads to the following asymptotic behavior
of the anomalous dimension
\be
\gamma_{\bit{\scriptstyle q}}(\lambda) = \Gamma_{\rm cusp} (\lambda) \ln q +
O(N^0) \, .
\ee
For $N\gg 1$, the conserved charge scales in the lower part of the spectrum as
$q \sim \mathcal{O} (N^2)$ and in the upper one as $q \sim \mathcal{O} (N^3)$
so that
\be
\gamma_{\rm min} (\lambda) =2\Gamma_{\rm cusp} (\lambda) \ln N
\,,\qquad\gamma_{\rm max} (\lambda) =3\Gamma_{\rm cusp} (\lambda) \ln N \,.
\ee
These relations are in agreement with the results of Ref.~\cite{BelGorKor03}. So
far we studied the dilatation operator for nonlocal light-cone operator of length
$L=3$. It worth mentioning that the relation \re{H-large spin} holds true for
arbitrary $L$.

%%%%%%%%%%%%%%%%%%%%%%%%%%%%%%%%%%%%%%%%%%%%%%%%%%%%%%%%%%%%%%%%%%%%%
\subsubsection{Large $\beta_0-$limit}
\label{LargeBetaLimitSect}
%%%%%%%%%%%%%%%%%%%%%%%%%%%%%%%%%%%%%%%%%%%%%%%%%%%%%%%%%%%%%%%%%%%%%

As follows from \re{pair1} and \re{pair2}, the corrections to the two-loop
anomalous dimensions coming from the terms involving the $\beta-$function
preserve integrability. This may appear counterintuitive if one lands on the
assumption that existence of integrability is intertwined with conformal
symmetry. However, our findings demonstrate that the two phenomena are not
related to one another.

We argued in Sect.\ \ref{Tw2LimitingCases} that one can enhance the contribution
of conformal symmetry breaking terms $\sim \beta_0$ by going over to the formal
limit $\beta_0 \to\infty$ with $\lambda \beta_0 = \ $fixed. In this limit, the
all-loop dilatation kernel is given by one-loop Feynman diagrams with the gluon
propagator dressed by one-loop vacuum polarization bubble insertions. The
resulting expression for the dilatation operator,
Eq.~\re{KernelLargeBetaRenormConfSpin}, is proportional to the one-loop
dilatation operator in which the ``bare'' conformal spin of quark/gaugino fields
$j=1$ is replaced by its renormalized value $j = 1 +
\beta_0 g^2/(32 \pi^2)$.

The same effect can be understood within the $\varepsilon-$expansion as follows
\cite{Vas04}. According to \re{beta-epsilon}, for $D = 4 - 2 \varepsilon$, the
beta-function receives an additional contribution $\sim \varepsilon$. Therefore,
there exists a value of $D = D_{\rm c}$ for which the $D_{\rm c}-$dimensional
$\beta$-function vanishes and the theory becomes conformal for this space-time
dimension. It follows from \re{beta-epsilon} that this occurs for
\be
\varepsilon_{\rm c} = \frac{\beta (g^2)}{2}
\, .
\ee
Since the canonical dimensions of fields depends on $D$, it will propagate into
the value of the conformal spin, Eq.~\re{conf-spin}. For quarks/gauginos, one
gets
\be
\label{RenromalizationConfSpin}
j = \frac{D_{\rm c}}{4} = 1 - \frac{\beta (g^2)}{4} \simeq 1 + \frac{\beta_0
g^2}{32 \pi^2} \, ,
\ee
where in the last step we used the perturbative expansion \re{DefBetaExpansion}.
This precisely matches our finding in Eq.\ \re{KernelLargeBetaRenormConfSpin} for
the renormalized conformal spin. Thus, the evolution kernel in the
large$-\beta_0$ limit coincides with the evolution kernel in the conformal
invariant field theory living in the ``critical'' $D_{\rm c}-$dimensions.%
\footnote{We are grateful to A.~Manashov for proposing this interpretation.}

We conclude that in the large$-\beta_0$ limit, the all-loop dilatation operator
$\mathbb{H}(\lambda)$ is given by the one-loop dilatation operator with the shifted
conformal spin \re{RenromalizationConfSpin}. We recall that the latter operator
coincides with the Hamiltonian of the Heisenberg $SL(2;\mathbb{R})$ spin chain.
From the point of view of integrable models, renormalization of the conformal
spin corresponds to deformation of unitary irreducible spin $j=1$ representation
of the $SL(2) \sim SO(2, 1)$ group to yet another irreducible but not unitary
representation with spin $j=1 + {\beta_0 g^2}/{(32 \pi^2)}$. One can easily
construct the ``all loop-order'' integrals of motion $q_2(\lambda)$ and
$q_3(\lambda)$ for the corresponding system by merely substituting $j$ in Eqs.\
\re{ThreePartConfSpinChange} and \re{Q3chargeLO} by its renormalized value
\re{RenromalizationConfSpin}
\be\label{q-n-rel}
q_n(\lambda) = q_n^{(0)}\big|_{j=1 + \frac{\beta_0 g^2}{32 \pi^2}} = \e^{\lambda
\mathbb{Z}}q_n^{(0)}\e^{-\lambda \mathbb{Z}} \qquad (n=2,3)\,,
\ee
where the notation was introduced for the operator
\be
\mathbb{Z} = \frac{\beta_0}{4 c_{\rm \scriptscriptstyle R}} \frac{d}{d j} \, .
\ee
Expanding \re{q-n-rel} in powers of $\lambda$ one reproduces \re{q3-rot}.

The large $\beta_0-$limit provides a nontrivial example of all-loop dilatation
operator and allows one to trace the deformation of integrable structures to
higher loops. As in the previous section, the above analysis can be immediately
generalized to the nonlocal light-cone operators of arbitrary length $L$.

%%%%%%%%%%%%%%%%%%%%%%%%%%%%%%%%%%%%%%%%%%%%%%%%%%%%%%%%%%%%%%%%%%%%%
\section{Discussion and conclusions}
\label{DiscussionConcl}
%%%%%%%%%%%%%%%%%%%%%%%%%%%%%%%%%%%%%%%%%%%%%%%%%%%%%%%%%%%%%%%%%%%%%

In the present work we have derived the two-loop dilatation operator in the
noncompact $SL(2)$ sector of QCD and SYM theories with $\mathcal{N} = 1, 2, 4$
supercharges. We have chosen the three-quark and three-gaugino maximal-helicity
operators, respectively, as a case of study. The choice was driven by the
closure of the sector under two-loop renormalization and the simplicity of
calculations involved. Supersymmetry allows one to extend our findings to
the entire supermultiplet embodying operators with different particle content.

It is known that the dilatation operator in the aforementioned gauge theories is
integrable to one-loop order in the sector under consideration. The major
motivation for the current study was to verify whether integrability survives to
two loops. To this end we had to use a guiding principle which will definitely
point to its existence. We have used a criterion of double degeneracy of
eigenvalues of the dilatation operator with zero quasimomentum. The eigenvalues
with nonzero quasimomentum are always degenerate due to invariance of the
dilatation operator under discrete symmetries---cyclic and parity permutations.
We found that in QCD the degeneracy is lifted while in SYM theories it persists
to two loop order for arbitrary number of supercharges $\mathcal{N}$. In the
latter case, the spectrum of the two-loop dilatation operator depends on
$\mathcal{N}$. We demonstrated that for the light-cone operator of arbitrary
length $L=2,3,\ldots$ the $\mathcal{N}-$dependence has a remarkably simple form,
Eq.~\re{N-factorization}. This implies that the dilatation operators in the
$SL(2)$ sector of all SYM theories share the same integrability properties.

Making use of the degeneracy to two loops, we constructed the conserved charges
$q_2(\lambda)$ and $q_3(\lambda)$ which commute with the two-loop Hamiltonian and
among themselves. The first term in the perturbative expansion of these charges,
$q_2^{(0)}$ and $q_3^{(0)}$, coincide with the integrals of motion of the $SL(2)$
Heisenberg magnet of length $L=3$ and spin $j=1$ \cite{BraDerMan98,Bel99,DerKorMan00}.
As such, they are given by conformal invariant polynomials in two-particle $SL(2)$
Casimirs $L_{kn}^2$ (with $k\neq n=1,2,3$), Eq.~\re{Casimir}, and can be obtained
from the auxiliary transfer matrix in a standard manner~\cite{TakFad79}.
The interpretation of higher order corrections to $q_2(\lambda)$ and $q_3(\lambda)$
is an extremely nontrivial task both due to breaking of conformal symmetry in two
loops and due to the fact that the $SL(2)$ spin magnet is noncompact. Namely, the
two-particle Casimirs $L_{kn}^2$ are given by infinite-dimensional matrices in the
conventional oscillatory basis and their powers are independent on each other. This
should be contrasted with the compact $SU(2)$ spin$-1/2$ magnet in which case the
spin operators are given by Pauli matrices and all powers of two-particle Casimirs
$L_{kn}^2$ are expressed in terms of unity operators and its first power. As we
noticed in Sect.\ \ref{CriteriumSection}, another possibility to identify
integrable structures to two loops could be to perform a unitary transformation
of the dilatation operator $\mathbb{H}(\lambda)$ and conserved charges $q_n(\lambda)$
with the operator $1 + \lambda \mathbb{Z}$. It removes higher order corrections
from the conserved charges and yields the Hamiltonian $\widetilde{\mathbb{H}}
= \mathbb{H}^{(0)}+\lambda \widetilde{\mathbb{H}}^{(1)}$ which commutes with
$q_n^{(0)}$ (with $n=2,3$)
\be
[q_n^{(0)}, \widetilde{\mathbb{H}}(\lambda)] = 0 + \mathcal{O} (\lambda^2) \, .
\ee
Since the one-loop Hamiltonian is a function of the conserved charges
$\mathbb{H}^{(0)} = \mathbb{H}^{(0)} (q_2^{(0)}, q_3^{(0)})$, we conclude that
the leading and the next-to-leading Hamiltonians are mutually commuting
\be
[ \mathbb{H}^{(0)}_{\phantom{i}} , \widetilde{\mathbb{H}}_{\phantom{i}}^{(1)} ] =
0 \, .
\ee
Thus, $\widetilde{\mathbb{H}}_{\phantom{i}}^{(1)}$ is a ``higher'' Hamiltonian of
the $SL(2)$ Heisenberg spin chain. In principle, it can be constructed from
$\mathbb{H}^{(0)}_{\phantom{i}}$ by applying the boost operator
\cite{Tet82,Skl92} ${B} = \sum_k k \mathbb{H}^{(0)}_{k, k+1}$, that is $[{B} ,
\dots [ {B}, \mathbb{H}^{(0)}] ]$, and forming their linear combination with
suitably adjusted coefficients. In order to meet the parity requirement, only
even powers of ${B}$ may appear in the expansion.

We demonstrated that the integrability is not tied to the survival of the
conformal symmetry at two-loop order. On the contrary, in QCD case, the
beta-function term in the dilatation operator preserves the pairing of
eigenvalues contrary to the conformal invariant part of the Hamiltonian. This
phenomenon carries on to SYM models where the conformal invariant part is
integrable as well.

We found that the nonplanar diagrams provide vanishing contribution to the
dilatation operator for $L=3$-particle operators in SYM models. For $L > 3$, they
induce next-to-nearest neighbor interactions at leading order, which are however
suppressed by $1/N_c^2$ with respect to the leading nearest-neighbor
contributions. These $1/N_c^2-$corrections explicitly break integrability but
this phenomenon can be restored by going to the large-$N_c$ limit. Therefore, the
large-$N_c$ limit is a necessary condition for integrability to survive. Note
that to two-loop order there are nonplanar Feynman graphs (like in Figs.\
\ref{gluonthreeparticle} (b) and \ref{twoloopkernel} (b)). These diagrams were
not considered presently since they are accompanied by a vanishing color
structure for $L = 3$ and also die out in the multi-color limit for $L > 3$. For
the QCD baryon operator the multi-color counting is not applicable since the
corresponding operator exists only for $N_c = 3$. Even if one sets by hand the
nonplanar two-particle graph to zero, the pairing would not be restored due an
intricate interplay between color factors of different Feynman diagrams in the
fundamental and adjoint representations.

The two-loop dilatation operator has a rather complex structure in the momentum
representation. It is however in this representation that higher-loop computation
machinery is developed the best. It would be interesting to translate the
obtained expressions into the coordinate, light-cone position representation and,
eventually, write down the two-loop dilatation operator in terms of light-cone
superfields as it was done to one-loop order in Ref.~\cite{BelDerKorMan04}. In
this form, all symmetries of the dilatation operator are manifest and they may
well hint on its possible structure to higher loops.

%%%%%%%%%%%%%%%%%%%%%%%%%%%%%%%%%%%%%%%%%%%%%%%%%%%%%%%%%%%%%%%%%%%%%%%%%%%%%%%
\section*{Acknowledgements}
%%%%%%%%%%%%%%%%%%%%%%%%%%%%%%%%%%%%%%%%%%%%%%%%%%%%%%%%%%%%%%%%%%%%%%%%%%%%%%%

This work was supported by the U.S.\ National Science Foundation under grant no.\
PHY-0456520 (A.B. and D.M.). We are indebted to W. Vogelsang and F.-M. Dittes for
providing us their notes on two-loop calculations which were indispensable at
early stages of the project. We would also like to thank V. Braun, S. Derkachov
and A. Manashov for useful discussions. A.B. and D.M. would like to thank
Laboratoire de Physique Th\'eorique (Orsay) for hospitality extended to them
during their visit where a part of this work has been done.

\vspace{0.5cm}

\appendix

\setcounter{section}{0} \setcounter{equation}{0}
\renewcommand{\theequation}{\Alph{section}.\arabic{equation}}

%%%%%%%%%%%%%%%%%%%%%%%%%%%%%%%%%%%%%%%%%%%%%%%%%%%%%%%%%%%%%%%%%%%%%
\section{Forward limit}
\label{ForwardLimitAppendix}
%%%%%%%%%%%%%%%%%%%%%%%%%%%%%%%%%%%%%%%%%%%%%%%%%%%%%%%%%%%%%%%%%%%%%

In this Appendix we describe the limiting procedure which allows one to obtain
the three-particle kernel in the forward limit $u_1 + u_2 + u_3 = v_1 + v_2 + v_3
= 0$ from the expression \re{Def-Vcon} defined on the simplices \re{simplex}
and \re{u-simplex}.

As the first step, one extends the expression for the evolution kernel
\re{Def-Vcon} to the entire domain $- \infty < u_1 + u_2 + u_3 = v_1 + v_2 + v_3
< \infty$ by making use of the support properties of defining Feynman diagrams.
In the two-particle subchannels, the procedure is well-known and it amounts to
substituting \cite{MulDitRobGeyHor98}
\begin{eqnarray}
\label{Def-GenTheSubCha}
\theta \left(v_j - u_j \right) \quad \rightarrow \quad \Theta \left(u_a,
v_a\right) \, ,
\end{eqnarray}
with generalized step function defined in \re{step-function}. The analogous
procedure can be formulated for the three-particle irreducible diagram too. As we
found in Sect.~\ref{ThreeParticleIrrContribution}, the three-particle kernel
\re{Def-Vcon} is given by the sum of various $\mathbb{F}_i-$functions which are
accompanied by the following step-function structures
\ba
\vartheta^{(1)}_{123}
\!\!\!&\equiv&\!\!\! \theta (u_1) \theta (u_3 - v_2 - v_3) \theta (1 - u_1 - u_3)
\, , \\
\vartheta^{(2)}_{123}
\!\!\!&\equiv&\!\!\! \theta (u_1) \theta (v_1 - u_1) \theta (u_3 - v_3) \theta (v_2 + v_3 - u_3)
\, , \\
\vartheta^{(4)}_{123}
\!\!\!&\equiv&\!\!\! \theta (u_1 - v_1) \theta (u_3 - v_3) \theta (1 - u_1 - u_3)
\, .
\ea
The remaining two structures $\vartheta^{(i)}_{123}$ for $i = 5,6$ are found from
the above by permutation of arguments as in Eq.~\re{SymmetryRelations}. It turns
out that the generalized step-function structure can be restored through the
following substitutions
\begin{eqnarray}
\label{def-Gen-theta1}
\vartheta_{123}^{(1)}
\!\!\!&\rightarrow&\!\!\!
\Theta_{123}^{(1)}
=
\Theta (u_1, v_1)
\left[
\Theta(u_2, 1 - u_1 - v_3) - \Theta (1 - v_1 - u_3, v_2)
\right]
\, , \\
\label{def-Gen-theta2}
\vartheta_{123}^{(2)}
\!\!\!&\rightarrow&\!\!\!
\Theta_{123}^{(2)}
=
\Theta (u_1, v_1) \Theta(1 - v_1 - u_3, v_2)
\, , \\
\label{def-Gen-theta4}
\vartheta_{123}^{(4)}
\!\!\!&\rightarrow&\!\!\!
\Theta_{123}^{(4)}
=
\Theta (u_2, 1 - v_1 - u_3) \Theta (1 - v_1 - u_3, v_2)
\, .
\end{eqnarray}
Having defined a generalized three-particle kernel in this way, the reduction to
the forward limit can be easily performed with the help of \re{Def-ForKer}. In the
following we assume without loss of generality that $v_1 \equiv y_1/\tau$ and
$v_3 \equiv y_3/\tau$ are positive in \re{Def-ForKer}. Consequently, in the
forward case we find for the building block (\ref{Def-GenTheSubCha}):
\be
{\rm LIM} \, \Theta \left(u_a, v_a\right)
=
\left\{
\begin{array}{ll}
\theta (x_a) \theta (y_a - x_a)
\, ,
&\quad
a = 1, 3
\\
- \theta(- x_1 - x_3) \theta(x_1 + x_3 - y_1 - y_3)
\, ,
&\quad
a = 2
\end{array}
\right.
\, .
\ee
The forward limit of Eqs.\ (\ref{def-Gen-theta1}), (\ref{def-Gen-theta2}), and
(\ref{def-Gen-theta4}) can be easily calculated and, one finds
\begin{eqnarray}
{\rm LIM}\,
\Theta_{123}^{(1)}
\!\!\!&=&\!\!\!
\theta (x_1) \, \theta (y_1 - x_1)\,
\left[ \theta (x_3 + y_1) - \theta (x_1 + x_3) \right] \theta (y_3 - x_3)
\, , \\
{\rm LIM}\, \Theta_{123}^{(2)}
\!\!\!&=&\!\!\!
- \theta (x_1) \, \theta (y_1 - x_1) \, \theta(x_3 + y_1) \, \theta (y_3 - x_3)
\, , \nonumber\\
{\rm LIM}\, \Theta_{123}^{(4)}
\!\!\!&=&\!\!\!
\theta (x_3) \, \theta (y_1 - x_1) \, \theta (x_1) \, \theta(y_3 - x_3)
\, . \nonumber
\end{eqnarray}

%%%%%%%%%%%%%%%%%%%%%%%%%%%%%%%%%%%%%%%%%%%%%%%%%%%%%%%%%%%%%%%%%%%%%
\section{Subtraction terms}
\label{DoublePlusAppendix}
%%%%%%%%%%%%%%%%%%%%%%%%%%%%%%%%%%%%%%%%%%%%%%%%%%%%%%%%%%%%%%%%%%%%%

The two-particle kernel \re{V12-full} receives contribution from the subtraction
terms \re{sub-term} involving the integral of three-particle irreducible kernel
\be
W(u_1,u_2| v_1, v_2) = \int_0^1 du_3'\, \mathbb{V}^{(1)}_{123} (u_1, 1-u_1-u_3',
u_3' | v_1, v_2, v_3)
\ee
The evaluation of the integral is straightforward and leads to
\begin{eqnarray}
\label{Def-DelVcon}
W(u_1,u_2 | v_1,v_2) \!\!\!&=&\!\!\! \theta(\bar{u}_1) \theta (u_1 - v_2 - v_1)
s_1 (u_1, u_2 | v_1, v_2)
\\
&+&\!\!\! \theta(u_2)\theta(v_2 - u_2) s_2 (u_1, u_2 | v_1, v_2) + \theta(u_1)
\theta(v_1 - u_1) s_3 (u_1, u_2 | v_1, v_2) \, , \nonumber
\end{eqnarray}
where the notation was introduced for the functions
\begin{eqnarray}
\label{Def-DelF1}
s_1 (u_1, u_2 | v_1, v_2) \!\!\!&=&\!\!\! - \frac{2 \bar{u}_1 }{v_3 (v_1 - u_1) }
- L(u_1 - \bar{v_3}, v_3, v_1 - u_1)
\\
&-&\!\!\! L(u_1 - \bar{v_3}, v_3, v_2) + L(u_1 - v_1, \bar{v}_1, v_3) + L(u_1 -
v_1, \bar{v}_1, v_2) \, ,
\nonumber\\
\label{Def-DelF2}
s_2 (u_1, u_2 | v_1, v_2) \!\!\!&=&\!\!\! \frac{1}{2 v_2} \left( 4 -
\ln^2\frac{v_2 - u_2}{v_2} \right)
\\
&-&\!\!\! L(u_1 - v_1, v_2, v_3) + L(u_1 - v_1, \bar{v}_1, v_3) + L(u_1 - v_1,
\bar{v}_1, v_2)
\nonumber\\
&-&\!\!\! 2 f (u_2, v_2) \left[ {\rm Li}_2 \left(\frac{u_2}{v_2} \right) -
\zeta(2) \right] + \frac{v_2 + v_1}{v_2 v_1} \left[ {\rm Li}_2 \left(
\frac{u_2}{v_2 + v_1} \right) - {\rm Li}_2 \left( \frac{u_2}{v_2} \right) \right]
\, , \nonumber\\
\label{Def-DelF3}
s_3 (u_1, u_2 | v_1, v_2) \!\!\!&=&\!\!\! f (u_1,v_1) \left[ \frac{1}{2}
\ln^2\frac{u_2}{v_2} + \ln \frac{u_2}{v_2} \ln\frac{u_1}{v_1} + 2 \, {\rm
Li}_2\left(\frac{v_1 - u_1}{u_2}\right) \right]
\\
&+&\!\!\! \frac{v_2 + v_1}{v_2 v_1} \left[- \ln\frac{u_2}{v_2} \ln\frac{v_1}{v_2
+v_1} + {\rm Li}_2\left(\frac{v_1-u_1}{u_2}\right) + {\rm
Li}_2\left(\frac{v_2}{v_2 +v_1}\right) - \zeta(2) \right] \, , \nonumber
\end{eqnarray}
with the $f-$ and $L-$functions defined in \re{f-def} and \re{Def-F1TwoPar},
respectively.

The subtraction term \re{sub-term} provides a genuine contribution to the
two-particle irreducible kernel $\mathbb{V}_{12}$, Eq.~\re{V12-full}. Similar
contribution also comes from the gluon kernel \re{Def-FurSubCon} which involves
the $\mathbb{V}^{(1)}_{\scriptscriptstyle\rm SV}(u,v)-$term. It is instructive to
examine the sum of three terms
\begin{eqnarray}
\Delta \mathbb{V}_{12} (\bit{u} | \bit{v}) \!\!\!&=&\!\!\! W(u_1,u_2| v_1,
v_2)+W(u_2,u_1| v_2, v_1) + \frac{1}{u_1 + u_2}
\mathbb{V}^{(1)}_{\scriptscriptstyle\rm SV}
 \left( \frac{u_1}{u_1 + u_2}, \frac{v_1}{v_1 + v_2} \right) \, .
\end{eqnarray}
Taking into account  Eqs.\ (\ref{Def-DelVcon})--(\ref{Def-DelF3}) and
(\ref{SelEneVer-tra}) we find that various terms cancel against each other
leading to
\begin{eqnarray}
\Delta \mathbb{V}_{12} (\bit{u} | \bit{v}) = \theta (\bar{u}_2) \theta (u_2 - v_1
- v_2) \Delta \mathcal{F}_1 (u_1, u_2 | v_1, v_2) \!\!\!&+&\!\!\! \theta (u_1)
\theta (v_1 - u_1) \Delta \mathcal{F}_2 (u_1, u_2 | v_1, v_2)
\qquad\\
&+&\!\!\! \left\{u_1 \leftrightarrow u_2 \atop v_1 \leftrightarrow v_2\right\} \,
, \nonumber
\end{eqnarray}
with ($v_3 = 1 - v_1 - v_2$)
\begin{eqnarray}
\label{Def-Delf1}
\Delta \mathcal{F}_1 (u_1, u_2 | v_1, v_2) \!\!\!&=&\!\!\! - \frac{u_1 \bar{u}_2
}{v_3 (v_2 - u_2) } - L(u_3 - \bar{v}_1, v_3, v_2 - u_2) - L(u_3 - \bar{v}_1,
v_3, v_2)
\\
&+&\!\!\! L(u_2 - v_2, \bar{v}_2, v_3) + L(u_2 - v_2, \bar{v}_2, v_1)
\, , \nonumber\\
\label{Def-Delfa}
\Delta \mathcal{F}_2 (u_1, u_2 | v_1, v_2) \!\!\!&=&\!\!\! - L(u_2 - v_2, v_1,
v_3) + L (u_2 - v_2, \bar{v}_2, v_3) + L(u_2 - v_2, \bar{v}_2, v_1)
\nonumber\\
&+&\!\!\! \frac{1}{2 v_1} \left( 4 - \ln^2\frac{v_1 - u_1}{v_1} \right) -
\frac{1}{2} f(u_1,v_2) \left[ 3 \ln\frac{u_1}{v_1} + \ln^2\frac{u_1}{v_1}\right]
\, . \nonumber\\
\end{eqnarray}
We observe that it does not involve dilogarithms.

%%%%%%%%%%%%%%%%%%%%%%%%%%%%%%%%%%%%%%%%%%%%%%%%%%%%%%%%%%%%%%%%%%%%%

%%%%%%%%%%%%%%%%%%%%%%%%%%%%%%%%%%%%%%%%%%%%%%%%%%%%%%%%%%%%%%%%%%%%%

\end{document}